\documentclass[%
reprint,
twocolumn,
superscriptaddress,
showpacs,preprintnumbers,
nofootinbib,
aps,
prl,
]{revtex4-1}

\usepackage{subfigure}

\usepackage{amsmath}
\usepackage{amssymb}
\usepackage{mathtools}

\usepackage{hyperref}

\usepackage{graphicx}
\usepackage{dcolumn}
\usepackage{color}

\usepackage{ulem}

\usepackage{comment}
\usepackage{placeins}

\graphicspath{
  {./}
  {./figures/}
}

\def\al{\alpha}

\def\om{\omega}

\newcommand{\ben}{\begin{equation}}
\newcommand{\een}{\end{equation}}
\newcommand{\bea}{\begin{eqnarray}}
\newcommand{\eea}{\end{eqnarray}}
\newcommand{\ba}{\begin{array}}
\newcommand{\ea}{\end{array}}
\newcommand{\bit}{\begin{itemize}}
\newcommand{\eit}{\end{itemize}}

\newcommand{\bv}{\textbf{v}}

\newcommand{\cs}{c_\text{s}} 
\newcommand{\pdof}{g}  
\newcommand{\pdofc}{\pdof_*} 
\newcommand{\TN}{T_\text{n}} 
\newcommand{\Tc}{T_\text{c}} 
\newcommand{\vw}{v_\text{w}} 
\newcommand{\vwass}{v_{\text{w}}} 
\newcommand{\vProfMax}{\vel_\mathrm{p}} 

\newcommand{\Ubp}{\overline{U}_\phi} 
\newcommand{\Ubf}{\overline{U}_\text{f}} 
\newcommand{\Ubfmax}{\overline{U}_\text{f\text{,max}}} 
\newcommand{\UbfExp}{\overline{U}_{\text{f,exp}}}

\newcommand{\VbPerp}{\overline{\vel}_\perp} 
\newcommand{\VbPar}{\overline{\vel}_\parallel} 
\newcommand{\VbTot}{\overline{\vel}} 
\newcommand{\VbTotMax}{\overline{\vel}_\text{max}} 
\newcommand{\VbPerpMax}{\overline{\vel}_{\perp\text{,max}}} 

\newcommand{\potZeroT}{V_\text{0}} 
\newcommand{\potT}{V} 
\newcommand{\potConst}{V_\text{c}} 
\newcommand{\potDelta}{\Delta V_\text{0}} 

\newcommand{\Rc}{R_\text{c}} 
\newcommand{\Rbc}{R_*} 
\newcommand{\HN}{H_\text{n}} 
\newcommand{\Nb}{N_\text{b}}

\newcommand{\quadPar}{M^2}
\newcommand{\cubPar}{\mu}
\newcommand{\relgamma}{\gamma}

\newcommand{\StrParB}{\alpha}

\newcommand{\StrParBMax}{\alpha_{\mathrm{max}}}

\newcommand{\rGW}{\rho_\text{gw}}
\newcommand{\OmGW}{\Omega_\text{gw}}
\newcommand{\OmGWExp}{\Omega_\text{gw,exp}}

\newcommand{\OmGWscaled}{\tilde\Omega_\text{gw}}

\newcommand\vel{v}

\newcommand{\phiAtMin}{\phi_\text{b}}

\newcommand{\EOS}{\om} 

\newcommand{\Vol}{{\mathcal V}}

\newcommand{\tPC}{t_\text{pc}} 

\definecolor{newgreen}{RGB}{10,100,20}
\definecolor{purple}{rgb}{0.5,0,0.5}
\definecolor{BLUE}{rgb}{0,0,1}

\def\lsi{\raise0.3ex\hbox{$<$\kern-0.75em\raise-1.1ex\hbox{$\sim$}}}

\graphicspath{
  {./figures/}
  {./}
}

\begin{document}

\newcommand{\Sussex}{\affiliation{
    Department of Physics and Astronomy,
    University of Sussex, Falmer, Brighton BN1 9QH,
    U.K.
}}

\newcommand{\HIPetc}{\affiliation{
    Department of Physics and Helsinki Institute of Physics,
    PL 64,
    FI-00014 University of Helsinki,
    Finland
}}

\newcommand{\Nottingham}{\affiliation{
    School of Physics and Astronomy,
    University of Nottingham,
    Nottingham NG7 2RD,
    U.K.
}}

\title{Vorticity, kinetic energy, and suppressed gravitational wave production in strong first-order phase transitions
}

\author{Daniel Cutting}
\email{d.cutting@sussex.ac.uk}
\Sussex
\HIPetc
\author{Mark Hindmarsh}
\email{m.b.hindmarsh@sussex.ac.uk}
\Sussex
\HIPetc
\author{David J. Weir}
\email{david.weir@nottingham.ac.uk}
\HIPetc
\Nottingham

\date{\today}

\begin{abstract}
  We have performed the first 3-dimensional simulations of strong
  first-order thermal phase transitions in the early Universe.  For
  deflagrations, we find that the rotational component of the fluid
  velocity increases as the transition strength is increased.  For
  detonations, however, the rotational velocity component remains
  constant and small.  We also find that the efficiency with which
  kinetic energy is transferred to the fluid falls below theoretical
  expectations as we increase the transition strength. The probable
  origin of the kinetic energy deficit is the formation of reheated
  droplets of the metastable phase during the collision, slowing the
  bubble walls.  The rate of increase in the gravitational wave energy
  density for deflagrations in strong transitions is suppressed
  compared to that predicted in earlier work. This is largely
  accounted for by the reduction in kinetic energy.  Current modelling
  therefore substantially overestimates the gravitational wave signal
  for strong transitions with deflagrations, in the most extreme case
  by a factor of $10^{3}$. Detonations are less affected.
\end{abstract}

\preprint{HIP-2019-15/TH}

\maketitle

The Laser Interferometer Space Antenna (LISA), scheduled for launch in
2034, will open the mHz band of the emerging field of gravitational
wave astronomy~\cite{Audley:2017drz}. One of the most exciting goals
of LISA is to probe the early universe by searching for gravitational
wave signals from a first-order phase transition.

While the Standard Model is a cross-over
\cite{Kajantie:1996mn,Kajantie:1996qd}, there are many extensions with
first-order phase transitions. These range from adding a scalar
singlet~\cite{Profumo:2007wc,Espinosa:2011ax,Cline:2012hg,Profumo:2014opa,Beniwal:2018hyi}
or doublet~\cite{Kakizaki:2015wua,Dorsch:2016nrg,Basler:2016obg}, to
models with spontaneously broken conformal
symmetry~\cite{Randall:2006py,PhysRevD.82.083513,Konstandin:2011dr,vonHarling:2017yew,Dillon:2017ctw,Megias:2018sxv,Bruggisser:2018mrt}.
There are also models with phase transitions in hidden
sectors~\cite{Schwaller:2015tja,Addazi:2017gpt,Aoki:2017aws,Croon:2018erz,Breitbach:2018ddu,Okada:2018xdh,Hasegawa:2019amx}. Non-perturbative
methods are sometimes necessary to establish the order of the
phase
transition~\cite{Gorda:2018hvi,Gould:2019qek,Kainulainen:2019kyp}.

An important parameter of a first-order phase transition is the trace
anomaly difference, which quantifies the energy available for
conversion to shear stress, and hence the power of the gravitational
wave signal.  If the trace anomaly difference is comparable to the
radiation energy density of the universe, we call the transition
`strong'.  We denote the ratio of the trace anomaly to the thermal
energy $\StrParB$, in which case a strong transition has
$\StrParB \sim 1$. We call $\StrParB \gg 1$ `very strong'; our results
do not access this region.

Substantial progress has been made in understanding gravitational
wave production from first-order transitions with weak
($\StrParB \sim 10^{-2}$) to intermediate ($\StrParB \sim 10^{-1}$)
strength using numerical
simulations~\cite{Hindmarsh:2013xza,Giblin:2014qia,Hindmarsh:2015qta,Hindmarsh:2017gnf},
as well as
modelling~\cite{Hindmarsh:2016lnk,Jinno:2016vai,Konstandin:2017sat}.
While the fluid motion is well-described as a linear superposition of
sound waves after a weak transition \cite{Hindmarsh:2013xza},
rotational modes and turbulence are expected in stronger transitions
\cite{Witten:1984rs,KurkiSuonio:1984ba}, which could substantially
affect the gravitational wave
signal~\cite{Kamionkowski:1993fg,Caprini:2007xq,Gogoberidze:2007an,Caprini:2009yp,Caprini:2009fx,Niksa:2018ofa}.

At the same time, investigation of the underlying particle physics
models indicates that intermediate to strong transitions are
common in conservative extensions of the Standard
Model~\cite{Ellis:2018mja,Ellis:2019oqb}, and very strong transitions
are possible in models of composite Higgs and nearly conformal
potentials~\cite{Randall:2006py,PhysRevD.82.083513,Konstandin:2011dr,vonHarling:2017yew,Dillon:2017ctw,Megias:2018sxv,Bruggisser:2018mrt}.
It is also clear that LISA will be most likely to observe transitions
where nonlinear effects like shocks and turbulence become
important~\cite{Hindmarsh:2017gnf}.  Recent work tackling the
non-linear regime includes gravitational wave production from
magnetohydrodynamic turbulence~\cite{Pol:2019yex} and studies of shock
collisions using a mixture of 1-dimensional simulations and
modelling~\cite{Jinno:2019jhi}.

In this paper, we present results from the first numerical simulations
of strong first-order phase transitions.  We measure the fraction of
the fluid kinetic energy in rotational modes, as traced by the
mean-square velocity.  As we increase the strength of the transition,
this proportion grows substantially for deflagrations, with up to
$ 65 \% $ of the mean square velocity found in rotational motion. The
rotational proportion is far less for detonations, remaining roughly
constant for all transition strengths.

As the transition strength $\al$ is increased, the
efficiency of fluid kinetic energy production decreases below
expectation.  For deflagrations, this is associated with reduced wall
speeds for expanding bubbles and reheating of the region in front
of the walls, reducing the pressure difference
\cite{KurkiSuonio:1984ba,Konstandin:2010dm,Megevand:2017vtb}.  The
kinetic energy loss leads to a suppression in the gravitational wave
power, by a factor which can be as small as O($10^{-3}$).  This means
that current models substantially overestimate gravitational wave
production from strong transitions with deflagrations. Detonations are
less affected.

{We model the phase transition with a real scalar field $\phi$,
  coupled to a perfect fluid.  We assume that there is no extra
  physics generating a magnetic field either before or during the
  phase transition.}  The model follows that used in previous
work~\cite{Ignatius:1993qn,Hindmarsh:2015qta,Hindmarsh:2017gnf},
differing by a change in the effective potential and therefore the
equation of state. Our previous work used the high-temperature
expansion of the one-loop thermal effective potential, and we found
that in stronger transitions, the total energy could drop below the
scalar potential energy, which is unphysical. In this scenario, our
algorithm would compute the temperature to be imaginary, causing a
crash. Indeed, the high-temperature expansion is known to fail well
below $\Tc$; for example, the speed of sound diverges and then becomes
imaginary. To fix this we have introduced a simpler bag model equation
of state, described below.  The new equation of state {changes only
  how the relevant thermodynamic parameters $\alpha$ and $\vw$ are
  realised in terms of the parameters of the potential and field-fluid
  coupling term.  The flows around the expanding bubbles, and hence
  the gravitational wave spectrum, depend on the underlying theory
  only through $\alpha$ and the wall speed $\vw$, with the overall
  frequency scale set by the redshifted mean bubble separation.  }

Our coupled field-fluid system has energy-momentum tensor
\begin{equation}
  T^{\mu\nu} = \partial^\mu \phi \partial^\nu \phi - \frac{1}{2}
  g^{\mu\nu}(\partial\phi)^2 +(\epsilon + p) U^\mu U^\nu + g^{\mu\nu} p
\end{equation}
where $U=\relgamma (1,{\bv})$, with fluid 3-velocity $\bv$ and
associated Lorentz factor $\relgamma$. The internal energy
$\epsilon$ and pressure $p$ are
\begin{equation}
  \epsilon = 3 a(\phi)T^4 + \potZeroT(\phi), \qquad  p = a(\phi)T^4 -
  \potZeroT(\phi),
\end{equation}
and the enthalpy is $w=\epsilon + p$.

The zero-temperature effective potential is
\begin{equation}
  \potZeroT(\phi) = \frac{1}{2}\quadPar \phi^2 - \frac{1}{3}\cubPar \phi^3 +
  \frac{1}{4} \lambda \phi^4 - \potConst\text,
\end{equation}
where $\potConst$ is chosen such that $\potZeroT(\phiAtMin) = 0$, and
$\phiAtMin$ is the value of $\phi$ in the broken phase at $T=0$.  We
denote the potential energy difference between the
vacua by $ \potDelta = \potZeroT(0)-\potZeroT(\phiAtMin)$.

We write the thermal effective potential of our bag model as
\begin{equation}
  \potT(\phi,T)=\potZeroT(\phi) - T^4\left(a(\phi)-a_0\right), 
\end{equation}
where $a(\phi)$ models the change in degrees of freedom
during the transition. We take
\begin{equation}
    a(\phi)= a_0 - \frac{\potDelta}{\Tc^4}\left[3
      \left(\frac{\phi}{\phiAtMin}\right)^2 -
    2 \left(\frac{\phi}{\phiAtMin}\right)^3\right],
\end{equation}
where $a_0=(\pi^2/90)\pdofc$ with $\pdofc$ the effective number of
relativistic degrees of freedom in the symmetric phase. Both $\phi=0$ and $\phi=\phiAtMin$ are stationary points of the
function for all $T$.  For our choice of $a(\phi)$ the minima of
$\potT$ become degenerate at $T=\Tc$, as required.

The energy-momentum tensor can be decomposed into field and fluid
parts, coupled through a friction term,
\begin{equation}
  \partial_\mu T_\phi^{\mu\nu} = - \partial_\mu T_\text{f}^{\mu\nu} =
\eta U^\mu \partial_\mu\phi \partial^\nu \phi.
\end{equation}
Ref.~\cite{Hindmarsh:2017gnf} used a field- and temperature-dependent
friction parameter $\eta=\tilde{\eta}{\phi^2}/{T}$. Although this
models high temperature physics more accurately~\cite{Liu:1992tn},
strong transitions can reach small temperatures and again the
high-temperature approximation fails.  With small temperatures
we also find numerical instabilities and so revert to
using a constant $\eta$.

The phase transition strength is parametrised by 
the trace anomaly difference 
\begin{equation}
  \Delta\theta(T)=\frac{1}{4} \frac{d}{dT}\Delta\potT - \Delta\potT\text,
\end{equation}
where $\Delta\potT=\potT(0,T) - \potT(\phiAtMin,T)$.  The strength
parameter is then 
\begin{equation}
  \StrParB={\Delta\theta(\TN)}/{\epsilon_\text{r}(\TN)}.
\end{equation}
where $\TN$ is the nucleation temperature and
$\epsilon_\text{r} = 3w/4$ the radiation energy density.

We assume that the duration of the phase transition is much less than
the Hubble time $\HN^{-1}$, and neglect the effect of
expansion. This is comparable to the statement that
$\HN\Rbc \ll 1$, where $\Rbc$ is the mean bubble separation. In this
regime the contribution of bubble collisions to the gravitational wave
signal is negligible. 
{To neglect expansion the final
  simulation time $t_\mathrm{fin}$ must also be much smaller than
  $\HN^{-1}$. For all our simulations $t_\mathrm{fin} \leq 10\,\Rbc$.}

The mean gravitational wave energy density is 
\begin{equation}
  \rGW = \frac{1}{32 \pi G}  
  \frac{1}{\Vol}\int_\Vol \mathrm{d}^3x\, \overline{\dot h^{\text{TT}}_{ij} \dot h^{\text{TT}}_{ij}}\text,
\end{equation}
where $\Vol$ is the simulation volume, $h^{\text{TT}}_{ij}$ is the
transverse traceless metric perturbation and the line indicates
averaging over a characteristic period of the gravitational waves.  We
find $h^{\text{TT}}_{ij}$ in Fourier space by a standard technique
\cite{GarciaBellido:2007af,Hindmarsh:2013xza,Hindmarsh:2015qta},
sourced only by the fluid, the dominant contribution when
$\al \lesssim 1$ and
$\HN\Rbc \ll
1$~\cite{Hindmarsh:2013xza,Hindmarsh:2015qta,Hindmarsh:2017gnf}.

We express the gravitational wave energy density in terms of the
parameter $\OmGW = {\rGW}/{\rho_\text{c}}$, with $\rho_\text{c}$ the
critical energy density. Our assumptions on $\al$ and $\HN\Rbc$ ensure
that $\OmGW \ll 1$ at all times. They also ensure that the
gravitational backreaction is negligible compared to the pressure
forces, as the wavelength of the density perturbations $\sim \Rbc$ is
much less than the Jeans length $\sim \cs/ \HN \sqrt{\delta}$, where
$\delta$ is the energy density contrast averaged over the wavelength
being considered\footnote{For the parameter space we consider, we
  determined that the variation of the energy density is at most a
  factor of twenty in the asymptotic fluid profile that develops
  around an expanding bubble; $\delta$ will be less than this due to
  averaging over a given wavelength.}.  We leave a deeper analysis of
gravitational backreaction in the case
$\HN \Rbc \sim \cs/ \sqrt{\delta}$ to a later study.

We perform a series of three-dimensional simulations of the
field-fluid system. The simulation code is the same as used in
Ref.~\cite{Hindmarsh:2017gnf} except for the above changes.

We scan over $\StrParB$ for three subsonic deflagrations with
asymptotic wall speeds $\vwass=\{ 0.24,\,0.44,\,0.56 \}$, and two
detonations with $\vwass=\{0.82,\,0.92\}$. The asymptotic wall speeds,
and their fluid profiles, are found with a spherically symmetric version of the code
\cite{Hindmarsh:2015qta,Hindmarsh:2017gnf,
  KurkiSuonio:1995vy,KurkiSuonio:1995pp}, run with the same parameters
until $t=10000\Tc^{-1}$.  As we increase $\StrParB$, the maximum
velocity of the asymptotic fluid profile $\vProfMax$ increases.
For each $\vwass$, there is a maximum
$\vProfMax$, and hence a maximum strength $\StrParBMax$, above which
solutions either do not exist (subsonic deflagrations), or change into
hybrids.  We do not consider hybrids here.

The values of $\eta$ needed for these wall speeds are given in the
supplemental material. By comparison, the Standard Model estimate is
$\eta\simeq 3\,\phiAtMin^2/\Tc$\,
\cite{John:2000zq,Liu:1992tn,Moore:1995si}.

All simulations have the number of bubbles $\Nb=8$, lattice spacing
$\delta x=1.0 \,\Tc^{-1}$, timestep $\delta t=0.2 \,\Tc^{-1}$, and
$L^3=960^3$ lattice sites, giving a mean bubble separation
$\Rbc=L\,\delta x/\Nb^{1/3}=480 \,\Tc^{-1}$. All bubbles are nucleated
simultaneously {with a gaussian profile} at the same locations at the
start of each simulation. {The initial profile of the bubbles is
  insignificant as they approach the same asymptotic
  profile.}

We fix $\pdofc=106.75$, $\quadPar=0.0427 \,\Tc^2$,
$\cubPar=0.168 \,\Tc$ and $\lambda=0.0732$, in turn fixing
$\phiAtMin=2.0 \,\Tc$. This sets the relative change in degrees of
freedom to $[a(\phiAtMin) -a_0]/a_0=5.9 \times 10^{-3}$.  To change
the transition strength we vary $\TN$.

We output slices of the temperature $T$, fluid speed $v$ and vorticity
magnitude $\left| \nabla \times \bv\right|$. Movies created from these
slices are available at
\cite{external}. Selected stills are included in the supplemental
material.

We measure the RMS fluid 3-velocity $\VbTot$, and its irrotational and
rotational parts $\VbPar$ and $\VbPerp$.  We also track the
enthalpy-weighted RMS four-velocity $\Ubf$ defined as
\begin{equation}
\label{e:UbarfDef}
\Ubf^2 = \frac{1}{\overline{w}\mathcal{V}}\int_\mathcal{V} \mathrm{d}^3 x
\, w\relgamma^2 \vel^2,
\end{equation}
where $\overline{w}$ the mean enthalpy density. This gives an
indication of the magnitude of the shear stress, the source of
gravitational waves.

A similar quantity $\Ubp$ can be constructed to track
the progress of the phase transition 
\begin{equation}
\label{e:UbarPhiDef}
\Ubp^2 = \frac{1}{\overline{w}\mathcal{V}}\int_\mathcal{V} \mathrm{d}^3x
\, \partial_i \phi \partial_i \phi,
\end{equation}
proportional to the total area of the phase boundary.  We
call the time when $\Ubp$ reaches its maximum the peak collision
time, $\tPC$.  Note that $\tPC \propto \Rbc/\vw$. To see how these
global quantities evolve during a detonation and a deflagration see
Fig.~\ref{fig:Vs-single-alph} in the supplemental material.

To check the dependence of our key observables on lattice spacing,
{we perform simulations with the same physical volume and various
  lattice spacings} $\delta x\, \Tc=\{2.0,1.5,1.0,0.75,0.5\}$ for
$\vwass=0.24$ and $\vwass=0.92$ and $\StrParB=0.5$.  { We find that
  $\VbPerpMax^2$, $\Ubfmax$, and $\OmGW$ converge with lattice
  spacing. We perform a quadratic fit with $\delta x$ for each
  quantity, finding that $\Ubfmax$ and $\OmGW$ differ from the
  continuum limit by $\mathrm{O}(1\%)$. The quantity that is most
  sensitive to the grid is $\VbPerpMax^2$ which we underestimate from
  the continuum limit by $\simeq 25\%$.  We also test convergence
  of key observables with timestep, finding in all cases that
  convergence is better for $\delta t$ than for $\delta x$. It is
  important to check how close the colliding bubbles are to their
  asymptotic profile. We find that spherically symmetric bubbles with diameter $\Rc$
  have at worst $\Ubf$ within $20\%$ of the asymptotic $\Ubf$. In the
  supplemental material we show our convergence test results and the
  deviation from asymptotic $\Ubf$ for all $\vw$ and $\StrParB$
  considered here.}

\begin{figure} 
  \centering 
  \includegraphics[width=0.48\textwidth]{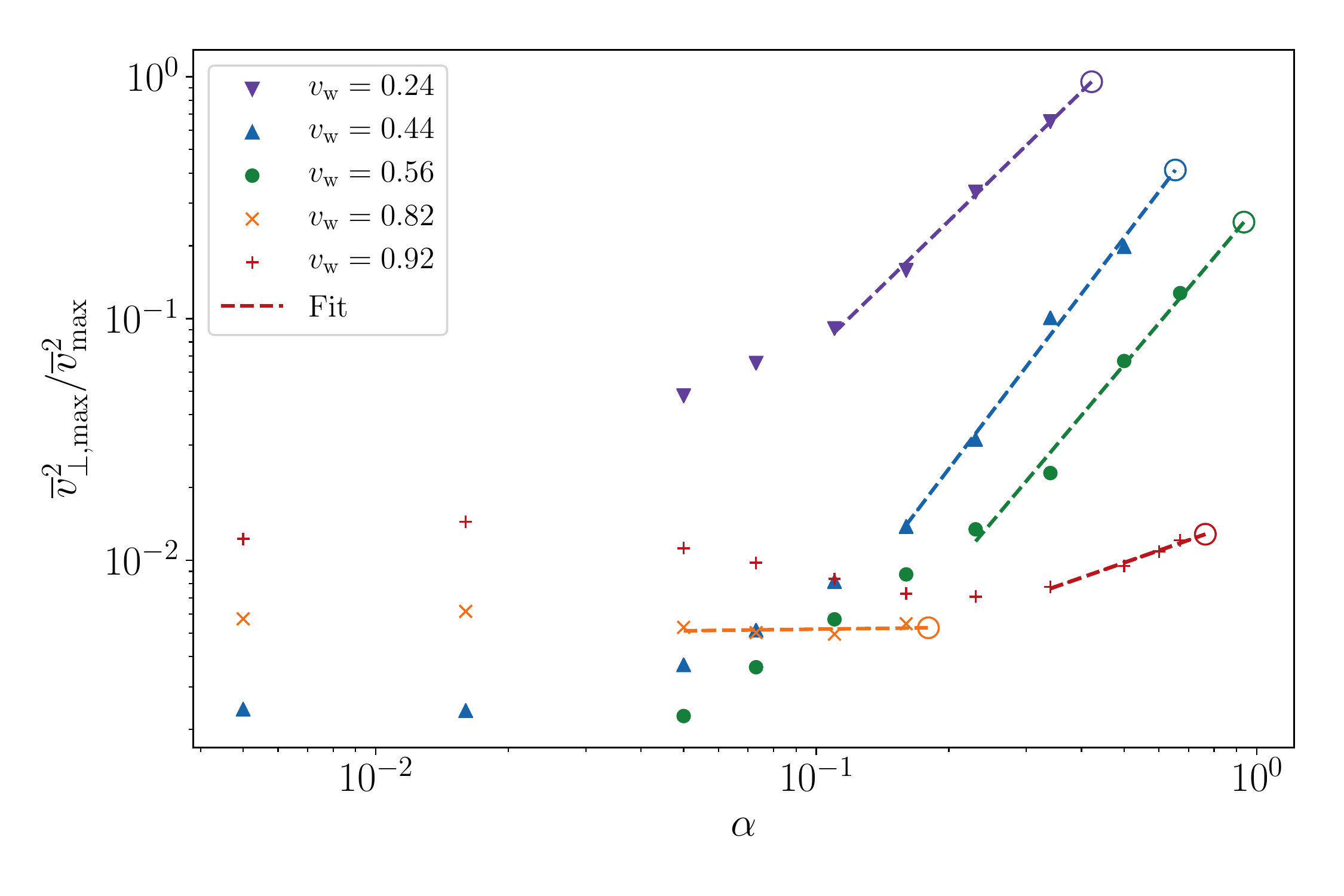}
  \caption{Proportion of mean square fluid velocity 
    in the rotational modes. We plot the ratio of $\VbPerpMax$ to
    $\VbTotMax$ against $\StrParB$.  Dashed lines give a linear fit
    for the last four simulation points. The fits are extrapolated
    to $\StrParBMax$ for deflagrations, or to the largest
    $\StrParB$ for which a wall speed corresponds to a detonation
    (hollow circles).}
  \label{fig:vperp-vs-alpha}
\end{figure}

From our simulations we see that a rotational component of velocity is
generated during the bubble collision phase. In order to gauge the
relative amount of kinetic energy in the rotational component of
velocity, we consider the ratio of the maxima of mean square
3-velocities $\VbPerpMax^2 / \VbTotMax^2$. We plot this in
Fig.~\ref{fig:vperp-vs-alpha}. As we increase $\StrParB$ for the
deflagrations, we see that the proportion of the velocity found in
rotational modes increases dramatically, whereas for detonations it
stays constant. The deflagrations with smaller wall velocities have a
larger proportion of the velocity in rotational modes. For
$\vwass=0.24$, $\StrParB=0.34$ the ratio
$\VbPerpMax^2 / \VbTotMax^2=0.65$, and if we naively extrapolate the
trend in the last few points up to $\StrParBMax$ this increases to
$0.95$.

Fig.~\ref{fig:slices-def-supl} of the supplemental material shows that
the vorticity is generated inside the bubbles, not outside where the
fluid shells first interact.

\begin{figure} 
  \centering
  \subfigure{\includegraphics[width=0.48\textwidth]{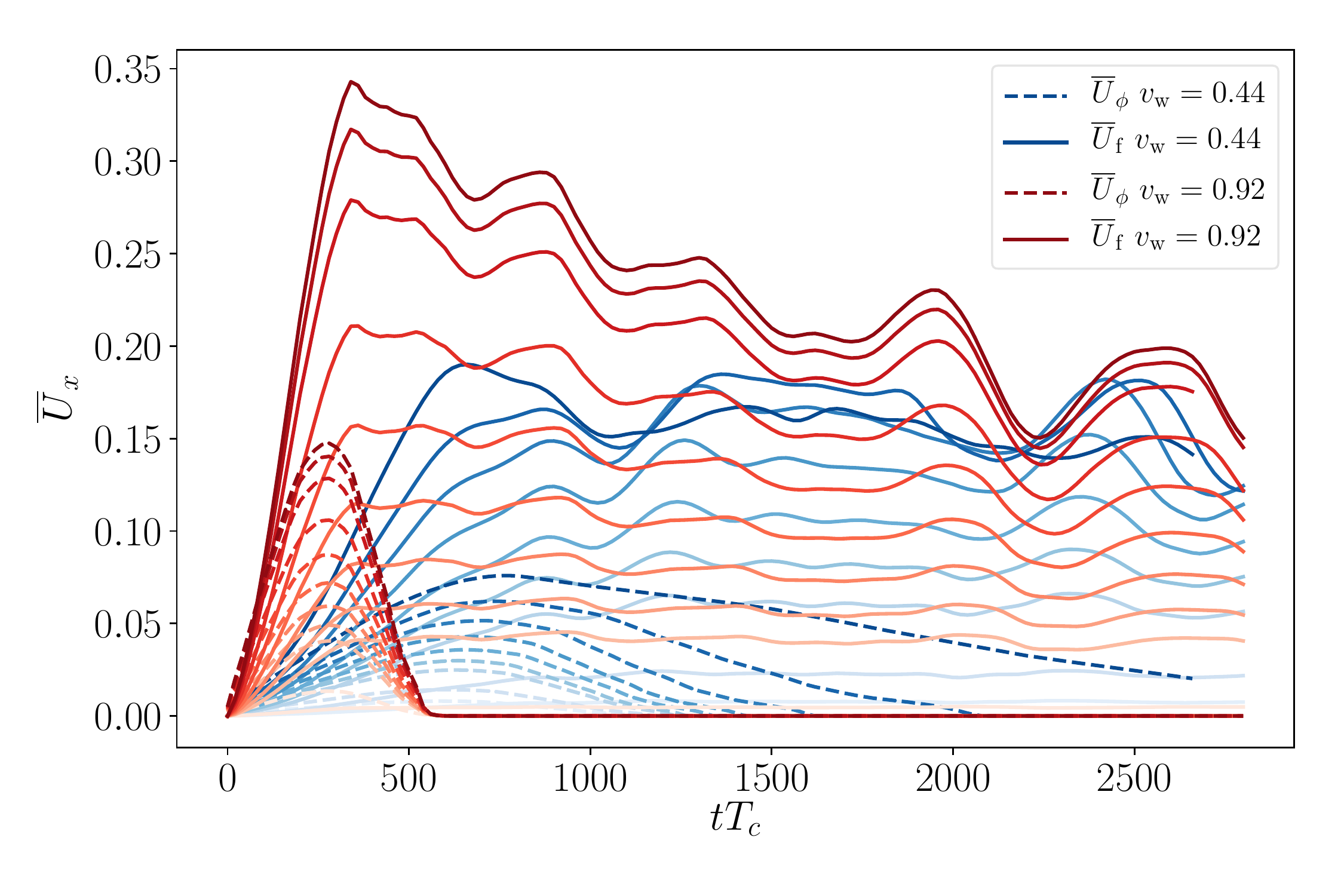}}
  \caption{The evolution of $\Ubp$ (dashed lines) and $\Ubf$ (solid
    lines) for simulations with increasing $\StrParB$ (darker shades). In
     blue we show deflagrations with $\vwass=0.44$ whereas red lines show
     detonations with $\vwass=0.92$.}
  \label{fig:multi-ubarphi}
\end{figure}

To better understand transfer of energy from the scalar field to the
fluid, we plot how $\Ubp$ and $\Ubf$ change as we increase $\StrParB$
for detonations with $\vwass=0.92$ and deflagrations with
$\vwass=0.44$ (Fig.~\ref{fig:multi-ubarphi}).  When $\Ubp$ reaches its
maximum, the volumes in each phase are approximately equal.  As the
phase boundary sweeps out the remaining regions of metastable phase,
$\Ubp$ relaxes to zero.  It is striking that for deflagrations the
relaxation takes longer as we increase $\StrParB$, whereas for
detonations the shape of $\Ubp$ remains unchanged.  The phase
boundaries in a deflagration must therefore move more slowly in the
later stages, as the transition strength increases.

The reason for the slowing is that the metastable phase is reheated by
the fluid shells in front of the bubble walls
\cite{KurkiSuonio:1984ba,Konstandin:2010dm,Megevand:2017vtb}.  Towards
the end of the transition the remaining metastable phase forms into
hot droplets (see Fig.~\ref{fig:slices-def-supl} in the supplemental
material).  The higher pressure inside the droplets opposes their
collapse.

For detonations, where the fluid shell develops behind the bubble
wall, shrinking regions of the metastable phase are not reheated (see
Fig.~\ref{fig:slices-det-supl} in the supplemental material).

Fig.~\ref{fig:multi-ubarphi} also shows that $\Ubf$ increases with
$\al$, as one expects from the increasing scalar potential
energy.  However, the maximum is below that expected from a
single bubble, which is a good estimate of $\Ubf$ at low $\StrParB$
\cite{Hindmarsh:2015qta,Hindmarsh:2017gnf}.

\begin{figure} 
  \centering 
  \includegraphics[width=0.48\textwidth]{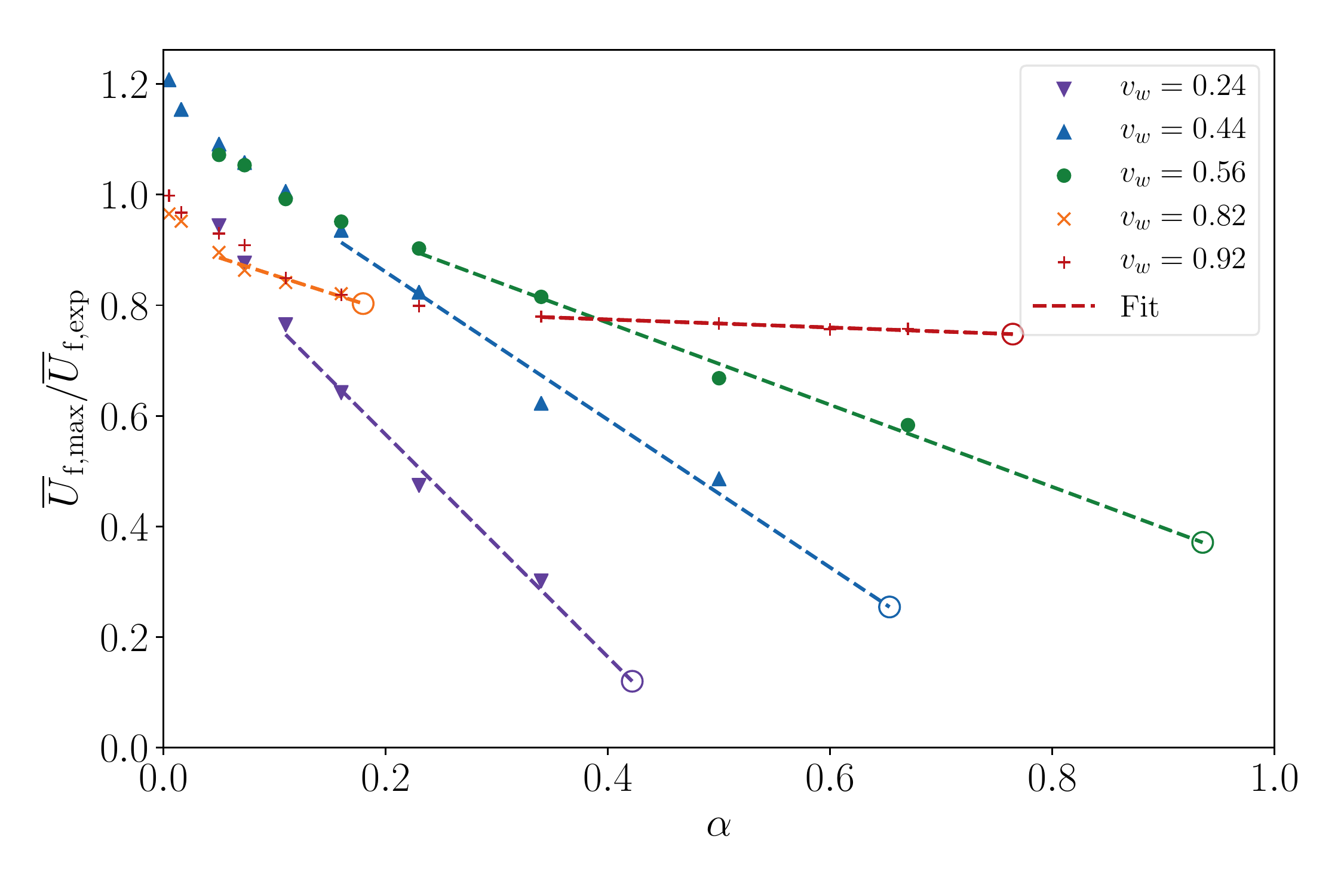}
  \caption{Comparison between the maximum value of $\Ubf$ in each
    simulation and that predicted by \cite{Espinosa:2010hh} for the
    given $\vwass$ and $\StrParB$. Dashed lines give a linear fit for
    the last four simulation points. Hollow circles show the
    extrapolation to $\StrParBMax$ for deflagrations, or up to to the
    largest $\StrParB$ for which the wall speed corresponds to a
    detonation.}
  \label{fig:ubarf-max-comp}
\end{figure}

To obtain the single-bubble estimate, simulations of expanding
spherical bubbles are performed, and the expected enthalpy-weighted
RMS velocity $\UbfExp$ is that of the fluid shell
when the wall reaches a diameter of $\Rbc$.  We then take the ratio
with the maximum of $\Ubf$ in each simulation, shown in
Fig.~\ref{fig:ubarf-max-comp}. Note that due to finite volume effects
$\Ubf$ oscillates in our simulations, giving an O(10\%) uncertainty to
this estimate.
  
For all wall speeds, the ratio of $\Ubfmax$ to $\UbfExp$
decreases as we increase the transition strength. However, for
deflagrations the decrease in the kinetic efficiency is
more dramatic, and more rapid for slower walls: in the slowest
deflagration ($\vw = 0.24$), $\Ubfmax/\UbfExp$ reaches $0.3$.  The
decrease is approximately linear; a naive linear extrapolation to the
maximum possible strength is indicated by open circles.  The loss of
kinetic energy is probably a result of the slowing
discussed above, limiting the transfer of energy.

\begin{figure} 
 \centering 
 \includegraphics[width=0.48\textwidth]{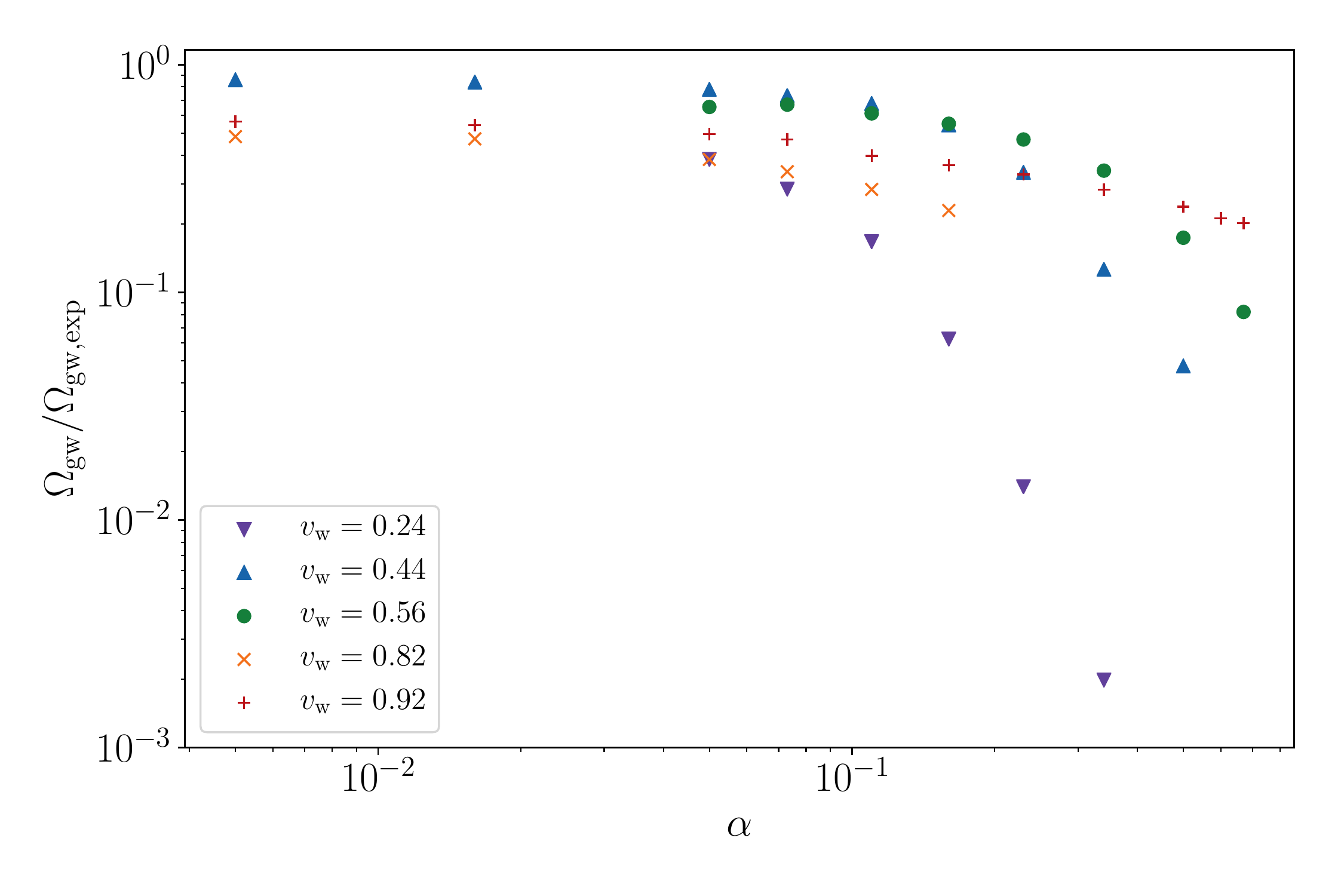}
 \caption{Comparison of the gravitational waves produced in our
   simulations against that predicted by Eq.~(\ref{eqn:OmGWExp}) using
   $\UbfExp$ found from $\vwass$ and $\StrParB$.}
 \label{fig:Omega-gw}
\end{figure}

The deficit in kinetic energy can be expected to reduce the
gravitational wave signal.  In current modelling~\cite{Caprini:2015zlo,Hindmarsh:2017gnf}, the expected
gravitational wave density parameter from a flow with $\UbfExp$ at time
$t \ll \HN^{-1}$ is 
\begin{equation}
  \label{eqn:OmGWExp}
  \OmGWExp= 3\, \OmGWscaled
  \left(\dfrac{\overline{w}}{\overline{\epsilon}}\right)^2 \UbfExp^4
  (\HN t) (\HN \Rbc)\text,
\end{equation}
where $\OmGWscaled$ has been shown to be a constant of
$\mathrm{O}(10^{-2})$ in weak and intermediate transitions. 
Here, we take
$\OmGWscaled=10^{-2}$.  
In Fig.~\ref{fig:Omega-gw} we plot the ratio of $\OmGW/t$ to
$\OmGWExp/t$, where $\OmGW/t$ is averaged over the final
$\Delta t= 2 \Rbc$ of the simulation. 
In the most extreme case,
$\vwass=0.24$ and $\StrParB=0.34$, the ratio is $2 \times 10^{-3}$.
This is even less than the kinetic energy suppression suggests, a
factor of $(\Ubfmax/\UbfExp)^4 \simeq 8\times10^{-3}$.

A table of simulation parameters and measured quantities can be found
in the supplemental material.

We have performed the first 3-dimensional simulations of
strong first-order phase transitions, with the strength
parameter $\al$ up to an order of magnitude larger than those
previously studied~\cite{Hindmarsh:2017gnf}.

A rotational component of velocity $\VbPerp$ is generated during the
collision phase.  For deflagrations, the ratio
$\VbPerpMax^2/\VbTotMax^2$ grows rapidly with $\al$, reaching $0.65$
for $\vwass=0.24$.  For detonations, the ratio is
O($10^{-2}$)---showing no consistent trend with $\al$.

For stronger phase transitions a smaller proportion of the scalar
potential energy is transferred into fluid kinetic energy than is
expected from the behaviour of isolated bubbles.  For
deflagrations, we suppose that the deficit is due to reheating of
the metastable phase slowing the bubble walls during the collision
phase. The deficit can be substantial, with $\Ubfmax/\UbfExp$ falling
to $\sim0.3$ for $\vwass=0.24$ in our simulations, and could fall as
low as $0.1$ using a naive linear extrapolation to the maximum
possible strength at that wall speed.
  
The gravitational wave intensity is lower than expected, by a factor
of order $10^{-3}$ for the strongest deflagration with the lowest wall
speed.  This can mostly be accounted for by the kinetic energy
deficit.  Detonations do not suffer such a dramatic suppression, with
the smallest suppression factor about $0.2$ for $\vwass=0.92$.

Our results have important consequences for gravitational waves from
phase transitions.  They indicate that the current model
\cite{Caprini:2015zlo,Hindmarsh:2017gnf} overestimates the
gravitational wave power spectrum for strong transitions, by a factor
of a few for detonations, and by an order of magnitude or more for
deflagrations. We estimate\footnote{We use PTPlot \texttt{v1.01},
  \url{http://www.ptplot.org/ptplot/}\cite{Caprini:2019egz}, to compute and plot signal to
  noise ratio (SNR) curves from first-order phase transitions for
  LISA. We choose $\pdofc=106.75$ and an optimistic
  $\TN = 100\, \mathrm{GeV}$. The resulting plot shows that an SNR of
  10 requires $\Ubf$ of at least 0.07 for all $\HN \Rbc$.}  that to
obtain a signal to noise ratio of 10 $\Ubf$ must be at least
0.07. Therefore the kinetic energy suppression we observe will probably
render transitions with $\vw=0.24$ unobservable except for within a very small
region of parameter space. Faster walls suffer less suppression,
though the observable parameter space is still reduced.

We plan larger simulations to characterise more precisely the
suppression, {and its effect on observability}.

\FloatBarrier
\begin{acknowledgments}
  The authors would like to thank Chiara Caprini, Kari Rummukainen,
  and Dani\`ele Steer for helpful discussions. Our simulations made
  use of the resources of the Finnish Centre for Scientific Computing
  CSC. DC (ORCID ID 0000-0002-7395-7802) is supported by an STFC
  Studentship. MH (ORCID ID 0000-0002-9307-437X) acknowledges support
  from the Science and Technology Facilities Council, grant
  no. ST/P000819/1.  DJW (ORCID ID 0000-0001-6986-0517) is supported
  by an Science and Technology Facilities Council Ernest Rutherford
  Fellowship, grant no. ST/R003904/1, by the Research Funds of the
  University of Helsinki, and by the Academy of Finland, grant
  no. 286769.
\end{acknowledgments}

\bibliography{strong-pt-short}

\begin{thebibliography}{59}%
\makeatletter
\providecommand \@ifxundefined [1]{%
 \@ifx{#1\undefined}
}%
\providecommand \@ifnum [1]{%
 \ifnum #1\expandafter \@firstoftwo
 \else \expandafter \@secondoftwo
 \fi
}%
\providecommand \@ifx [1]{%
 \ifx #1\expandafter \@firstoftwo
 \else \expandafter \@secondoftwo
 \fi
}%
\providecommand \natexlab [1]{#1}%
\providecommand \enquote  [1]{``#1''}%
\providecommand \bibnamefont  [1]{#1}%
\providecommand \bibfnamefont [1]{#1}%
\providecommand \citenamefont [1]{#1}%
\providecommand \href@noop [0]{\@secondoftwo}%
\providecommand \href [0]{\begingroup \@sanitize@url \@href}%
\providecommand \@href[1]{\@@startlink{#1}\@@href}%
\providecommand \@@href[1]{\endgroup#1\@@endlink}%
\providecommand \@sanitize@url [0]{\catcode `\\12\catcode `\$12\catcode
  `\&12\catcode `\#12\catcode `\^12\catcode `\_12\catcode `\%12\relax}%
\providecommand \@@startlink[1]{}%
\providecommand \@@endlink[0]{}%
\providecommand \url  [0]{\begingroup\@sanitize@url \@url }%
\providecommand \@url [1]{\endgroup\@href {#1}{\urlprefix }}%
\providecommand \urlprefix  [0]{URL }%
\providecommand \Eprint [0]{\href }%
\providecommand \doibase [0]{http://dx.doi.org/}%
\providecommand \selectlanguage [0]{\@gobble}%
\providecommand \bibinfo  [0]{\@secondoftwo}%
\providecommand \bibfield  [0]{\@secondoftwo}%
\providecommand \translation [1]{[#1]}%
\providecommand \BibitemOpen [0]{}%
\providecommand \bibitemStop [0]{}%
\providecommand \bibitemNoStop [0]{.\EOS\space}%
\providecommand \EOS [0]{\spacefactor3000\relax}%
\providecommand \BibitemShut  [1]{\csname bibitem#1\endcsname}%
\let\auto@bib@innerbib\@empty
\bibitem [{\citenamefont {Audley}\ \emph {et~al.}(2017)\citenamefont {Audley}
  \emph {et~al.}}]{Audley:2017drz}%
  \BibitemOpen
  \bibfield  {author} {\bibinfo {author} {\bibfnamefont {H.}~\bibnamefont
  {Audley}} \emph {et~al.},\ }\href@noop {} {\  (\bibinfo {year} {2017})},\
  \Eprint {http://arxiv.org/abs/1702.00786} {arXiv:1702.00786 [astro-ph.IM]}
  \BibitemShut {NoStop}%
\bibitem [{\citenamefont {Kajantie}\ \emph {et~al.}(1996)\citenamefont
  {Kajantie}, \citenamefont {Laine}, \citenamefont {Rummukainen},\ and\
  \citenamefont {Shaposhnikov}}]{Kajantie:1996mn}%
  \BibitemOpen
  \bibfield  {author} {\bibinfo {author} {\bibfnamefont {K.}~\bibnamefont
  {Kajantie}}, \bibinfo {author} {\bibfnamefont {M.}~\bibnamefont {Laine}},
  \bibinfo {author} {\bibfnamefont {K.}~\bibnamefont {Rummukainen}}, \ and\
  \bibinfo {author} {\bibfnamefont {M.~E.}\ \bibnamefont {Shaposhnikov}},\
  }\href {\doibase 10.1103/PhysRevLett.77.2887} {\bibfield  {journal} {\bibinfo
   {journal} {Phys.Rev.Lett.}\ }\textbf {\bibinfo {volume} {77}},\ \bibinfo
  {pages} {2887} (\bibinfo {year} {1996})},\ \Eprint
  {http://arxiv.org/abs/hep-ph/9605288} {arXiv:hep-ph/9605288 [hep-ph]}
  \BibitemShut {NoStop}%
\bibitem [{\citenamefont {Kajantie}\ \emph {et~al.}(1997)\citenamefont
  {Kajantie}, \citenamefont {Laine}, \citenamefont {Rummukainen},\ and\
  \citenamefont {Shaposhnikov}}]{Kajantie:1996qd}%
  \BibitemOpen
  \bibfield  {author} {\bibinfo {author} {\bibfnamefont {K.}~\bibnamefont
  {Kajantie}}, \bibinfo {author} {\bibfnamefont {M.}~\bibnamefont {Laine}},
  \bibinfo {author} {\bibfnamefont {K.}~\bibnamefont {Rummukainen}}, \ and\
  \bibinfo {author} {\bibfnamefont {M.~E.}\ \bibnamefont {Shaposhnikov}},\
  }\href {\doibase 10.1016/S0550-3213(97)00164-8} {\bibfield  {journal}
  {\bibinfo  {journal} {Nucl.Phys.}\ }\textbf {\bibinfo {volume} {B493}},\
  \bibinfo {pages} {413} (\bibinfo {year} {1997})},\ \Eprint
  {http://arxiv.org/abs/hep-lat/9612006} {arXiv:hep-lat/9612006 [hep-lat]}
  \BibitemShut {NoStop}%
\bibitem [{\citenamefont {Profumo}\ \emph {et~al.}(2007)\citenamefont
  {Profumo}, \citenamefont {Ramsey-Musolf},\ and\ \citenamefont
  {Shaughnessy}}]{Profumo:2007wc}%
  \BibitemOpen
  \bibfield  {author} {\bibinfo {author} {\bibfnamefont {S.}~\bibnamefont
  {Profumo}}, \bibinfo {author} {\bibfnamefont {M.~J.}\ \bibnamefont
  {Ramsey-Musolf}}, \ and\ \bibinfo {author} {\bibfnamefont {G.}~\bibnamefont
  {Shaughnessy}},\ }\href {\doibase 10.1088/1126-6708/2007/08/010} {\bibfield
  {journal} {\bibinfo  {journal} {JHEP}\ }\textbf {\bibinfo {volume} {08}},\
  \bibinfo {pages} {010} (\bibinfo {year} {2007})},\ \Eprint
  {http://arxiv.org/abs/0705.2425} {arXiv:0705.2425 [hep-ph]} \BibitemShut
  {NoStop}%
\bibitem [{\citenamefont {Espinosa}\ \emph {et~al.}(2012)\citenamefont
  {Espinosa}, \citenamefont {Konstandin},\ and\ \citenamefont
  {Riva}}]{Espinosa:2011ax}%
  \BibitemOpen
  \bibfield  {author} {\bibinfo {author} {\bibfnamefont {J.~R.}\ \bibnamefont
  {Espinosa}}, \bibinfo {author} {\bibfnamefont {T.}~\bibnamefont
  {Konstandin}}, \ and\ \bibinfo {author} {\bibfnamefont {F.}~\bibnamefont
  {Riva}},\ }\href {\doibase 10.1016/j.nuclphysb.2011.09.010} {\bibfield
  {journal} {\bibinfo  {journal} {Nucl. Phys.}\ }\textbf {\bibinfo {volume}
  {B854}},\ \bibinfo {pages} {592} (\bibinfo {year} {2012})},\ \Eprint
  {http://arxiv.org/abs/1107.5441} {arXiv:1107.5441 [hep-ph]} \BibitemShut
  {NoStop}%
\bibitem [{\citenamefont {Cline}\ and\ \citenamefont
  {Kainulainen}(2013)}]{Cline:2012hg}%
  \BibitemOpen
  \bibfield  {author} {\bibinfo {author} {\bibfnamefont {J.~M.}\ \bibnamefont
  {Cline}}\ and\ \bibinfo {author} {\bibfnamefont {K.}~\bibnamefont
  {Kainulainen}},\ }\href {\doibase 10.1088/1475-7516/2013/01/012} {\bibfield
  {journal} {\bibinfo  {journal} {JCAP}\ }\textbf {\bibinfo {volume} {1301}},\
  \bibinfo {pages} {012} (\bibinfo {year} {2013})},\ \Eprint
  {http://arxiv.org/abs/1210.4196} {arXiv:1210.4196 [hep-ph]} \BibitemShut
  {NoStop}%
\bibitem [{\citenamefont {Profumo}\ \emph {et~al.}(2015)\citenamefont
  {Profumo}, \citenamefont {Ramsey-Musolf}, \citenamefont {Wainwright},\ and\
  \citenamefont {Winslow}}]{Profumo:2014opa}%
  \BibitemOpen
  \bibfield  {author} {\bibinfo {author} {\bibfnamefont {S.}~\bibnamefont
  {Profumo}}, \bibinfo {author} {\bibfnamefont {M.~J.}\ \bibnamefont
  {Ramsey-Musolf}}, \bibinfo {author} {\bibfnamefont {C.~L.}\ \bibnamefont
  {Wainwright}}, \ and\ \bibinfo {author} {\bibfnamefont {P.}~\bibnamefont
  {Winslow}},\ }\href {\doibase 10.1103/PhysRevD.91.035018} {\bibfield
  {journal} {\bibinfo  {journal} {Phys. Rev.}\ }\textbf {\bibinfo {volume}
  {D91}},\ \bibinfo {pages} {035018} (\bibinfo {year} {2015})},\ \Eprint
  {http://arxiv.org/abs/1407.5342} {arXiv:1407.5342 [hep-ph]} \BibitemShut
  {NoStop}%
\bibitem [{\citenamefont {Beniwal}\ \emph {et~al.}(2019)\citenamefont
  {Beniwal}, \citenamefont {Lewicki}, \citenamefont {White},\ and\
  \citenamefont {Williams}}]{Beniwal:2018hyi}%
  \BibitemOpen
  \bibfield  {author} {\bibinfo {author} {\bibfnamefont {A.}~\bibnamefont
  {Beniwal}}, \bibinfo {author} {\bibfnamefont {M.}~\bibnamefont {Lewicki}},
  \bibinfo {author} {\bibfnamefont {M.}~\bibnamefont {White}}, \ and\ \bibinfo
  {author} {\bibfnamefont {A.~G.}\ \bibnamefont {Williams}},\ }\href {\doibase
  10.1007/JHEP02(2019)183} {\bibfield  {journal} {\bibinfo  {journal} {JHEP}\
  }\textbf {\bibinfo {volume} {02}},\ \bibinfo {pages} {183} (\bibinfo {year}
  {2019})},\ \Eprint {http://arxiv.org/abs/1810.02380} {arXiv:1810.02380
  [hep-ph]} \BibitemShut {NoStop}%
\bibitem [{\citenamefont {Kakizaki}\ \emph {et~al.}(2015)\citenamefont
  {Kakizaki}, \citenamefont {Kanemura},\ and\ \citenamefont
  {Matsui}}]{Kakizaki:2015wua}%
  \BibitemOpen
  \bibfield  {author} {\bibinfo {author} {\bibfnamefont {M.}~\bibnamefont
  {Kakizaki}}, \bibinfo {author} {\bibfnamefont {S.}~\bibnamefont {Kanemura}},
  \ and\ \bibinfo {author} {\bibfnamefont {T.}~\bibnamefont {Matsui}},\ }\href
  {\doibase 10.1103/PhysRevD.92.115007} {\bibfield  {journal} {\bibinfo
  {journal} {Phys. Rev.}\ }\textbf {\bibinfo {volume} {D92}},\ \bibinfo {pages}
  {115007} (\bibinfo {year} {2015})},\ \Eprint
  {http://arxiv.org/abs/1509.08394} {arXiv:1509.08394 [hep-ph]} \BibitemShut
  {NoStop}%
\bibitem [{\citenamefont {Dorsch}\ \emph {et~al.}(2017)\citenamefont {Dorsch},
  \citenamefont {Huber}, \citenamefont {Konstandin},\ and\ \citenamefont
  {No}}]{Dorsch:2016nrg}%
  \BibitemOpen
  \bibfield  {author} {\bibinfo {author} {\bibfnamefont {G.~C.}\ \bibnamefont
  {Dorsch}}, \bibinfo {author} {\bibfnamefont {S.~J.}\ \bibnamefont {Huber}},
  \bibinfo {author} {\bibfnamefont {T.}~\bibnamefont {Konstandin}}, \ and\
  \bibinfo {author} {\bibfnamefont {J.~M.}\ \bibnamefont {No}},\ }\href
  {\doibase 10.1088/1475-7516/2017/05/052} {\bibfield  {journal} {\bibinfo
  {journal} {JCAP}\ }\textbf {\bibinfo {volume} {1705}},\ \bibinfo {pages}
  {052} (\bibinfo {year} {2017})},\ \Eprint {http://arxiv.org/abs/1611.05874}
  {arXiv:1611.05874 [hep-ph]} \BibitemShut {NoStop}%
\bibitem [{\citenamefont {Basler}\ \emph {et~al.}(2017)\citenamefont {Basler},
  \citenamefont {Krause}, \citenamefont {Muhlleitner}, \citenamefont
  {Wittbrodt},\ and\ \citenamefont {Wlotzka}}]{Basler:2016obg}%
  \BibitemOpen
  \bibfield  {author} {\bibinfo {author} {\bibfnamefont {P.}~\bibnamefont
  {Basler}}, \bibinfo {author} {\bibfnamefont {M.}~\bibnamefont {Krause}},
  \bibinfo {author} {\bibfnamefont {M.}~\bibnamefont {Muhlleitner}}, \bibinfo
  {author} {\bibfnamefont {J.}~\bibnamefont {Wittbrodt}}, \ and\ \bibinfo
  {author} {\bibfnamefont {A.}~\bibnamefont {Wlotzka}},\ }\href {\doibase
  10.1007/JHEP02(2017)121} {\bibfield  {journal} {\bibinfo  {journal} {JHEP}\
  }\textbf {\bibinfo {volume} {02}},\ \bibinfo {pages} {121} (\bibinfo {year}
  {2017})},\ \Eprint {http://arxiv.org/abs/1612.04086} {arXiv:1612.04086
  [hep-ph]} \BibitemShut {NoStop}%
\bibitem [{\citenamefont {Randall}\ and\ \citenamefont
  {Servant}(2007)}]{Randall:2006py}%
  \BibitemOpen
  \bibfield  {author} {\bibinfo {author} {\bibfnamefont {L.}~\bibnamefont
  {Randall}}\ and\ \bibinfo {author} {\bibfnamefont {G.}~\bibnamefont
  {Servant}},\ }\href {\doibase 10.1088/1126-6708/2007/05/054} {\bibfield
  {journal} {\bibinfo  {journal} {JHEP}\ }\textbf {\bibinfo {volume} {05}},\
  \bibinfo {pages} {054} (\bibinfo {year} {2007})},\ \Eprint
  {http://arxiv.org/abs/hep-ph/0607158} {arXiv:hep-ph/0607158 [hep-ph]}
  \BibitemShut {NoStop}%
\bibitem [{\citenamefont {Konstandin}\ \emph {et~al.}(2010)\citenamefont
  {Konstandin}, \citenamefont {Nardini},\ and\ \citenamefont
  {Quiros}}]{PhysRevD.82.083513}%
  \BibitemOpen
  \bibfield  {author} {\bibinfo {author} {\bibfnamefont {T.}~\bibnamefont
  {Konstandin}}, \bibinfo {author} {\bibfnamefont {G.}~\bibnamefont {Nardini}},
  \ and\ \bibinfo {author} {\bibfnamefont {M.}~\bibnamefont {Quiros}},\ }\href
  {\doibase 10.1103/PhysRevD.82.083513} {\bibfield  {journal} {\bibinfo
  {journal} {Phys. Rev. D}\ }\textbf {\bibinfo {volume} {82}},\ \bibinfo
  {pages} {083513} (\bibinfo {year} {2010})}\BibitemShut {NoStop}%
\bibitem [{\citenamefont {Konstandin}\ and\ \citenamefont
  {Servant}(2011)}]{Konstandin:2011dr}%
  \BibitemOpen
  \bibfield  {author} {\bibinfo {author} {\bibfnamefont {T.}~\bibnamefont
  {Konstandin}}\ and\ \bibinfo {author} {\bibfnamefont {G.}~\bibnamefont
  {Servant}},\ }\href {\doibase 10.1088/1475-7516/2011/12/009} {\bibfield
  {journal} {\bibinfo  {journal} {JCAP}\ }\textbf {\bibinfo {volume} {1112}},\
  \bibinfo {pages} {009} (\bibinfo {year} {2011})},\ \Eprint
  {http://arxiv.org/abs/1104.4791} {arXiv:1104.4791 [hep-ph]} \BibitemShut
  {NoStop}%
\bibitem [{\citenamefont {von Harling}\ and\ \citenamefont
  {Servant}(2018)}]{vonHarling:2017yew}%
  \BibitemOpen
  \bibfield  {author} {\bibinfo {author} {\bibfnamefont {B.}~\bibnamefont {von
  Harling}}\ and\ \bibinfo {author} {\bibfnamefont {G.}~\bibnamefont
  {Servant}},\ }\href {\doibase 10.1007/JHEP01(2018)159} {\bibfield  {journal}
  {\bibinfo  {journal} {JHEP}\ }\textbf {\bibinfo {volume} {01}},\ \bibinfo
  {pages} {159} (\bibinfo {year} {2018})},\ \Eprint
  {http://arxiv.org/abs/1711.11554} {arXiv:1711.11554 [hep-ph]} \BibitemShut
  {NoStop}%
\bibitem [{\citenamefont {Dillon}\ \emph {et~al.}(2018)\citenamefont {Dillon},
  \citenamefont {El-Menoufi}, \citenamefont {Huber},\ and\ \citenamefont
  {Manuel}}]{Dillon:2017ctw}%
  \BibitemOpen
  \bibfield  {author} {\bibinfo {author} {\bibfnamefont {B.~M.}\ \bibnamefont
  {Dillon}}, \bibinfo {author} {\bibfnamefont {B.~K.}\ \bibnamefont
  {El-Menoufi}}, \bibinfo {author} {\bibfnamefont {S.~J.}\ \bibnamefont
  {Huber}}, \ and\ \bibinfo {author} {\bibfnamefont {J.~P.}\ \bibnamefont
  {Manuel}},\ }\href {\doibase 10.1103/PhysRevD.98.086005} {\bibfield
  {journal} {\bibinfo  {journal} {Phys. Rev.}\ }\textbf {\bibinfo {volume}
  {D98}},\ \bibinfo {pages} {086005} (\bibinfo {year} {2018})},\ \Eprint
  {http://arxiv.org/abs/1708.02953} {arXiv:1708.02953 [hep-th]} \BibitemShut
  {NoStop}%
\bibitem [{\citenamefont {Meg{\'\i}as}\ \emph {et~al.}(2018)\citenamefont
  {Meg{\'\i}as}, \citenamefont {Nardini},\ and\ \citenamefont
  {Quir{\'o}s}}]{Megias:2018sxv}%
  \BibitemOpen
  \bibfield  {author} {\bibinfo {author} {\bibfnamefont {E.}~\bibnamefont
  {Meg{\'\i}as}}, \bibinfo {author} {\bibfnamefont {G.}~\bibnamefont
  {Nardini}}, \ and\ \bibinfo {author} {\bibfnamefont {M.}~\bibnamefont
  {Quir{\'o}s}},\ }\href {\doibase 10.1007/JHEP09(2018)095} {\bibfield
  {journal} {\bibinfo  {journal} {JHEP}\ }\textbf {\bibinfo {volume} {09}},\
  \bibinfo {pages} {095} (\bibinfo {year} {2018})},\ \Eprint
  {http://arxiv.org/abs/1806.04877} {arXiv:1806.04877 [hep-ph]} \BibitemShut
  {NoStop}%
\bibitem [{\citenamefont {Bruggisser}\ \emph {et~al.}(2018)\citenamefont
  {Bruggisser}, \citenamefont {Von~Harling}, \citenamefont {Matsedonskyi},\
  and\ \citenamefont {Servant}}]{Bruggisser:2018mrt}%
  \BibitemOpen
  \bibfield  {author} {\bibinfo {author} {\bibfnamefont {S.}~\bibnamefont
  {Bruggisser}}, \bibinfo {author} {\bibfnamefont {B.}~\bibnamefont
  {Von~Harling}}, \bibinfo {author} {\bibfnamefont {O.}~\bibnamefont
  {Matsedonskyi}}, \ and\ \bibinfo {author} {\bibfnamefont {G.}~\bibnamefont
  {Servant}},\ }\href {\doibase 10.1007/JHEP12(2018)099} {\bibfield  {journal}
  {\bibinfo  {journal} {JHEP}\ }\textbf {\bibinfo {volume} {12}},\ \bibinfo
  {pages} {099} (\bibinfo {year} {2018})},\ \Eprint
  {http://arxiv.org/abs/1804.07314} {arXiv:1804.07314 [hep-ph]} \BibitemShut
  {NoStop}%
\bibitem [{\citenamefont {Schwaller}(2015)}]{Schwaller:2015tja}%
  \BibitemOpen
  \bibfield  {author} {\bibinfo {author} {\bibfnamefont {P.}~\bibnamefont
  {Schwaller}},\ }\href {\doibase 10.1103/PhysRevLett.115.181101} {\bibfield
  {journal} {\bibinfo  {journal} {Phys. Rev. Lett.}\ }\textbf {\bibinfo
  {volume} {115}},\ \bibinfo {pages} {181101} (\bibinfo {year} {2015})},\
  \Eprint {http://arxiv.org/abs/1504.07263} {arXiv:1504.07263 [hep-ph]}
  \BibitemShut {NoStop}%
\bibitem [{\citenamefont {Addazi}\ and\ \citenamefont
  {Marciano}(2018)}]{Addazi:2017gpt}%
  \BibitemOpen
  \bibfield  {author} {\bibinfo {author} {\bibfnamefont {A.}~\bibnamefont
  {Addazi}}\ and\ \bibinfo {author} {\bibfnamefont {A.}~\bibnamefont
  {Marciano}},\ }\href {\doibase 10.1088/1674-1137/42/2/023107} {\bibfield
  {journal} {\bibinfo  {journal} {Chin. Phys.}\ }\textbf {\bibinfo {volume}
  {C42}},\ \bibinfo {pages} {023107} (\bibinfo {year} {2018})},\ \Eprint
  {http://arxiv.org/abs/1703.03248} {arXiv:1703.03248 [hep-ph]} \BibitemShut
  {NoStop}%
\bibitem [{\citenamefont {Aoki}\ \emph {et~al.}(2017)\citenamefont {Aoki},
  \citenamefont {Goto},\ and\ \citenamefont {Kubo}}]{Aoki:2017aws}%
  \BibitemOpen
  \bibfield  {author} {\bibinfo {author} {\bibfnamefont {M.}~\bibnamefont
  {Aoki}}, \bibinfo {author} {\bibfnamefont {H.}~\bibnamefont {Goto}}, \ and\
  \bibinfo {author} {\bibfnamefont {J.}~\bibnamefont {Kubo}},\ }\href {\doibase
  10.1103/PhysRevD.96.075045} {\bibfield  {journal} {\bibinfo  {journal} {Phys.
  Rev.}\ }\textbf {\bibinfo {volume} {D96}},\ \bibinfo {pages} {075045}
  (\bibinfo {year} {2017})},\ \Eprint {http://arxiv.org/abs/1709.07572}
  {arXiv:1709.07572 [hep-ph]} \BibitemShut {NoStop}%
\bibitem [{\citenamefont {Croon}\ \emph {et~al.}(2018)\citenamefont {Croon},
  \citenamefont {Sanz},\ and\ \citenamefont {White}}]{Croon:2018erz}%
  \BibitemOpen
  \bibfield  {author} {\bibinfo {author} {\bibfnamefont {D.}~\bibnamefont
  {Croon}}, \bibinfo {author} {\bibfnamefont {V.}~\bibnamefont {Sanz}}, \ and\
  \bibinfo {author} {\bibfnamefont {G.}~\bibnamefont {White}},\ }\href
  {\doibase 10.1007/JHEP08(2018)203} {\bibfield  {journal} {\bibinfo  {journal}
  {JHEP}\ }\textbf {\bibinfo {volume} {08}},\ \bibinfo {pages} {203} (\bibinfo
  {year} {2018})},\ \Eprint {http://arxiv.org/abs/1806.02332} {arXiv:1806.02332
  [hep-ph]} \BibitemShut {NoStop}%
\bibitem [{\citenamefont {Breitbach}\ \emph {et~al.}(2018)\citenamefont
  {Breitbach}, \citenamefont {Kopp}, \citenamefont {Madge}, \citenamefont
  {Opferkuch},\ and\ \citenamefont {Schwaller}}]{Breitbach:2018ddu}%
  \BibitemOpen
  \bibfield  {author} {\bibinfo {author} {\bibfnamefont {M.}~\bibnamefont
  {Breitbach}}, \bibinfo {author} {\bibfnamefont {J.}~\bibnamefont {Kopp}},
  \bibinfo {author} {\bibfnamefont {E.}~\bibnamefont {Madge}}, \bibinfo
  {author} {\bibfnamefont {T.}~\bibnamefont {Opferkuch}}, \ and\ \bibinfo
  {author} {\bibfnamefont {P.}~\bibnamefont {Schwaller}},\ }\href@noop {} {\
  (\bibinfo {year} {2018})},\ \Eprint {http://arxiv.org/abs/1811.11175}
  {arXiv:1811.11175 [hep-ph]} \BibitemShut {NoStop}%
\bibitem [{\citenamefont {Okada}\ and\ \citenamefont
  {Seto}(2018)}]{Okada:2018xdh}%
  \BibitemOpen
  \bibfield  {author} {\bibinfo {author} {\bibfnamefont {N.}~\bibnamefont
  {Okada}}\ and\ \bibinfo {author} {\bibfnamefont {O.}~\bibnamefont {Seto}},\
  }\href {\doibase 10.1103/PhysRevD.98.063532} {\bibfield  {journal} {\bibinfo
  {journal} {Phys. Rev.}\ }\textbf {\bibinfo {volume} {D98}},\ \bibinfo {pages}
  {063532} (\bibinfo {year} {2018})},\ \Eprint
  {http://arxiv.org/abs/1807.00336} {arXiv:1807.00336 [hep-ph]} \BibitemShut
  {NoStop}%
\bibitem [{\citenamefont {Hasegawa}\ \emph {et~al.}(2019)\citenamefont
  {Hasegawa}, \citenamefont {Okada},\ and\ \citenamefont
  {Seto}}]{Hasegawa:2019amx}%
  \BibitemOpen
  \bibfield  {author} {\bibinfo {author} {\bibfnamefont {T.}~\bibnamefont
  {Hasegawa}}, \bibinfo {author} {\bibfnamefont {N.}~\bibnamefont {Okada}}, \
  and\ \bibinfo {author} {\bibfnamefont {O.}~\bibnamefont {Seto}},\ }\href
  {\doibase 10.1103/PhysRevD.99.095039} {\bibfield  {journal} {\bibinfo
  {journal} {Phys. Rev.}\ }\textbf {\bibinfo {volume} {D99}},\ \bibinfo {pages}
  {095039} (\bibinfo {year} {2019})},\ \Eprint
  {http://arxiv.org/abs/1904.03020} {arXiv:1904.03020 [hep-ph]} \BibitemShut
  {NoStop}%
\bibitem [{\citenamefont {Gorda}\ \emph {et~al.}(2019)\citenamefont {Gorda},
  \citenamefont {Helset}, \citenamefont {Niemi}, \citenamefont {Tenkanen},\
  and\ \citenamefont {Weir}}]{Gorda:2018hvi}%
  \BibitemOpen
  \bibfield  {author} {\bibinfo {author} {\bibfnamefont {T.}~\bibnamefont
  {Gorda}}, \bibinfo {author} {\bibfnamefont {A.}~\bibnamefont {Helset}},
  \bibinfo {author} {\bibfnamefont {L.}~\bibnamefont {Niemi}}, \bibinfo
  {author} {\bibfnamefont {T.~V.~I.}\ \bibnamefont {Tenkanen}}, \ and\ \bibinfo
  {author} {\bibfnamefont {D.~J.}\ \bibnamefont {Weir}},\ }\href {\doibase
  10.1007/JHEP02(2019)081} {\bibfield  {journal} {\bibinfo  {journal} {JHEP}\
  }\textbf {\bibinfo {volume} {02}},\ \bibinfo {pages} {081} (\bibinfo {year}
  {2019})},\ \Eprint {http://arxiv.org/abs/1802.05056} {arXiv:1802.05056
  [hep-ph]} \BibitemShut {NoStop}%
\bibitem [{\citenamefont {Gould}\ \emph {et~al.}(2019)\citenamefont {Gould},
  \citenamefont {Kozaczuk}, \citenamefont {Niemi}, \citenamefont
  {Ramsey-Musolf}, \citenamefont {Tenkanen},\ and\ \citenamefont
  {Weir}}]{Gould:2019qek}%
  \BibitemOpen
  \bibfield  {author} {\bibinfo {author} {\bibfnamefont {O.}~\bibnamefont
  {Gould}}, \bibinfo {author} {\bibfnamefont {J.}~\bibnamefont {Kozaczuk}},
  \bibinfo {author} {\bibfnamefont {L.}~\bibnamefont {Niemi}}, \bibinfo
  {author} {\bibfnamefont {M.~J.}\ \bibnamefont {Ramsey-Musolf}}, \bibinfo
  {author} {\bibfnamefont {T.~V.~I.}\ \bibnamefont {Tenkanen}}, \ and\ \bibinfo
  {author} {\bibfnamefont {D.~J.}\ \bibnamefont {Weir}},\ }\href@noop {} {\
  (\bibinfo {year} {2019})},\ \Eprint {http://arxiv.org/abs/1903.11604}
  {arXiv:1903.11604 [hep-ph]} \BibitemShut {NoStop}%
\bibitem [{\citenamefont {Kainulainen}\ \emph {et~al.}(2019)\citenamefont
  {Kainulainen}, \citenamefont {Keus}, \citenamefont {Niemi}, \citenamefont
  {Rummukainen}, \citenamefont {Tenkanen},\ and\ \citenamefont
  {Vaskonen}}]{Kainulainen:2019kyp}%
  \BibitemOpen
  \bibfield  {author} {\bibinfo {author} {\bibfnamefont {K.}~\bibnamefont
  {Kainulainen}}, \bibinfo {author} {\bibfnamefont {V.}~\bibnamefont {Keus}},
  \bibinfo {author} {\bibfnamefont {L.}~\bibnamefont {Niemi}}, \bibinfo
  {author} {\bibfnamefont {K.}~\bibnamefont {Rummukainen}}, \bibinfo {author}
  {\bibfnamefont {T.~V.~I.}\ \bibnamefont {Tenkanen}}, \ and\ \bibinfo {author}
  {\bibfnamefont {V.}~\bibnamefont {Vaskonen}},\ }\href@noop {} {\  (\bibinfo
  {year} {2019})},\ \Eprint {http://arxiv.org/abs/1904.01329} {arXiv:1904.01329
  [hep-ph]} \BibitemShut {NoStop}%
\bibitem [{\citenamefont {Hindmarsh}\ \emph {et~al.}(2014)\citenamefont
  {Hindmarsh}, \citenamefont {Huber}, \citenamefont {Rummukainen},\ and\
  \citenamefont {Weir}}]{Hindmarsh:2013xza}%
  \BibitemOpen
  \bibfield  {author} {\bibinfo {author} {\bibfnamefont {M.}~\bibnamefont
  {Hindmarsh}}, \bibinfo {author} {\bibfnamefont {S.~J.}\ \bibnamefont
  {Huber}}, \bibinfo {author} {\bibfnamefont {K.}~\bibnamefont {Rummukainen}},
  \ and\ \bibinfo {author} {\bibfnamefont {D.~J.}\ \bibnamefont {Weir}},\
  }\href {\doibase 10.1103/PhysRevLett.112.041301} {\bibfield  {journal}
  {\bibinfo  {journal} {Phys.Rev.Lett.}\ }\textbf {\bibinfo {volume} {112}},\
  \bibinfo {pages} {041301} (\bibinfo {year} {2014})},\ \Eprint
  {http://arxiv.org/abs/1304.2433} {arXiv:1304.2433 [hep-ph]} \BibitemShut
  {NoStop}%
\bibitem [{\citenamefont {Giblin}\ and\ \citenamefont
  {Mertens}(2014)}]{Giblin:2014qia}%
  \BibitemOpen
  \bibfield  {author} {\bibinfo {author} {\bibfnamefont {J.~T.}\ \bibnamefont
  {Giblin}}\ and\ \bibinfo {author} {\bibfnamefont {J.~B.}\ \bibnamefont
  {Mertens}},\ }\href {\doibase 10.1103/PhysRevD.90.023532} {\bibfield
  {journal} {\bibinfo  {journal} {Phys.Rev.}\ }\textbf {\bibinfo {volume}
  {D90}},\ \bibinfo {pages} {023532} (\bibinfo {year} {2014})},\ \Eprint
  {http://arxiv.org/abs/1405.4005} {arXiv:1405.4005 [astro-ph.CO]} \BibitemShut
  {NoStop}%
\bibitem [{\citenamefont {Hindmarsh}\ \emph {et~al.}(2015)\citenamefont
  {Hindmarsh}, \citenamefont {Huber}, \citenamefont {Rummukainen},\ and\
  \citenamefont {Weir}}]{Hindmarsh:2015qta}%
  \BibitemOpen
  \bibfield  {author} {\bibinfo {author} {\bibfnamefont {M.}~\bibnamefont
  {Hindmarsh}}, \bibinfo {author} {\bibfnamefont {S.~J.}\ \bibnamefont
  {Huber}}, \bibinfo {author} {\bibfnamefont {K.}~\bibnamefont {Rummukainen}},
  \ and\ \bibinfo {author} {\bibfnamefont {D.~J.}\ \bibnamefont {Weir}},\
  }\href {\doibase 10.1103/PhysRevD.92.123009} {\bibfield  {journal} {\bibinfo
  {journal} {Phys. Rev.}\ }\textbf {\bibinfo {volume} {D92}},\ \bibinfo {pages}
  {123009} (\bibinfo {year} {2015})},\ \Eprint
  {http://arxiv.org/abs/1504.03291} {arXiv:1504.03291 [astro-ph.CO]}
  \BibitemShut {NoStop}%
\bibitem [{\citenamefont {Hindmarsh}\ \emph {et~al.}(2017)\citenamefont
  {Hindmarsh}, \citenamefont {Huber}, \citenamefont {Rummukainen},\ and\
  \citenamefont {Weir}}]{Hindmarsh:2017gnf}%
  \BibitemOpen
  \bibfield  {author} {\bibinfo {author} {\bibfnamefont {M.}~\bibnamefont
  {Hindmarsh}}, \bibinfo {author} {\bibfnamefont {S.~J.}\ \bibnamefont
  {Huber}}, \bibinfo {author} {\bibfnamefont {K.}~\bibnamefont {Rummukainen}},
  \ and\ \bibinfo {author} {\bibfnamefont {D.~J.}\ \bibnamefont {Weir}},\
  }\href {\doibase 10.1103/PhysRevD.96.103520} {\bibfield  {journal} {\bibinfo
  {journal} {Phys. Rev.}\ }\textbf {\bibinfo {volume} {D96}},\ \bibinfo {pages}
  {103520} (\bibinfo {year} {2017})},\ \Eprint
  {http://arxiv.org/abs/1704.05871} {arXiv:1704.05871 [astro-ph.CO]}
  \BibitemShut {NoStop}%
\bibitem [{\citenamefont {Hindmarsh}(2018)}]{Hindmarsh:2016lnk}%
  \BibitemOpen
  \bibfield  {author} {\bibinfo {author} {\bibfnamefont {M.}~\bibnamefont
  {Hindmarsh}},\ }\href {\doibase 10.1103/PhysRevLett.120.071301} {\bibfield
  {journal} {\bibinfo  {journal} {Phys. Rev. Lett.}\ }\textbf {\bibinfo
  {volume} {120}},\ \bibinfo {pages} {071301} (\bibinfo {year} {2018})},\
  \Eprint {http://arxiv.org/abs/1608.04735} {arXiv:1608.04735 [astro-ph.CO]}
  \BibitemShut {NoStop}%
\bibitem [{\citenamefont {Jinno}\ and\ \citenamefont
  {Takimoto}(2017)}]{Jinno:2016vai}%
  \BibitemOpen
  \bibfield  {author} {\bibinfo {author} {\bibfnamefont {R.}~\bibnamefont
  {Jinno}}\ and\ \bibinfo {author} {\bibfnamefont {M.}~\bibnamefont
  {Takimoto}},\ }\href {\doibase 10.1103/PhysRevD.95.024009} {\bibfield
  {journal} {\bibinfo  {journal} {Phys. Rev.}\ }\textbf {\bibinfo {volume}
  {D95}},\ \bibinfo {pages} {024009} (\bibinfo {year} {2017})},\ \Eprint
  {http://arxiv.org/abs/1605.01403} {arXiv:1605.01403 [astro-ph.CO]}
  \BibitemShut {NoStop}%
\bibitem [{\citenamefont {Konstandin}(2018)}]{Konstandin:2017sat}%
  \BibitemOpen
  \bibfield  {author} {\bibinfo {author} {\bibfnamefont {T.}~\bibnamefont
  {Konstandin}},\ }\href {\doibase 10.1088/1475-7516/2018/03/047} {\bibfield
  {journal} {\bibinfo  {journal} {JCAP}\ }\textbf {\bibinfo {volume} {1803}},\
  \bibinfo {pages} {047} (\bibinfo {year} {2018})},\ \Eprint
  {http://arxiv.org/abs/1712.06869} {arXiv:1712.06869 [astro-ph.CO]}
  \BibitemShut {NoStop}%
\bibitem [{\citenamefont {Witten}(1984)}]{Witten:1984rs}%
  \BibitemOpen
  \bibfield  {author} {\bibinfo {author} {\bibfnamefont {E.}~\bibnamefont
  {Witten}},\ }\href {\doibase 10.1103/PhysRevD.30.272} {\bibfield  {journal}
  {\bibinfo  {journal} {Phys.Rev.}\ }\textbf {\bibinfo {volume} {D30}},\
  \bibinfo {pages} {272} (\bibinfo {year} {1984})}\BibitemShut {NoStop}%
\bibitem [{\citenamefont {Kurki-Suonio}(1985)}]{KurkiSuonio:1984ba}%
  \BibitemOpen
  \bibfield  {author} {\bibinfo {author} {\bibfnamefont {H.}~\bibnamefont
  {Kurki-Suonio}},\ }\href {\doibase 10.1016/0550-3213(85)90135-X} {\bibfield
  {journal} {\bibinfo  {journal} {Nucl.Phys.}\ }\textbf {\bibinfo {volume}
  {B255}},\ \bibinfo {pages} {231} (\bibinfo {year} {1985})}\BibitemShut
  {NoStop}%
\bibitem [{\citenamefont {Kamionkowski}\ \emph {et~al.}(1994)\citenamefont
  {Kamionkowski}, \citenamefont {Kosowsky},\ and\ \citenamefont
  {Turner}}]{Kamionkowski:1993fg}%
  \BibitemOpen
  \bibfield  {author} {\bibinfo {author} {\bibfnamefont {M.}~\bibnamefont
  {Kamionkowski}}, \bibinfo {author} {\bibfnamefont {A.}~\bibnamefont
  {Kosowsky}}, \ and\ \bibinfo {author} {\bibfnamefont {M.~S.}\ \bibnamefont
  {Turner}},\ }\href {\doibase 10.1103/PhysRevD.49.2837} {\bibfield  {journal}
  {\bibinfo  {journal} {Phys.Rev.}\ }\textbf {\bibinfo {volume} {D49}},\
  \bibinfo {pages} {2837} (\bibinfo {year} {1994})},\ \Eprint
  {http://arxiv.org/abs/astro-ph/9310044} {arXiv:astro-ph/9310044 [astro-ph]}
  \BibitemShut {NoStop}%
\bibitem [{\citenamefont {Caprini}\ \emph {et~al.}(2008)\citenamefont
  {Caprini}, \citenamefont {Durrer},\ and\ \citenamefont
  {Servant}}]{Caprini:2007xq}%
  \BibitemOpen
  \bibfield  {author} {\bibinfo {author} {\bibfnamefont {C.}~\bibnamefont
  {Caprini}}, \bibinfo {author} {\bibfnamefont {R.}~\bibnamefont {Durrer}}, \
  and\ \bibinfo {author} {\bibfnamefont {G.}~\bibnamefont {Servant}},\ }\href
  {\doibase 10.1103/PhysRevD.77.124015} {\bibfield  {journal} {\bibinfo
  {journal} {Phys.Rev.}\ }\textbf {\bibinfo {volume} {D77}},\ \bibinfo {pages}
  {124015} (\bibinfo {year} {2008})},\ \Eprint {http://arxiv.org/abs/0711.2593}
  {arXiv:0711.2593 [astro-ph]} \BibitemShut {NoStop}%
\bibitem [{\citenamefont {Gogoberidze}\ \emph {et~al.}(2007)\citenamefont
  {Gogoberidze}, \citenamefont {Kahniashvili},\ and\ \citenamefont
  {Kosowsky}}]{Gogoberidze:2007an}%
  \BibitemOpen
  \bibfield  {author} {\bibinfo {author} {\bibfnamefont {G.}~\bibnamefont
  {Gogoberidze}}, \bibinfo {author} {\bibfnamefont {T.}~\bibnamefont
  {Kahniashvili}}, \ and\ \bibinfo {author} {\bibfnamefont {A.}~\bibnamefont
  {Kosowsky}},\ }\href {\doibase 10.1103/PhysRevD.76.083002} {\bibfield
  {journal} {\bibinfo  {journal} {Phys.Rev.}\ }\textbf {\bibinfo {volume}
  {D76}},\ \bibinfo {pages} {083002} (\bibinfo {year} {2007})},\ \Eprint
  {http://arxiv.org/abs/0705.1733} {arXiv:0705.1733 [astro-ph]} \BibitemShut
  {NoStop}%
\bibitem [{\citenamefont {Caprini}\ \emph
  {et~al.}(2009{\natexlab{a}})\citenamefont {Caprini}, \citenamefont {Durrer},\
  and\ \citenamefont {Servant}}]{Caprini:2009yp}%
  \BibitemOpen
  \bibfield  {author} {\bibinfo {author} {\bibfnamefont {C.}~\bibnamefont
  {Caprini}}, \bibinfo {author} {\bibfnamefont {R.}~\bibnamefont {Durrer}}, \
  and\ \bibinfo {author} {\bibfnamefont {G.}~\bibnamefont {Servant}},\ }\href
  {\doibase 10.1088/1475-7516/2009/12/024} {\bibfield  {journal} {\bibinfo
  {journal} {JCAP}\ }\textbf {\bibinfo {volume} {0912}},\ \bibinfo {pages}
  {024} (\bibinfo {year} {2009}{\natexlab{a}})},\ \Eprint
  {http://arxiv.org/abs/0909.0622} {arXiv:0909.0622 [astro-ph.CO]} \BibitemShut
  {NoStop}%
\bibitem [{\citenamefont {Caprini}\ \emph
  {et~al.}(2009{\natexlab{b}})\citenamefont {Caprini}, \citenamefont {Durrer},
  \citenamefont {Konstandin},\ and\ \citenamefont {Servant}}]{Caprini:2009fx}%
  \BibitemOpen
  \bibfield  {author} {\bibinfo {author} {\bibfnamefont {C.}~\bibnamefont
  {Caprini}}, \bibinfo {author} {\bibfnamefont {R.}~\bibnamefont {Durrer}},
  \bibinfo {author} {\bibfnamefont {T.}~\bibnamefont {Konstandin}}, \ and\
  \bibinfo {author} {\bibfnamefont {G.}~\bibnamefont {Servant}},\ }\href
  {\doibase 10.1103/PhysRevD.79.083519} {\bibfield  {journal} {\bibinfo
  {journal} {Phys.Rev.}\ }\textbf {\bibinfo {volume} {D79}},\ \bibinfo {pages}
  {083519} (\bibinfo {year} {2009}{\natexlab{b}})},\ \Eprint
  {http://arxiv.org/abs/0901.1661} {arXiv:0901.1661 [astro-ph.CO]} \BibitemShut
  {NoStop}%
\bibitem [{\citenamefont {Niksa}\ \emph {et~al.}(2018)\citenamefont {Niksa},
  \citenamefont {Schlederer},\ and\ \citenamefont {Sigl}}]{Niksa:2018ofa}%
  \BibitemOpen
  \bibfield  {author} {\bibinfo {author} {\bibfnamefont {P.}~\bibnamefont
  {Niksa}}, \bibinfo {author} {\bibfnamefont {M.}~\bibnamefont {Schlederer}}, \
  and\ \bibinfo {author} {\bibfnamefont {G.}~\bibnamefont {Sigl}},\ }\href
  {\doibase 10.1088/1361-6382/aac89c} {\bibfield  {journal} {\bibinfo
  {journal} {Class. Quant. Grav.}\ }\textbf {\bibinfo {volume} {35}},\ \bibinfo
  {pages} {144001} (\bibinfo {year} {2018})},\ \Eprint
  {http://arxiv.org/abs/1803.02271} {arXiv:1803.02271 [astro-ph.CO]}
  \BibitemShut {NoStop}%
\bibitem [{\citenamefont {Ellis}\ \emph {et~al.}(2018)\citenamefont {Ellis},
  \citenamefont {Lewicki},\ and\ \citenamefont {No}}]{Ellis:2018mja}%
  \BibitemOpen
  \bibfield  {author} {\bibinfo {author} {\bibfnamefont {J.}~\bibnamefont
  {Ellis}}, \bibinfo {author} {\bibfnamefont {M.}~\bibnamefont {Lewicki}}, \
  and\ \bibinfo {author} {\bibfnamefont {J.~M.}\ \bibnamefont {No}},\ }\href
  {\doibase 10.1088/1475-7516/2019/04/003} {\  (\bibinfo {year} {2018}),\
  10.1088/1475-7516/2019/04/003},\ \bibinfo {note} {[JCAP1904,003(2019)]},\
  \Eprint {http://arxiv.org/abs/1809.08242} {arXiv:1809.08242 [hep-ph]}
  \BibitemShut {NoStop}%
\bibitem [{\citenamefont {Ellis}\ \emph {et~al.}(2019)\citenamefont {Ellis},
  \citenamefont {Lewicki}, \citenamefont {No},\ and\ \citenamefont
  {Vaskonen}}]{Ellis:2019oqb}%
  \BibitemOpen
  \bibfield  {author} {\bibinfo {author} {\bibfnamefont {J.}~\bibnamefont
  {Ellis}}, \bibinfo {author} {\bibfnamefont {M.}~\bibnamefont {Lewicki}},
  \bibinfo {author} {\bibfnamefont {J.~M.}\ \bibnamefont {No}}, \ and\ \bibinfo
  {author} {\bibfnamefont {V.}~\bibnamefont {Vaskonen}},\ }\href@noop {} {\
  (\bibinfo {year} {2019})},\ \Eprint {http://arxiv.org/abs/1903.09642}
  {arXiv:1903.09642 [hep-ph]} \BibitemShut {NoStop}%
\bibitem [{\citenamefont {Pol}\ \emph {et~al.}(2019)\citenamefont {Pol},
  \citenamefont {Mandal}, \citenamefont {Brandenburg}, \citenamefont
  {Kahniashvili},\ and\ \citenamefont {Kosowsky}}]{Pol:2019yex}%
  \BibitemOpen
  \bibfield  {author} {\bibinfo {author} {\bibfnamefont {A.~R.}\ \bibnamefont
  {Pol}}, \bibinfo {author} {\bibfnamefont {S.}~\bibnamefont {Mandal}},
  \bibinfo {author} {\bibfnamefont {A.}~\bibnamefont {Brandenburg}}, \bibinfo
  {author} {\bibfnamefont {T.}~\bibnamefont {Kahniashvili}}, \ and\ \bibinfo
  {author} {\bibfnamefont {A.}~\bibnamefont {Kosowsky}},\ }\href@noop {} {\
  (\bibinfo {year} {2019})},\ \Eprint {http://arxiv.org/abs/1903.08585}
  {arXiv:1903.08585 [astro-ph.CO]} \BibitemShut {NoStop}%
\bibitem [{\citenamefont {Jinno}\ \emph {et~al.}(2019)\citenamefont {Jinno},
  \citenamefont {Seong}, \citenamefont {Takimoto},\ and\ \citenamefont
  {Um}}]{Jinno:2019jhi}%
  \BibitemOpen
  \bibfield  {author} {\bibinfo {author} {\bibfnamefont {R.}~\bibnamefont
  {Jinno}}, \bibinfo {author} {\bibfnamefont {H.}~\bibnamefont {Seong}},
  \bibinfo {author} {\bibfnamefont {M.}~\bibnamefont {Takimoto}}, \ and\
  \bibinfo {author} {\bibfnamefont {C.~M.}\ \bibnamefont {Um}},\ }\href@noop {}
  {\  (\bibinfo {year} {2019})},\ \Eprint {http://arxiv.org/abs/1905.00899}
  {arXiv:1905.00899 [astro-ph.CO]} \BibitemShut {NoStop}%
\bibitem [{\citenamefont {Konstandin}\ and\ \citenamefont
  {No}(2011)}]{Konstandin:2010dm}%
  \BibitemOpen
  \bibfield  {author} {\bibinfo {author} {\bibfnamefont {T.}~\bibnamefont
  {Konstandin}}\ and\ \bibinfo {author} {\bibfnamefont {J.~M.}\ \bibnamefont
  {No}},\ }\href {\doibase 10.1088/1475-7516/2011/02/008} {\bibfield  {journal}
  {\bibinfo  {journal} {JCAP}\ }\textbf {\bibinfo {volume} {1102}},\ \bibinfo
  {pages} {008} (\bibinfo {year} {2011})},\ \Eprint
  {http://arxiv.org/abs/1011.3735} {arXiv:1011.3735 [hep-ph]} \BibitemShut
  {NoStop}%
\bibitem [{\citenamefont {M{\'e}gevand}\ and\ \citenamefont
  {Ram{\'\i}rez}(2018)}]{Megevand:2017vtb}%
  \BibitemOpen
  \bibfield  {author} {\bibinfo {author} {\bibfnamefont {A.}~\bibnamefont
  {M{\'e}gevand}}\ and\ \bibinfo {author} {\bibfnamefont {S.}~\bibnamefont
  {Ram{\'\i}rez}},\ }\href {\doibase 10.1016/j.nuclphysb.2018.01.012}
  {\bibfield  {journal} {\bibinfo  {journal} {Nucl. Phys.}\ }\textbf {\bibinfo
  {volume} {B928}},\ \bibinfo {pages} {38} (\bibinfo {year} {2018})},\ \Eprint
  {http://arxiv.org/abs/1710.06279} {arXiv:1710.06279 [astro-ph.CO]}
  \BibitemShut {NoStop}%
\bibitem [{\citenamefont {Ignatius}\ \emph {et~al.}(1994)\citenamefont
  {Ignatius}, \citenamefont {Kajantie}, \citenamefont {Kurki-Suonio},\ and\
  \citenamefont {Laine}}]{Ignatius:1993qn}%
  \BibitemOpen
  \bibfield  {author} {\bibinfo {author} {\bibfnamefont {J.}~\bibnamefont
  {Ignatius}}, \bibinfo {author} {\bibfnamefont {K.}~\bibnamefont {Kajantie}},
  \bibinfo {author} {\bibfnamefont {H.}~\bibnamefont {Kurki-Suonio}}, \ and\
  \bibinfo {author} {\bibfnamefont {M.}~\bibnamefont {Laine}},\ }\href
  {\doibase 10.1103/PhysRevD.49.3854} {\bibfield  {journal} {\bibinfo
  {journal} {Phys.Rev.}\ }\textbf {\bibinfo {volume} {D49}},\ \bibinfo {pages}
  {3854} (\bibinfo {year} {1994})},\ \Eprint
  {http://arxiv.org/abs/astro-ph/9309059} {arXiv:astro-ph/9309059 [astro-ph]}
  \BibitemShut {NoStop}%
\bibitem [{\citenamefont {Liu}\ \emph {et~al.}(1992)\citenamefont {Liu},
  \citenamefont {McLerran},\ and\ \citenamefont {Turok}}]{Liu:1992tn}%
  \BibitemOpen
  \bibfield  {author} {\bibinfo {author} {\bibfnamefont {B.-H.}\ \bibnamefont
  {Liu}}, \bibinfo {author} {\bibfnamefont {L.~D.}\ \bibnamefont {McLerran}}, \
  and\ \bibinfo {author} {\bibfnamefont {N.}~\bibnamefont {Turok}},\ }\href
  {\doibase 10.1103/PhysRevD.46.2668} {\bibfield  {journal} {\bibinfo
  {journal} {Phys. Rev.}\ }\textbf {\bibinfo {volume} {D46}},\ \bibinfo {pages}
  {2668} (\bibinfo {year} {1992})}\BibitemShut {NoStop}%
\bibitem [{\citenamefont {Garcia-Bellido}\ \emph {et~al.}(2008)\citenamefont
  {Garcia-Bellido}, \citenamefont {Figueroa},\ and\ \citenamefont
  {Sastre}}]{GarciaBellido:2007af}%
  \BibitemOpen
  \bibfield  {author} {\bibinfo {author} {\bibfnamefont {J.}~\bibnamefont
  {Garcia-Bellido}}, \bibinfo {author} {\bibfnamefont {D.~G.}\ \bibnamefont
  {Figueroa}}, \ and\ \bibinfo {author} {\bibfnamefont {A.}~\bibnamefont
  {Sastre}},\ }\href {\doibase 10.1103/PhysRevD.77.043517} {\bibfield
  {journal} {\bibinfo  {journal} {Phys.Rev.}\ }\textbf {\bibinfo {volume}
  {D77}},\ \bibinfo {pages} {043517} (\bibinfo {year} {2008})},\ \Eprint
  {http://arxiv.org/abs/0707.0839} {arXiv:0707.0839 [hep-ph]} \BibitemShut
  {NoStop}%
\bibitem [{\citenamefont {Kurki-Suonio}\ and\ \citenamefont
  {Laine}(1996)}]{KurkiSuonio:1995vy}%
  \BibitemOpen
  \bibfield  {author} {\bibinfo {author} {\bibfnamefont {H.}~\bibnamefont
  {Kurki-Suonio}}\ and\ \bibinfo {author} {\bibfnamefont {M.}~\bibnamefont
  {Laine}},\ }\href {\doibase 10.1103/PhysRevD.54.7163} {\bibfield  {journal}
  {\bibinfo  {journal} {Phys.Rev.}\ }\textbf {\bibinfo {volume} {D54}},\
  \bibinfo {pages} {7163} (\bibinfo {year} {1996})},\ \Eprint
  {http://arxiv.org/abs/hep-ph/9512202} {arXiv:hep-ph/9512202 [hep-ph]}
  \BibitemShut {NoStop}%
\bibitem [{\citenamefont {Kurki-Suonio}\ and\ \citenamefont
  {Laine}(1995)}]{KurkiSuonio:1995pp}%
  \BibitemOpen
  \bibfield  {author} {\bibinfo {author} {\bibfnamefont {H.}~\bibnamefont
  {Kurki-Suonio}}\ and\ \bibinfo {author} {\bibfnamefont {M.}~\bibnamefont
  {Laine}},\ }\href {\doibase 10.1103/PhysRevD.51.5431} {\bibfield  {journal}
  {\bibinfo  {journal} {Phys.Rev.}\ }\textbf {\bibinfo {volume} {D51}},\
  \bibinfo {pages} {5431} (\bibinfo {year} {1995})},\ \Eprint
  {http://arxiv.org/abs/hep-ph/9501216} {arXiv:hep-ph/9501216 [hep-ph]}
  \BibitemShut {NoStop}%
\bibitem [{\citenamefont {John}\ and\ \citenamefont
  {Schmidt}(2001)}]{John:2000zq}%
  \BibitemOpen
  \bibfield  {author} {\bibinfo {author} {\bibfnamefont {P.}~\bibnamefont
  {John}}\ and\ \bibinfo {author} {\bibfnamefont {M.~G.}\ \bibnamefont
  {Schmidt}},\ }\href {\doibase 10.1016/S0550-3213(00)00768-9,
  10.1016/S0550-3213(02)01014-3} {\bibfield  {journal} {\bibinfo  {journal}
  {Nucl. Phys.}\ }\textbf {\bibinfo {volume} {B598}},\ \bibinfo {pages} {291}
  (\bibinfo {year} {2001})},\ \bibinfo {note} {[Erratum: Nucl.
  Phys.B648,449(2003)]},\ \Eprint {http://arxiv.org/abs/hep-ph/0002050}
  {arXiv:hep-ph/0002050 [hep-ph]} \BibitemShut {NoStop}%
\bibitem [{\citenamefont {Moore}\ and\ \citenamefont
  {Prokopec}(1995)}]{Moore:1995si}%
  \BibitemOpen
  \bibfield  {author} {\bibinfo {author} {\bibfnamefont {G.~D.}\ \bibnamefont
  {Moore}}\ and\ \bibinfo {author} {\bibfnamefont {T.}~\bibnamefont
  {Prokopec}},\ }\href {\doibase 10.1103/PhysRevD.52.7182} {\bibfield
  {journal} {\bibinfo  {journal} {Phys.Rev.}\ }\textbf {\bibinfo {volume}
  {D52}},\ \bibinfo {pages} {7182} (\bibinfo {year} {1995})},\ \Eprint
  {http://arxiv.org/abs/hep-ph/9506475} {arXiv:hep-ph/9506475 [hep-ph]}
  \BibitemShut {NoStop}%
\bibitem [{\citenamefont {Cutting}(2019)}]{external}%
  \BibitemOpen
  \bibfield  {author} {\bibinfo {author} {\bibfnamefont {D.}~\bibnamefont
  {Cutting}},\ }\href@noop {} {} (\bibinfo {year} {2019}),\ \bibinfo {note}
  {see movies of strong phase transitions available at
  {\url{https://vimeo.com/album/5968055}}}\BibitemShut {NoStop}%
\bibitem [{\citenamefont {Espinosa}\ \emph {et~al.}(2010)\citenamefont
  {Espinosa}, \citenamefont {Konstandin}, \citenamefont {No},\ and\
  \citenamefont {Servant}}]{Espinosa:2010hh}%
  \BibitemOpen
  \bibfield  {author} {\bibinfo {author} {\bibfnamefont {J.~R.}\ \bibnamefont
  {Espinosa}}, \bibinfo {author} {\bibfnamefont {T.}~\bibnamefont
  {Konstandin}}, \bibinfo {author} {\bibfnamefont {J.~M.}\ \bibnamefont {No}},
  \ and\ \bibinfo {author} {\bibfnamefont {G.}~\bibnamefont {Servant}},\ }\href
  {\doibase 10.1088/1475-7516/2010/06/028} {\bibfield  {journal} {\bibinfo
  {journal} {JCAP}\ }\textbf {\bibinfo {volume} {1006}},\ \bibinfo {pages}
  {028} (\bibinfo {year} {2010})},\ \Eprint {http://arxiv.org/abs/1004.4187}
  {arXiv:1004.4187 [hep-ph]} \BibitemShut {NoStop}%
\bibitem [{\citenamefont {Caprini}\ \emph {et~al.}(2016)\citenamefont {Caprini}
  \emph {et~al.}}]{Caprini:2015zlo}%
  \BibitemOpen
  \bibfield  {author} {\bibinfo {author} {\bibfnamefont {C.}~\bibnamefont
  {Caprini}} \emph {et~al.},\ }\href {\doibase 10.1088/1475-7516/2016/04/001}
  {\bibfield  {journal} {\bibinfo  {journal} {JCAP}\ }\textbf {\bibinfo
  {volume} {1604}},\ \bibinfo {pages} {001} (\bibinfo {year} {2016})},\ \Eprint
  {http://arxiv.org/abs/1512.06239} {arXiv:1512.06239 [astro-ph.CO]}
  \BibitemShut {NoStop}%
\end{thebibliography}%
\newpage
\section{Supplemental Material}

\twocolumngrid
\subsection{{Field and fluid equations of motion}}

{In order to obtain the equations of motion for our field and fluid
system we focus on the coupling between the field and fluid parts of
our energy momentum tensor.
The current of the energy-momentum tensor can be split into field and
fluid parts and coupled through a dissipative friction term,
\begin{align} \label{eqn:field_cur}
  [\partial_\mu T^{\mu \nu}]_\mathrm{field}& = (\partial_\mu
  \partial^\mu \phi) \partial^\nu \phi - \frac{\partial V}{\partial
                                             \phi} \partial^\nu \phi = \delta^\nu\text, \\
  \label{eqn:fluid_cur}
  [\partial_\mu T^{\mu \nu}]_\mathrm{fluid}& = \partial_\mu[(\epsilon
                                             + p)U^\mu U^\nu] +
                                             \partial^\nu p +\frac{\partial V}{\partial
                                             \phi} \partial^\nu \phi = -\delta^\nu\text.
\end{align}
We can then write this coupling term as
\begin{equation}
\delta^\nu = \eta U^\mu \partial_\mu \phi \partial^\nu \phi\text.
\end{equation}
}

{From these two equations we can extract the equation of motion for our
system. By taking Eq~(\ref{eqn:field_cur}) and dividing through by
$\delta^\nu \phi$ we obtain
\begin{equation}
  -\ddot{\phi} + \nabla^2 \phi - \frac{\partial V}{\partial \phi} =
  \eta \relgamma (\dot{\phi} + v^{i}\partial_i \phi).
\end{equation}
}

{
We find the equation of motion for the fluid energy density
$E=\relgamma \epsilon$ by contracting Eq~(\ref{eqn:fluid_cur}) with
$U_\nu$ giving 
\begin{align}
\dot{E} + \partial_i (E v^i) + p[\dot{\relgamma} + \partial(\relgamma
v^i)] - \frac{\partial V}{\partial \phi} \relgamma(\dot\phi + v^i
\partial_i \phi) \nonumber \\
= \eta \relgamma^2(\dot\phi + v^i \partial_i \phi)^2\text. 
\end{align}
}

{
Finally we obtain an expression for the fluid momentum density $Z_i =
\relgamma(\epsilon + p)U_i$ by considering the spatial components of Eq~(\ref{eqn:fluid_cur}),
\begin{equation}
 \dot{Z} + \partial_j (Z_i v^j) + \partial_i p + \frac{\partial
   V}{\partial \phi}\partial_i\phi = - \eta \relgamma(\dot\phi + v^j
 \partial_j \phi)\partial_i \phi\text.
\end{equation}
}

\subsection{{Gravitational waves}}

{
To obtain the gravitational wave energy density we must first
calculate the transverse traceless perturbations in the metric,
$h^{TT}_{ij}$. We operate in linearised gravity and therefore the equation
of motion for $h^{TT}_{ij}$ is
\begin{equation}
  \Box h^{TT}_{ij} = 16 \pi G T^{TT}_{ij},
\end{equation}
where $T^{TT}_{ij}$ is the transverse traceless projection of the energy-momentum tensor.
}

{
Due to the numerical cost of computing the transverse traceless
components of the energy-momentum tensor, it is useful to instead
track an auxiliary tensor $u_{ij}$ \cite{GarciaBellido:2007af} which evolves according to
\begin{equation}
  \Box u_{ij} = 16 \pi G T_{ij}.
\end{equation}
Then to obtain $h^{TT}_{ij}$ from $u_{ij}$ we apply the transverse
traceless projector in wave space,
\begin{equation}
\tilde{h}^{TT}_{ij}(\mathbf{k},t) = \Lambda_{ij,lm}(\mathbf{k}) \tilde{u}_{lm}(\mathbf{k},t)\text, 
\end{equation}
where
\begin{equation}
\Lambda_{ij,lm}(\mathbf{k} = P_{im}(\mathbf{k}) P_{jl}(\mathbf{k}) - \frac{1}{2}P_{ij}(\mathbf{k}) P_{lm}(\mathbf{k})\text,
\end{equation}
and
\begin{equation}
  P_{ij}(\mathbf{k})=\delta_{ij} - \hat{k}_i \hat{k}_j\text.
\end{equation}
This method then allows us to only need to perform the necessary
Fourier transforms and projections to calculate the gravitational wave
energy density at regular intervals rather than every timestep.
}

\subsection{{Resolution convergence}} 

\begin{figure}[htbp]
 \centering 
 \includegraphics[width=0.48\textwidth]{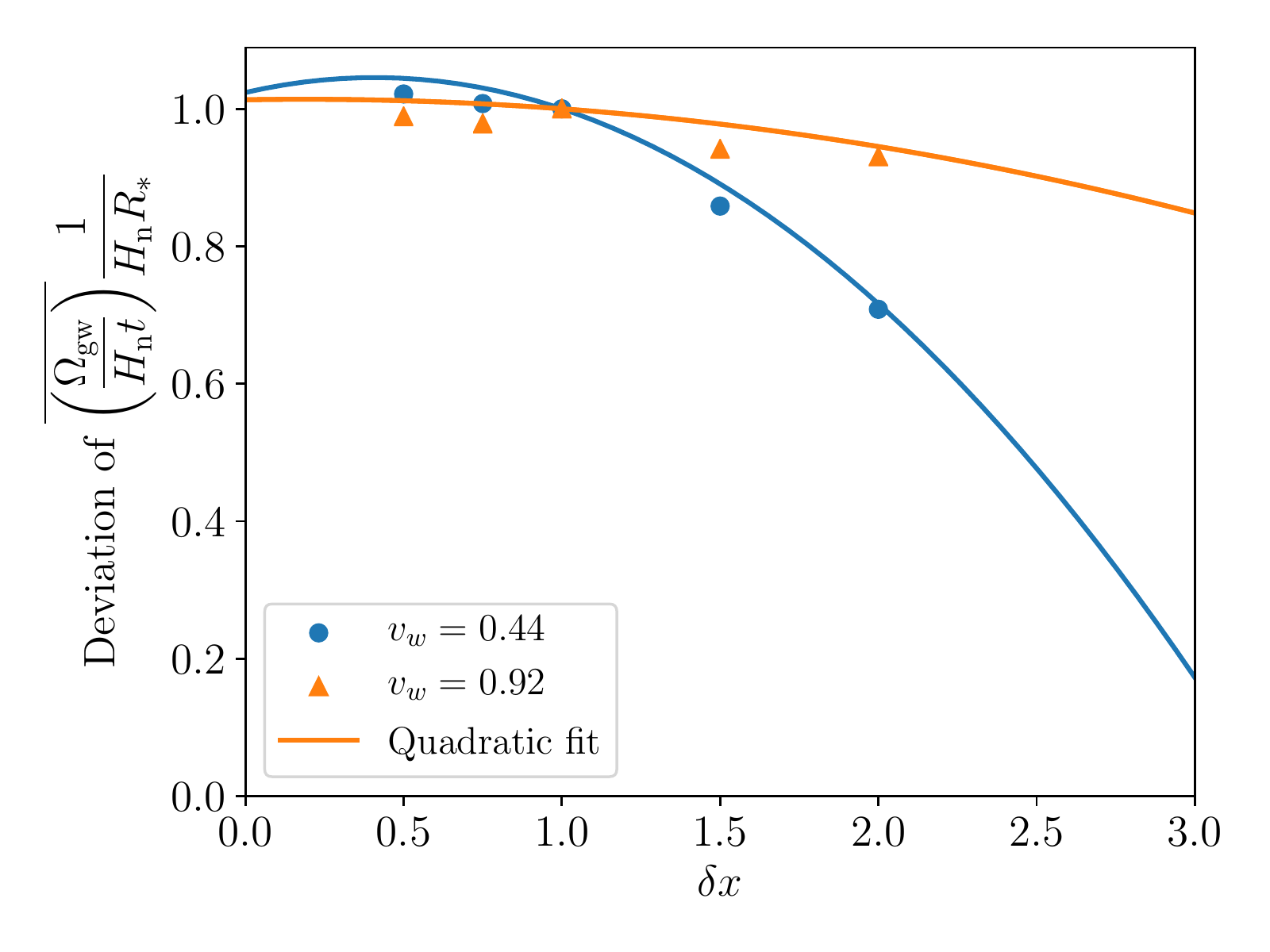}
 \caption{{Variation of
    gravitational wave energy density with
     $\delta x$ for $\vw=0.44$ and $\vw=0.92$ and transition strength
     of $\StrParB=0.5$. We normalise the $y$-axis by dividing by the result
     from the simulation presented in the paper ($\delta x=1.0$). Note
     that $\overline{\left(\OmGW/ \HN t \right)}$ signifies that we
     average the quantity inside the brackets over the final
     $\Delta t= 2 \Rbc$ of the simulation.}}
 \label{fig:lat-omgw}
\end{figure}

\begin{figure}[htbp]
 \centering 
 \includegraphics[width=0.48\textwidth]{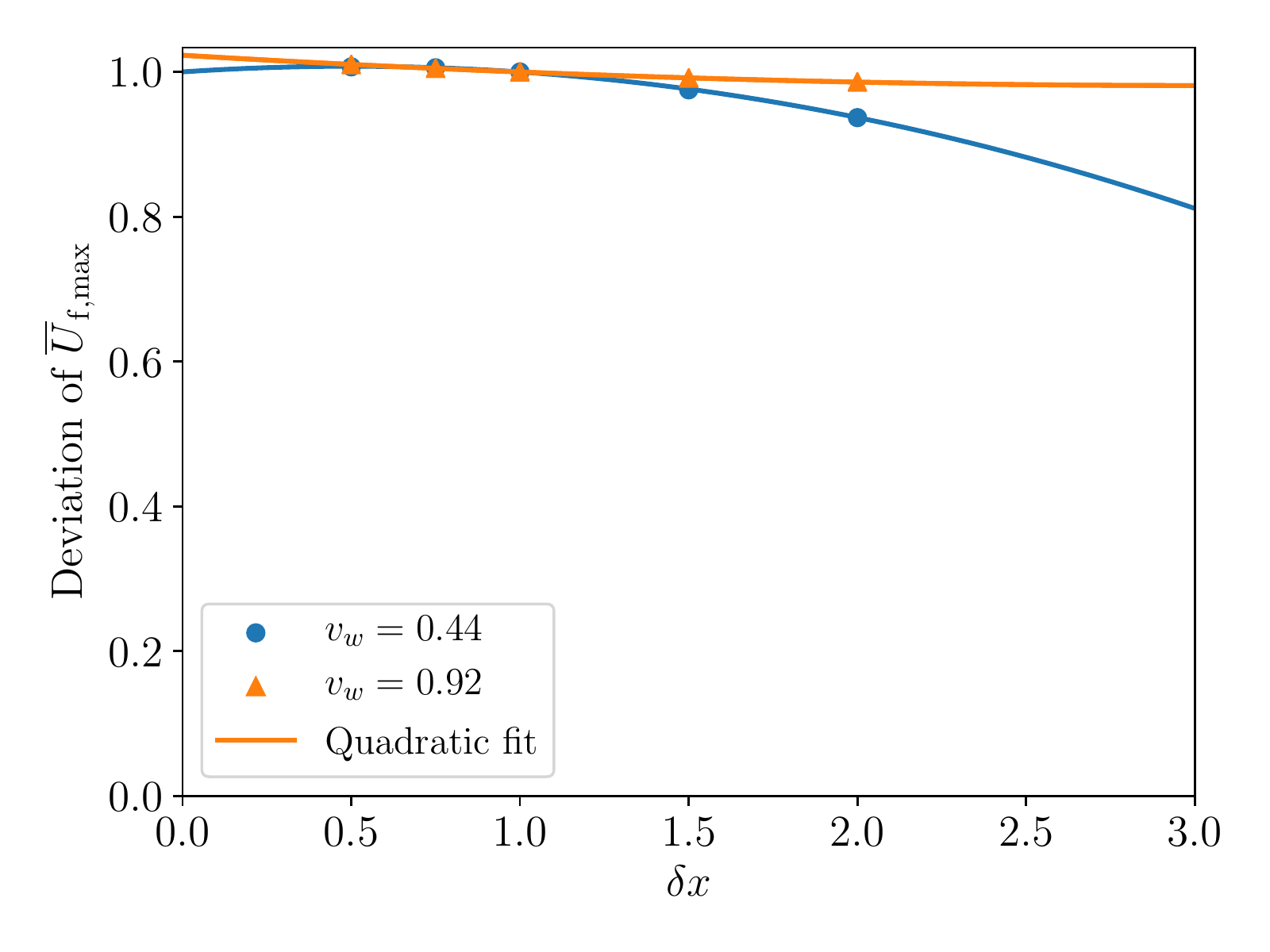}
 \caption{{Variation of
    $\Ubfmax$ with
     $\delta x$ for $\vw=0.44$ and $\vw=0.92$ and transition strength
     of $\StrParB=0.5$. We normalise the $y$-axis by dividing by the result
     from the simulation presented in the paper ($\delta x=1.0$).}}
 \label{fig:lat-ubarf}
\end{figure}

\begin{figure}[htbp]
 \centering 
 \includegraphics[width=0.48\textwidth]{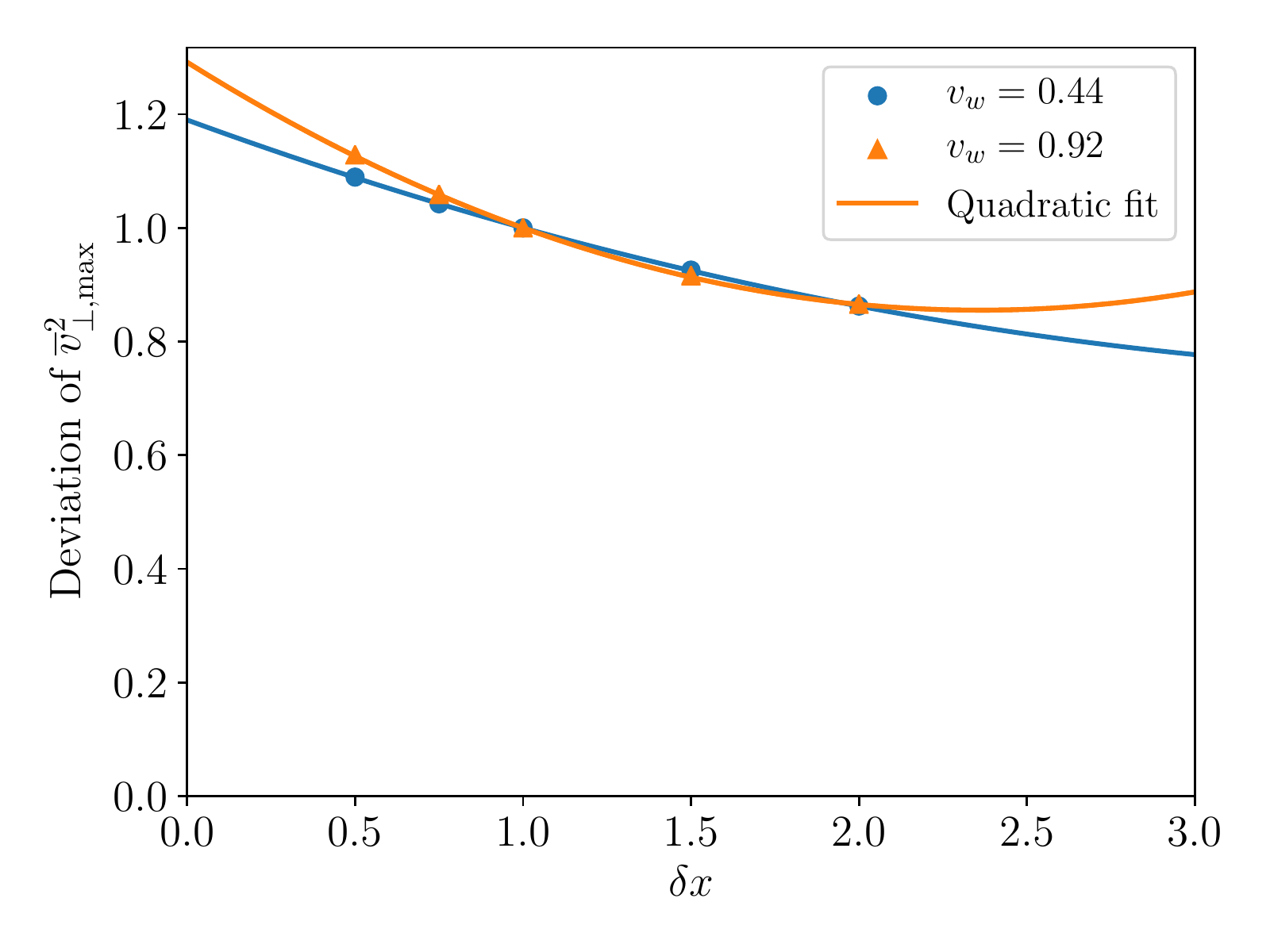}
 \caption{{Variation of
    $\VbPerpMax$ with
     $\delta x$ for $\vw=0.44$ and $\vw=0.92$ and transition strength
     of $\StrParB=0.5$. We normalise the $y$-axis by dividing by the result
     from the simulation presented in the paper ($\delta x=1.0$).}}
 \label{fig:lat-vperp}
\end{figure}

{To ensure the validity of our simulations we performed a series of
  lattice resolution checks. To do this we repeated two simulations
  with $\vw=0.44$ and $\vw=0.92$ and $\StrParB=0.5$ for a variety of
  different lattice spacings and timesteps while keeping the total
  physical volume and duration of the simulations fixed. We plot the
  convergence of several key quantities with $\delta x$ in
  Fig.~\ref{fig:lat-omgw} through Fig.~\ref{fig:lat-vperp}. We also
  plot a quadratic fit for the convergence of each quantity with
  $\delta x$. We can see that all quantities converge. The worst
  convergence is for $\VbPerpMax^2$ which for $\delta x=1.0\,\Tc^{-1}$ we underestimate by up to
  $25\%$ from the extrapolation to the continuum limit. We also
  performed tests for convergence of our simulations with $\delta
  t$. For $\delta t=0.2 \Tc^{-1}$ the error from our simulations is within
  $\sim1\%$ from the continuum limit for
  $\overline{\left(\OmGW/ \HN t \right)}(1/\HN\Rc)$ and $\Ubfmax$ and
  $\sim 5\%$ for $\VbPerpMax^2$.}

\subsection{{Convergence to asymptotic fluid flow}}

\begin{figure}[htbp]
 \centering 
 \includegraphics[width=0.48\textwidth]{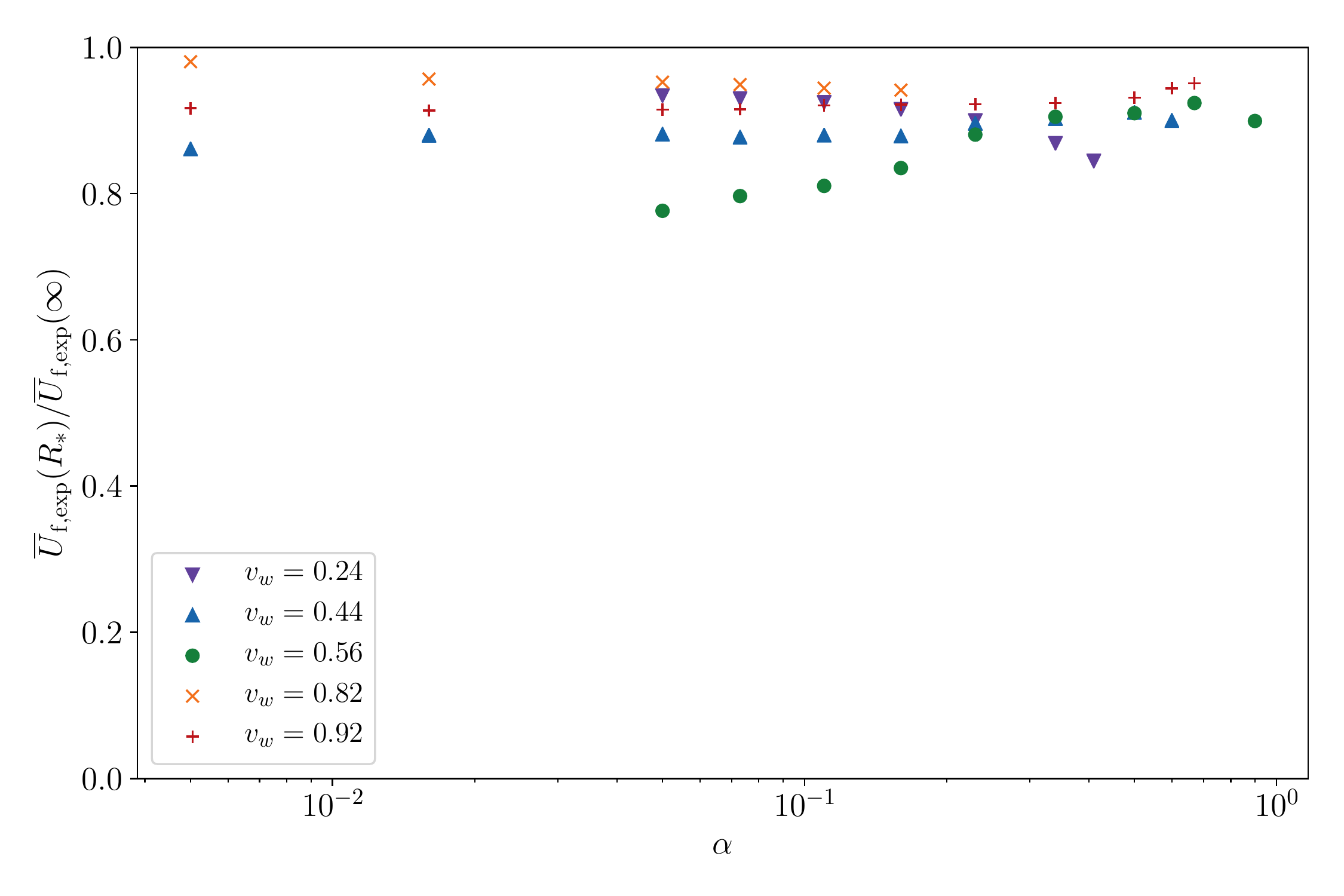}
 \caption{{Plot comparing $\UbfExp$ calculated for an isolated bubble
     when the diameter of
     the bubble is $\Rbc$ to late times ($t=10000\, T^{-1}_\mathrm{c}$)
     where it has reached the asymptotic profile. }}
 \label{fig:Ubar_f_conv}
\end{figure}

{ In addition to testing convergence with lattice spacing, we also
  check how close the fluid shells around colliding bubbles in our
  simulation are to the final asymptotic profiles. To do this
  we perform spherically symmetric 1D simulations of isolated bubbles
  and calculate $\UbfExp$ from the fluid shell when the bubble has
  diameter $\Rc$. We then compare this to $\UbfExp$ calculated from
  the fluid shell at $t=10000 T^{-1}_\mathrm{c}$, i.e when the
  diameter is $\gg \Rc$ and the profile has reached its asymptotic
  solution. We plot the ratio of these two quantities for all
  $\StrParB$ and $\vw$ in Fig.~\ref{fig:Ubar_f_conv}. We can see that
  the bubbles colliding with the diameter of the average bubble
  seperation are within $20\%$ of the asymptotic $\Ubf$ for all
  simulations. We believe this to be sufficient for this study, and
  save a further investigation on the convergence with increasing
  $\Rc$ for a future work. }

\subsection{Parameter space}

\begin{figure}[htbp]
\centering 
\includegraphics[width=0.48\textwidth]{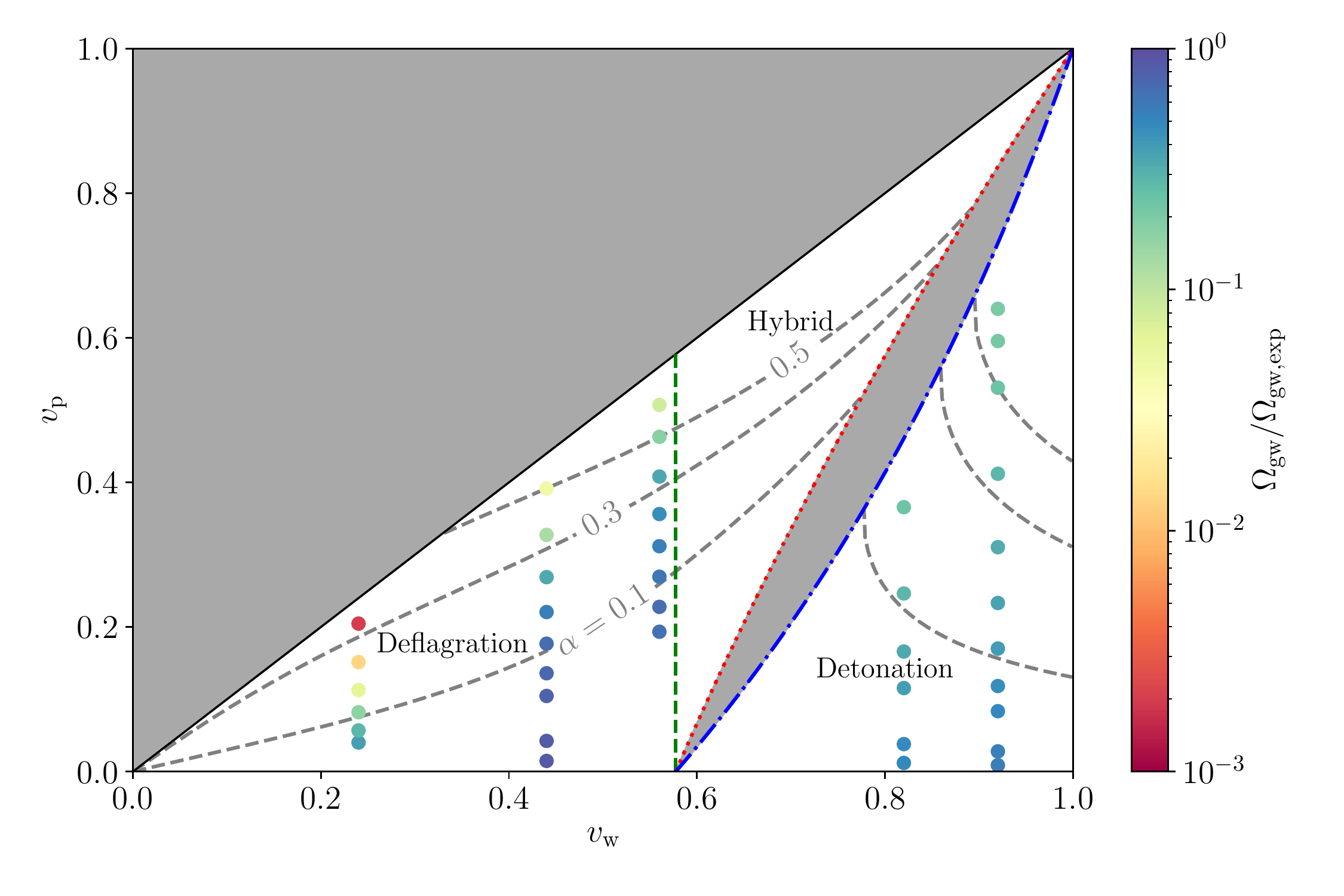}
\caption{Plot of maximum fluid flow velocity for the asymptotic
  profile $\vProfMax$ against the wall velocity. The green dashed line
  separates subsonic-deflagrations from supersonic-deflagrations. The
  blue dotted line gives the minimum $\vProfMax$ for a
  hybrid. Similarly the red dashed line shows the maximum $\vProfMax$
  for a detonation. In the grey regions there are no solutions. See
  Fig.~7 of \cite{Espinosa:2010hh} for more details. Each point has
  been coloured according to the suppression in gravitational waves
  given in Table~\ref{table:alldata}. Lines of constant $\StrParB$ are
  shown in dashed grey.}
\label{fig:vmax-vs-vw}
\end{figure}

In order to understand the regions of parameter space mapped out by
our simulations, it can be illuminating to plot the asymptotic maximum
fluid flow velocity $\vProfMax$ against the wall velocity
$\vw$
for each simulation point. We do this in
Fig.~\ref{fig:vmax-vs-vw}. Plotting the parameter space in this manner
separates subsonic-deflagrations, supersonic-deflagrations, and
detonations. Stronger phase transitions with the same wall velocity
have a larger value of $\vProfMax$. Transitions with $\vProfMax>\vw$
are forbidden as this would mean that in the wall frame fluid was
flowing out from the bubble. We additionally colour each simulation
point by the suppression factor in gravitational waves found in our study.

\subsection{Evolution of global quantities}

\begin{figure}[htbp]
  \centering
\subfigure{\includegraphics[width=0.48\textwidth]{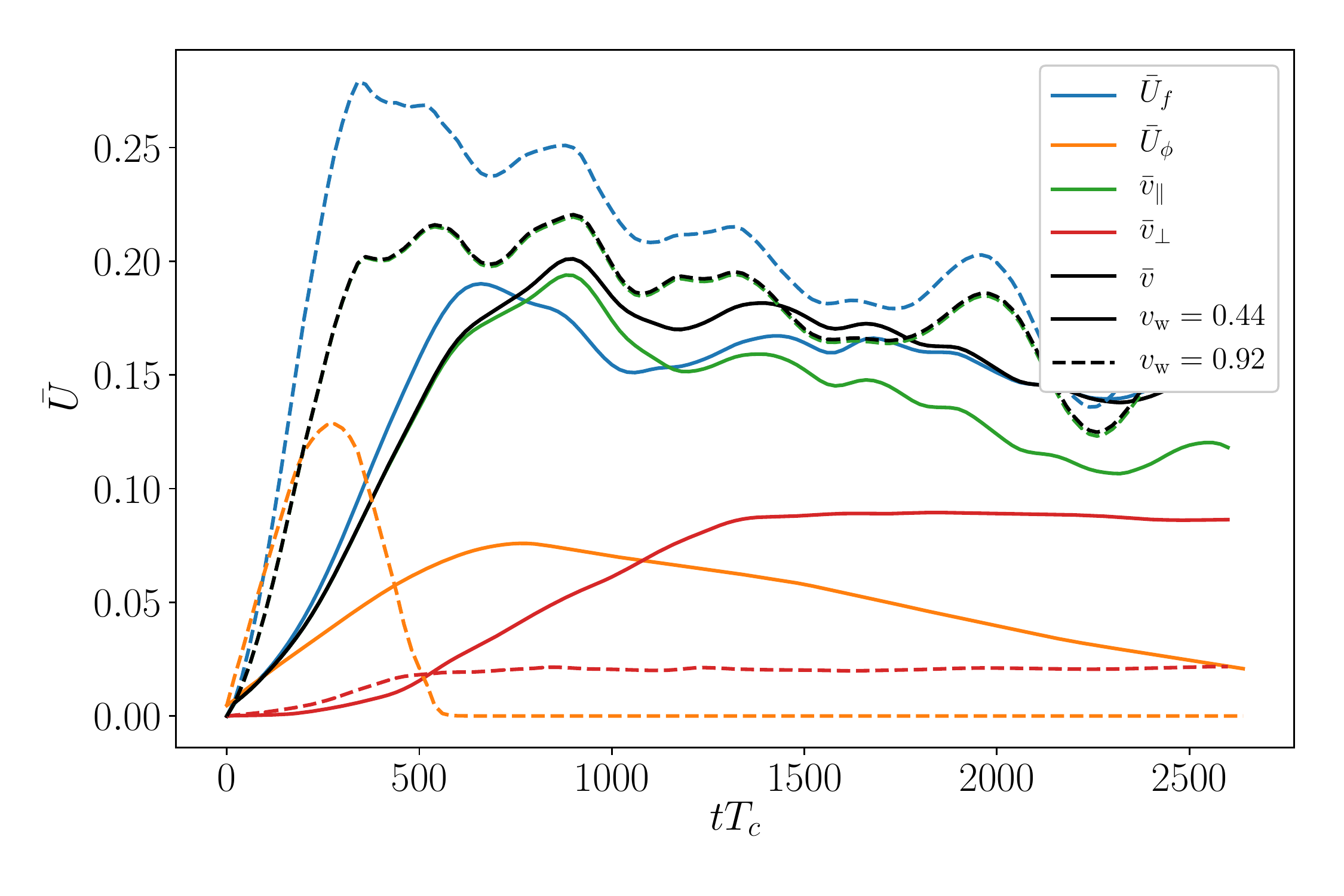}}
\caption{ The RMS fluid velocities decomposed into irrotational and
  rotational modes, plotted against time.  We also plot the quantities
  $\Ubf$ and $\Ubp$.  Solid
  lines show a subsonic deflagration with $\StrParB=0.5$,
  $\vwass=0.44$, and dashed lines a detonation with $\StrParB=0.5$,
  $\vwass=0.92$. }
  \label{fig:Vs-single-alph}
\end{figure}

In Fig.~\ref{fig:Vs-single-alph} we plot how $\Ubf$ and $\Ubp$ evolve
for a deflagration and a detonation, both with strength $\al = 0.5$.
We see that a rotational component of velocity $\VbPerp$ is generated
during the bubble collision phase, and that the deflagration generates
$\VbPerp$ more efficiently than the {detonation}.  We also see that,
for the deflagration, $\Ubp$ decreases more slowly than it increases,
indicating a slowing down of the phase boundary.

\subsection{Simulation slice stills}
\begin{figure}
  \centering
  \hfill
  \subfigure
  {\includegraphics[width=0.32\textwidth,clip]{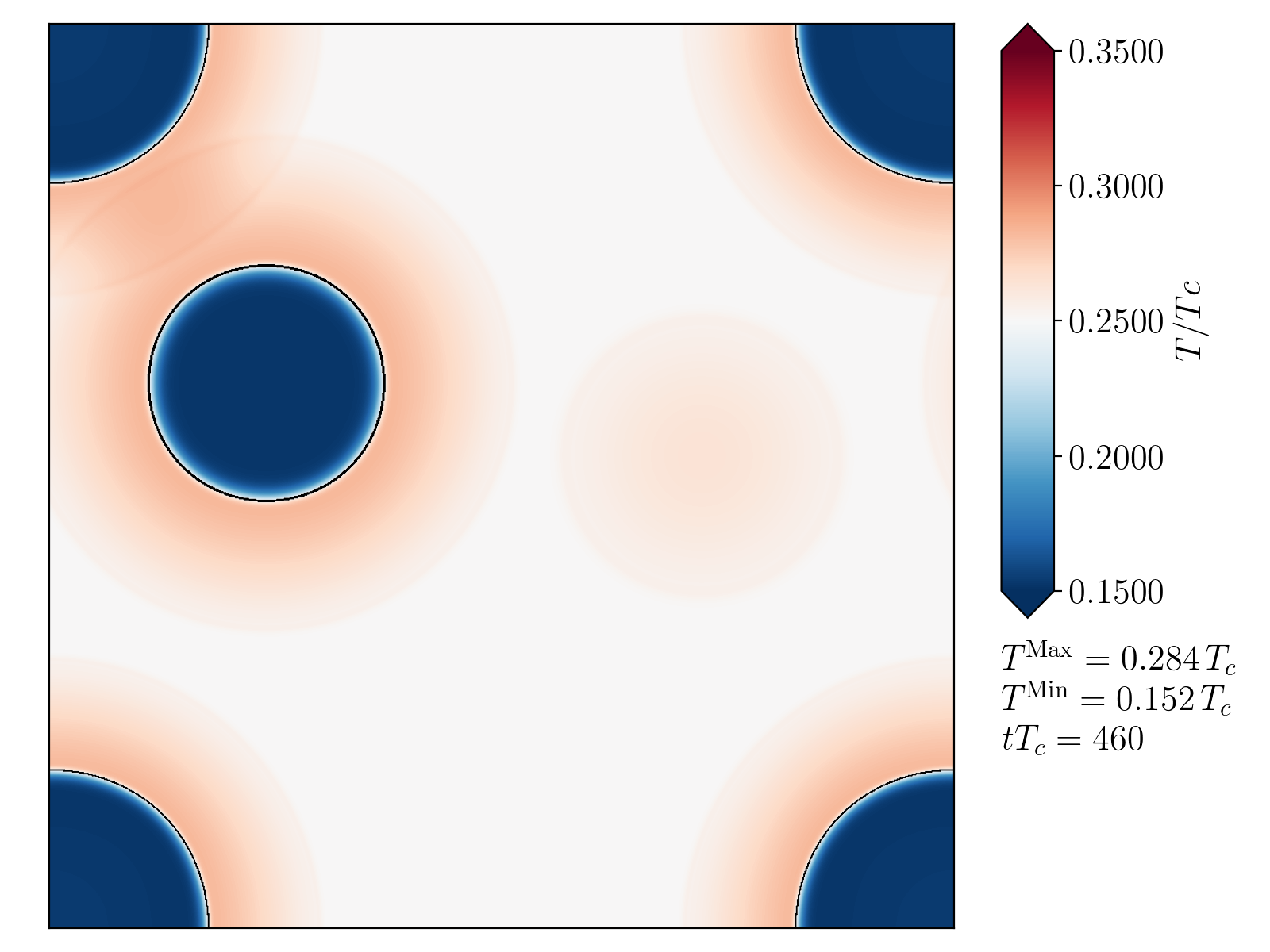}}
  \hfill
  \subfigure
  {\includegraphics[width=0.32\textwidth,clip]{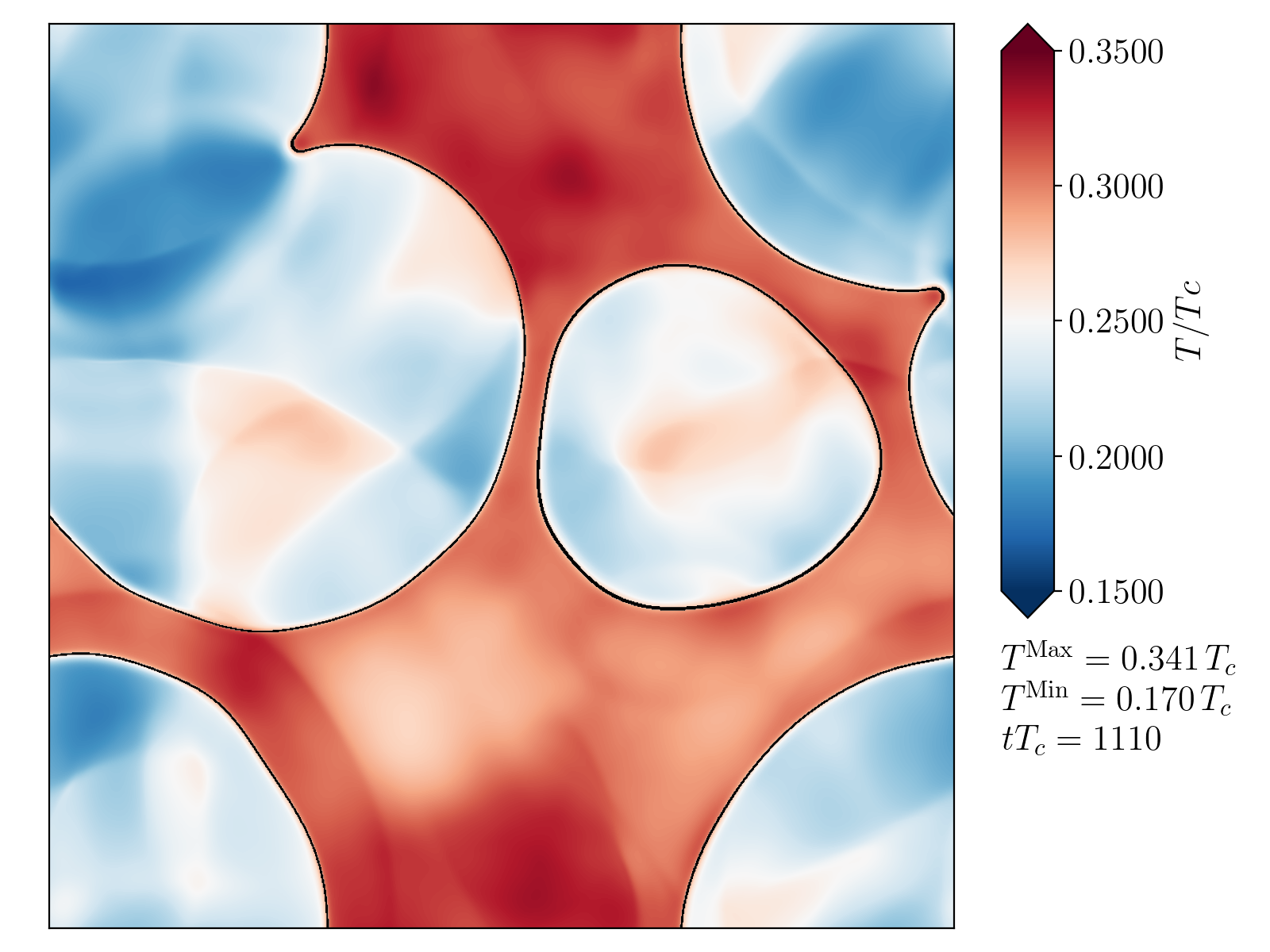}}
  \hfill
  \subfigure
  {\includegraphics[width=0.32\textwidth,clip]{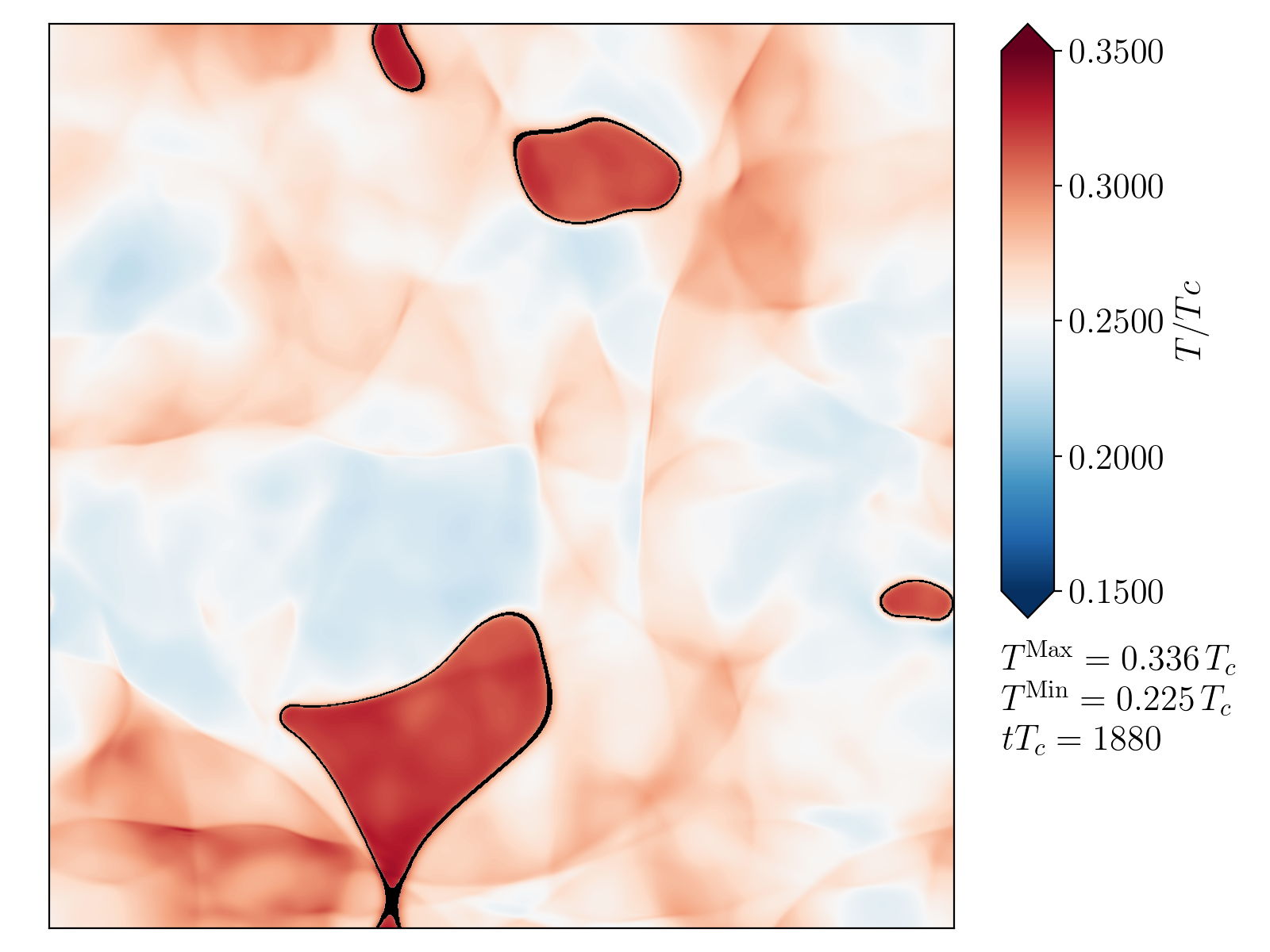}}
  \hfill
  \\
  \hfill
  \subfigure
  {\includegraphics[width=0.32\textwidth,clip]{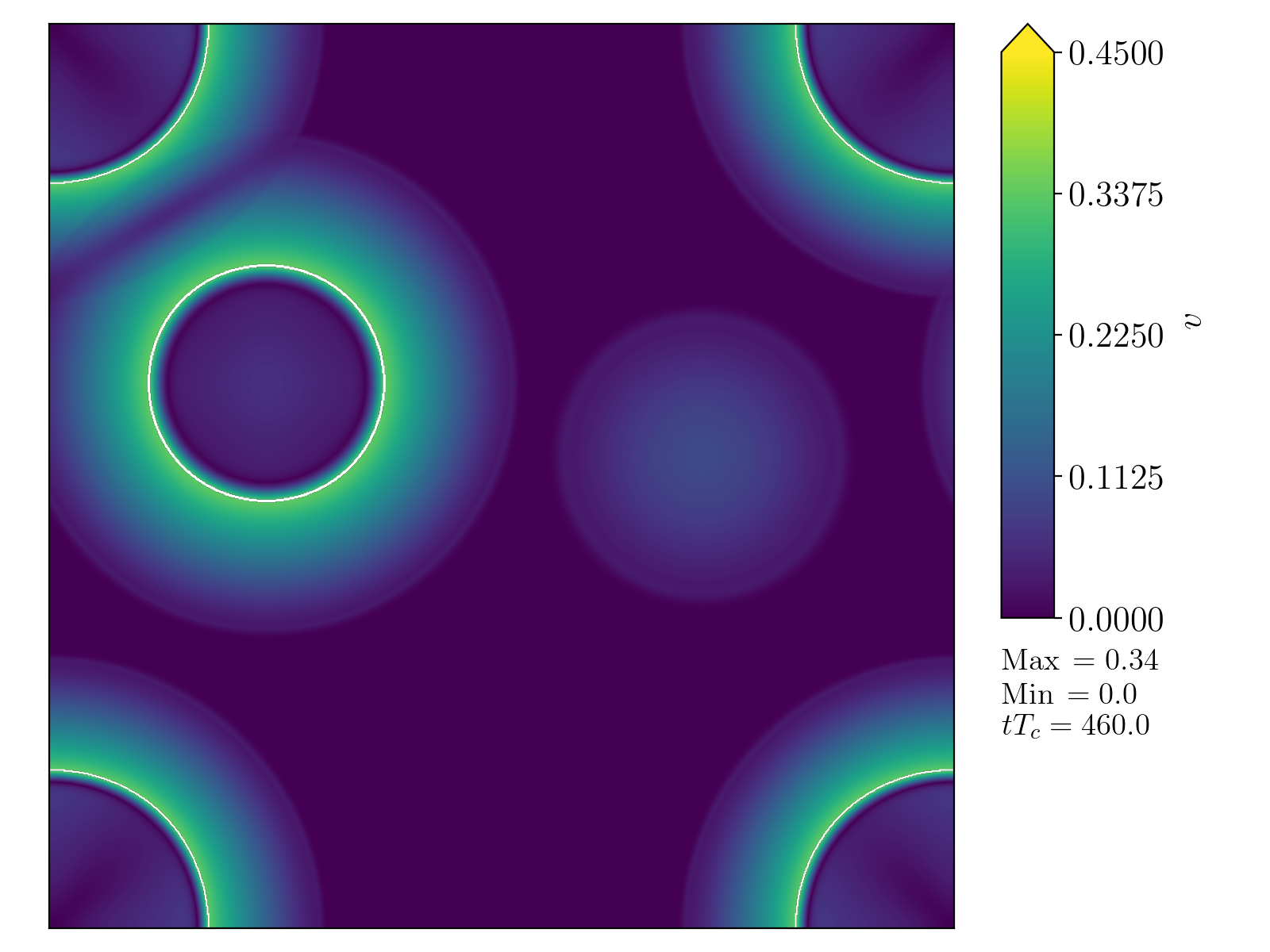}}
  \hfill
  \subfigure
  {\includegraphics[width=0.32\textwidth,clip]{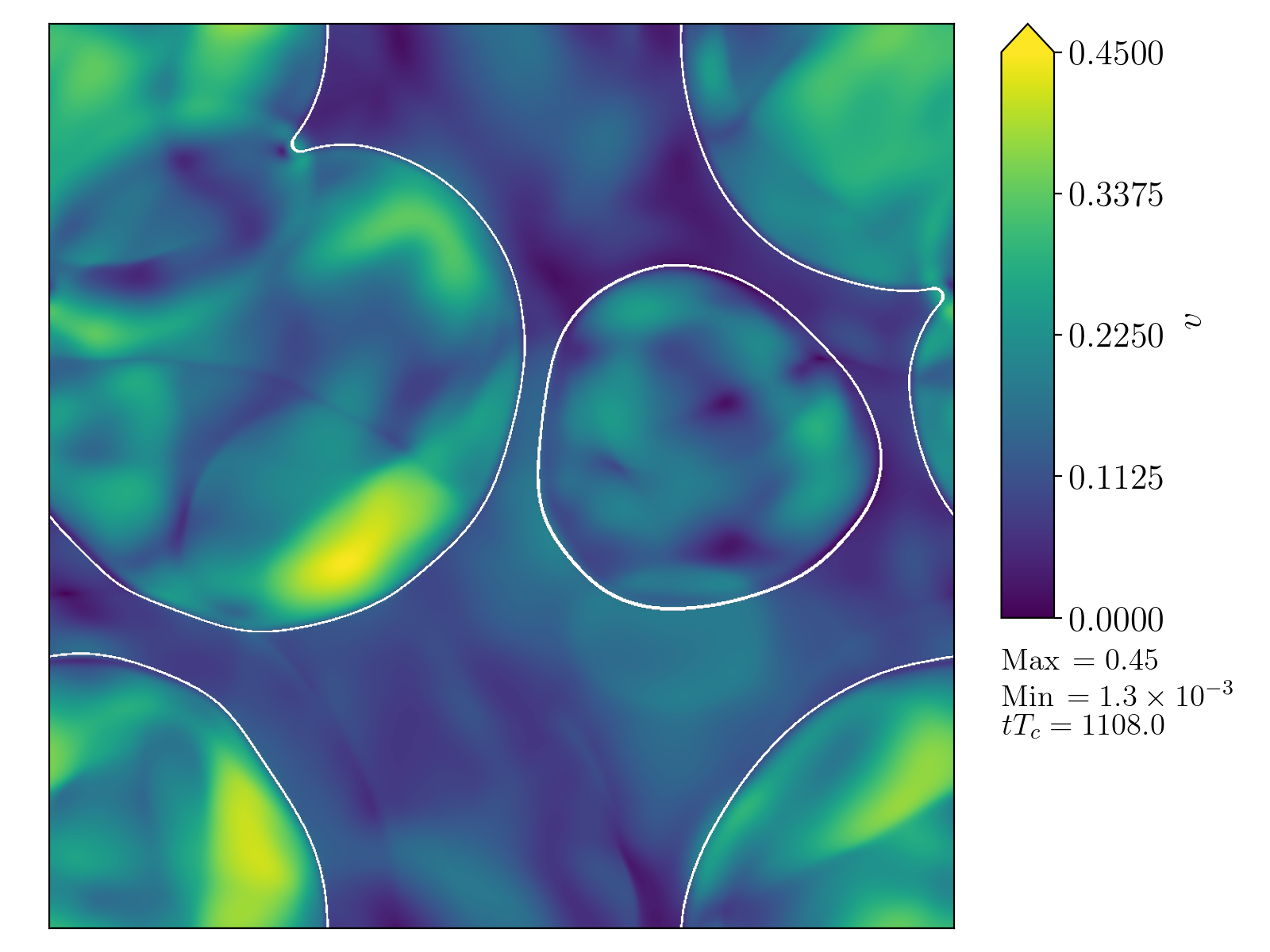}}
  \hfill
  \subfigure
  {\includegraphics[width=0.32\textwidth,clip]{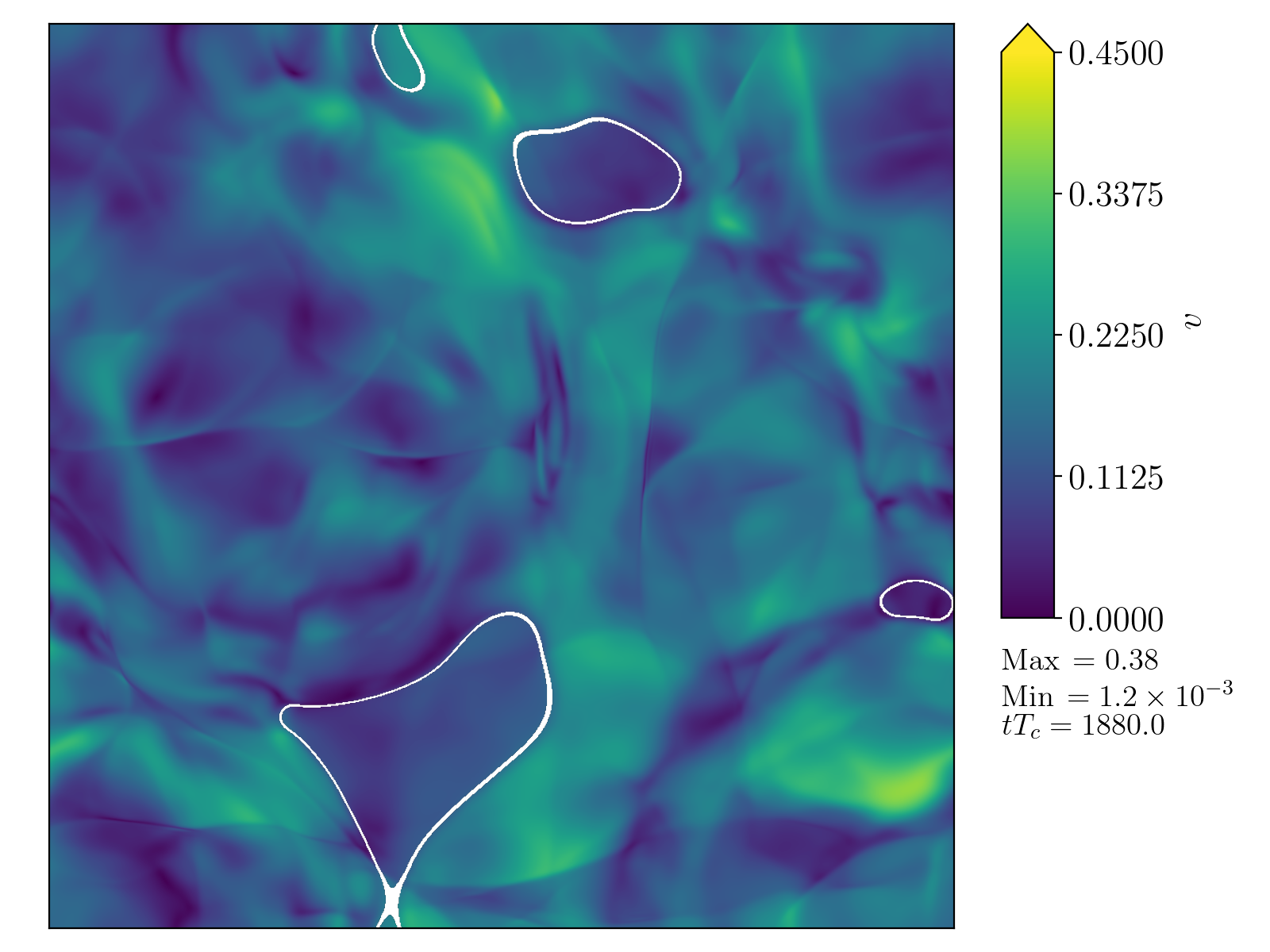}}
  \hfill
  \\
  \hfill
  \subfigure
  {\includegraphics[width=0.32\textwidth,clip]{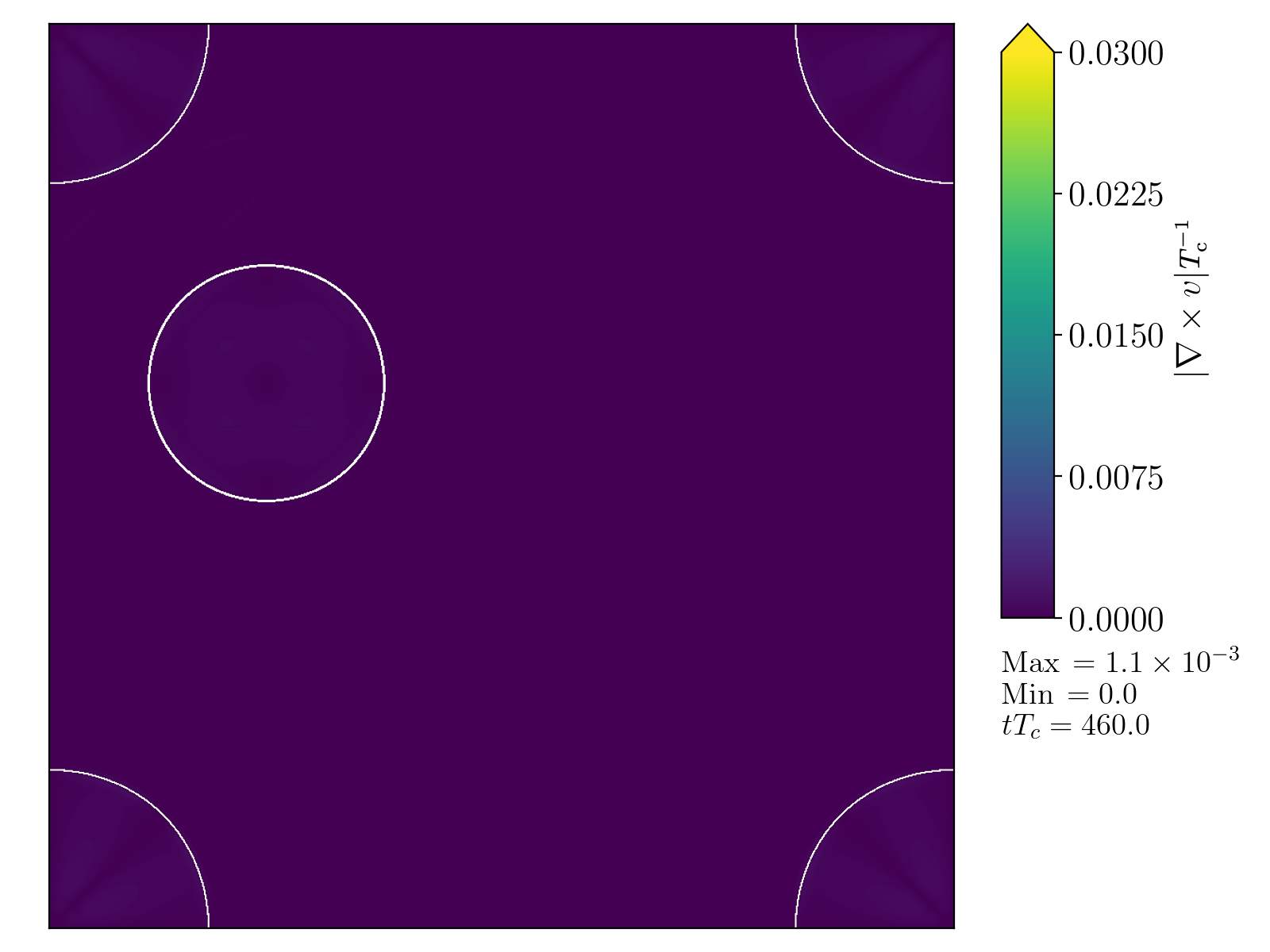}}
  \hfill
  \subfigure
  {\includegraphics[width=0.32\textwidth,clip]{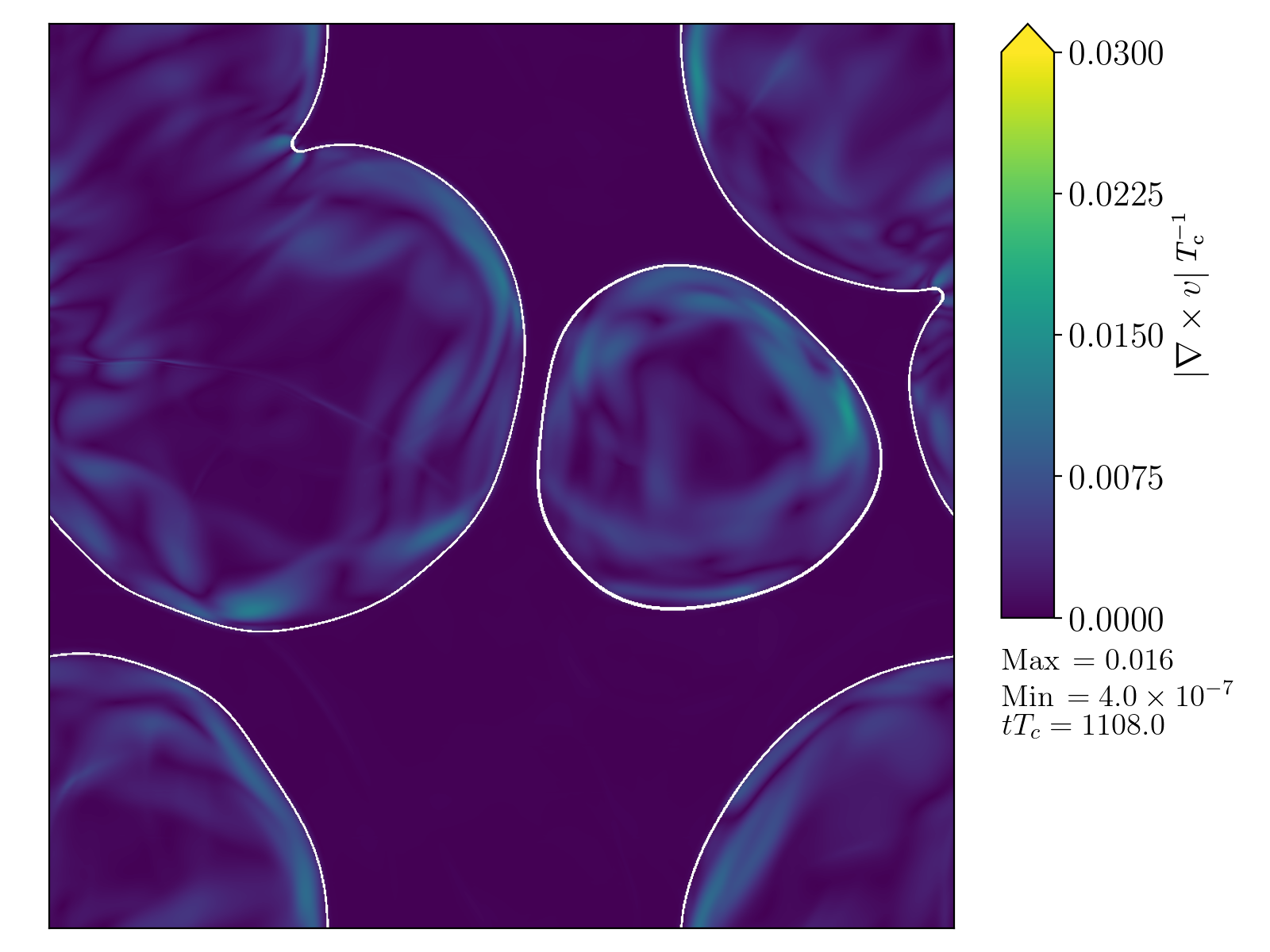}}
  \hfill
  \subfigure
  {\includegraphics[width=0.32\textwidth,clip]{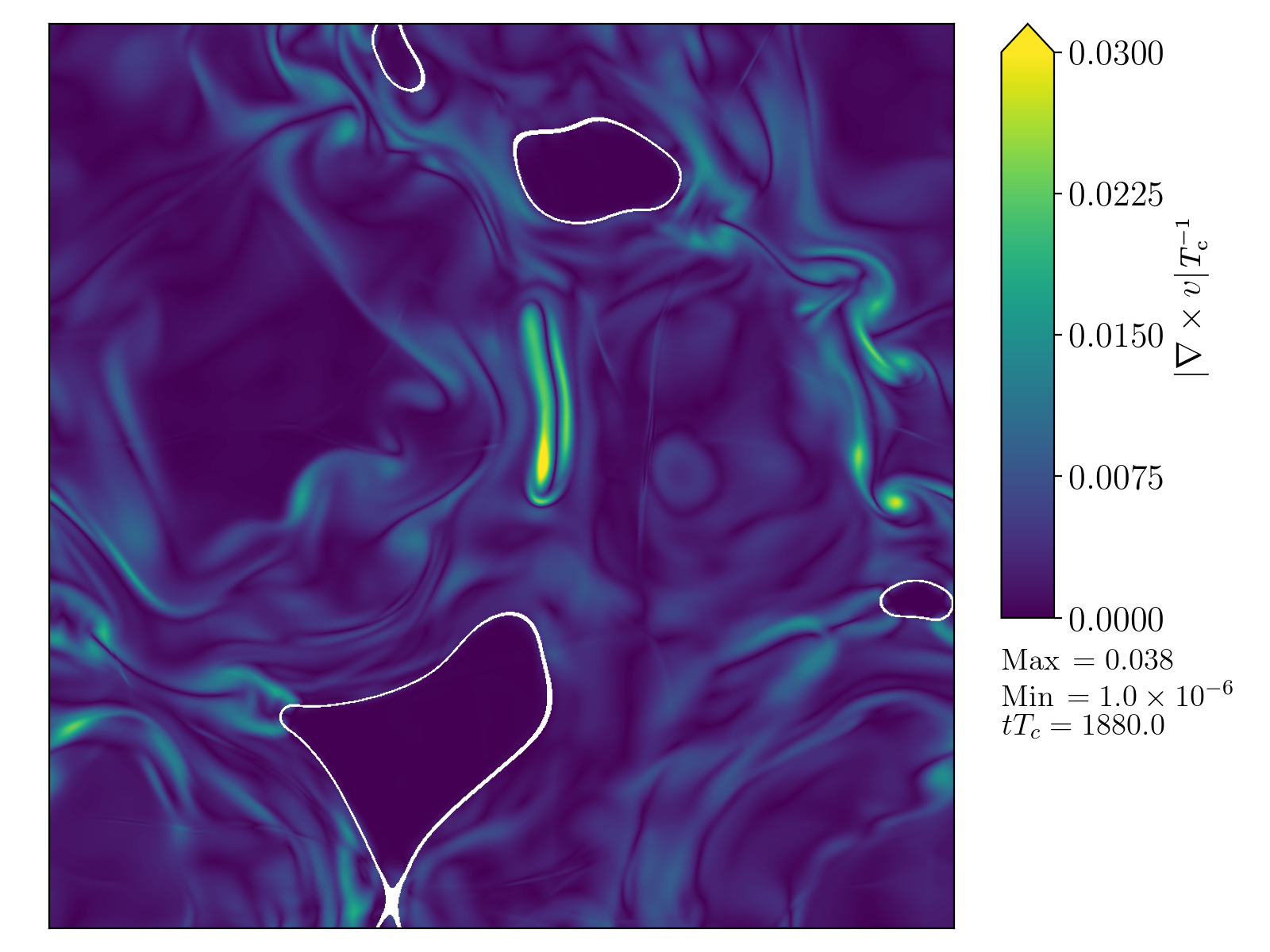}}
  \hfill
  \caption{Slices through $(0,y,z)$ for a simulation with $\vwass=0.44$,
    $\StrParB=0.5$, corresponding to a deflagration. In the top row we
    plot the temperature $T/\Tc$. The midpoint of this colormap
    corresponds to $\TN$. The middle row shows the fluid velocity
    $v$. The bottom row shows the vorticity
    $\left| \nabla \times v\right|$. The bubble walls are shaded in
    black for the top row, and white for the middle and bottom row. }
  \label{fig:slices-def-supl}
\end{figure}

\begin{figure}
  \centering
  \hfill
  \subfigure
  {\includegraphics[width=0.32\textwidth,clip]{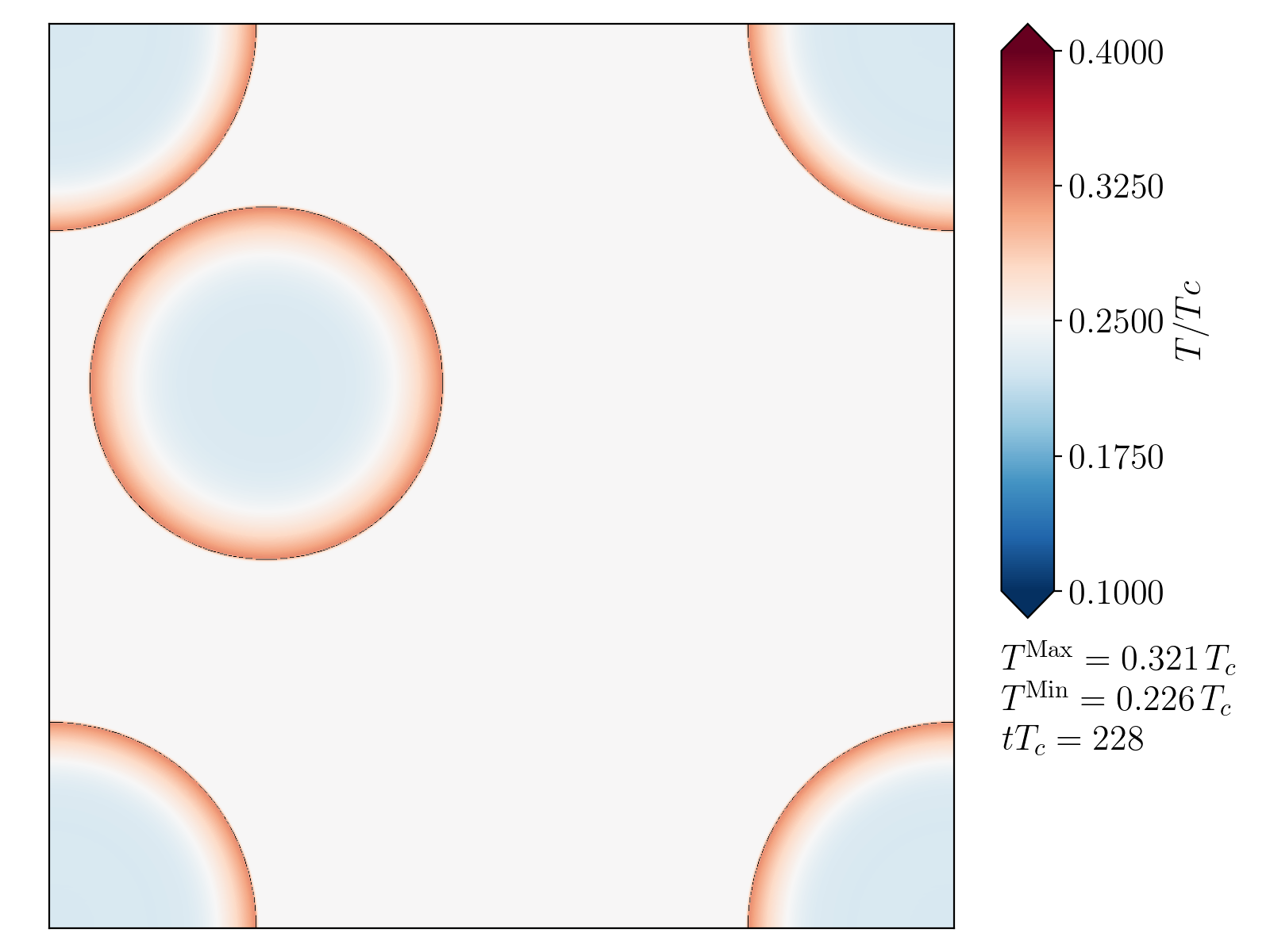}}
  \hfill
  \subfigure
  {\includegraphics[width=0.32\textwidth,clip]{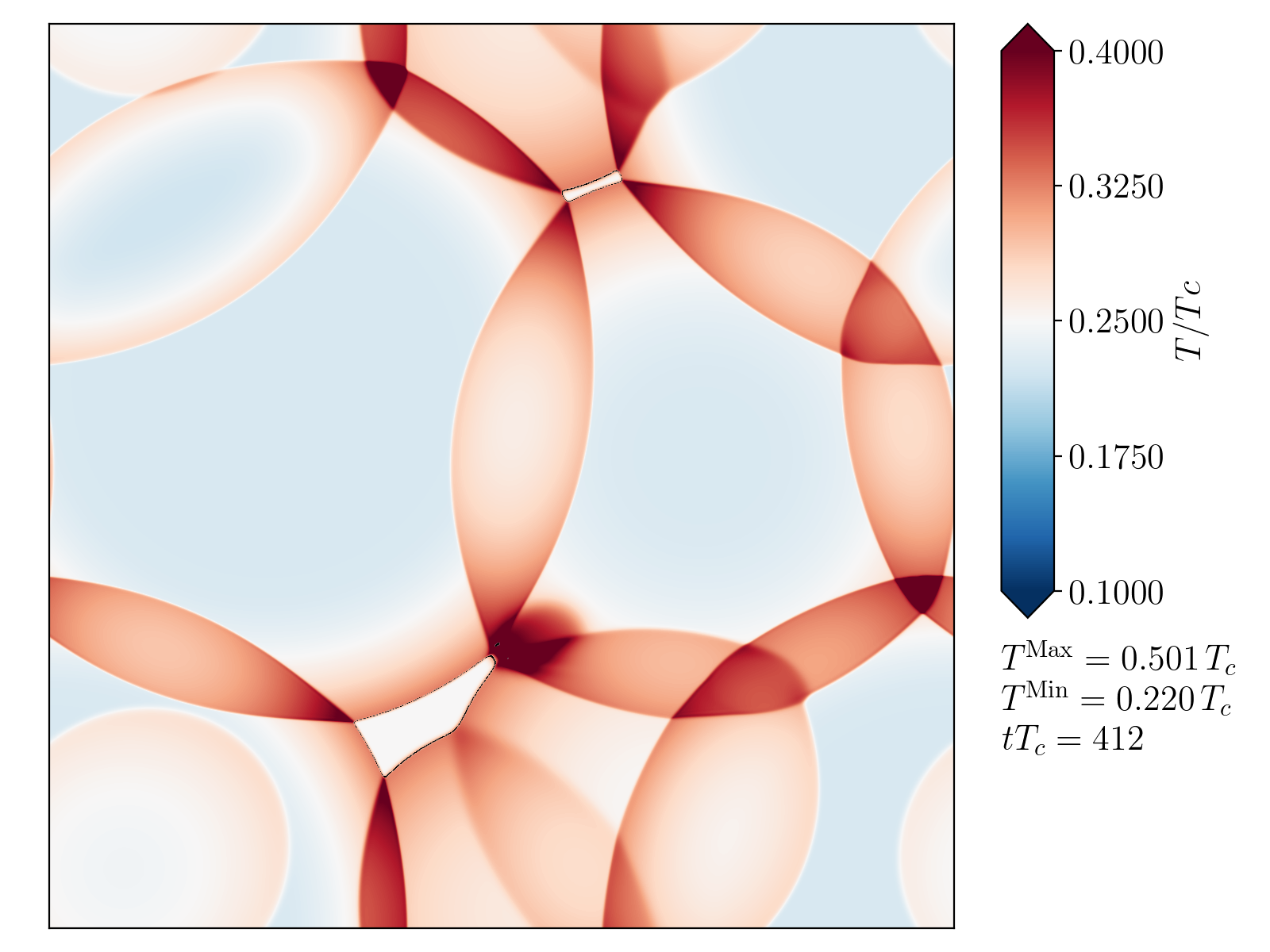}}
  \hfill
  \subfigure
  {\includegraphics[width=0.32\textwidth,clip]{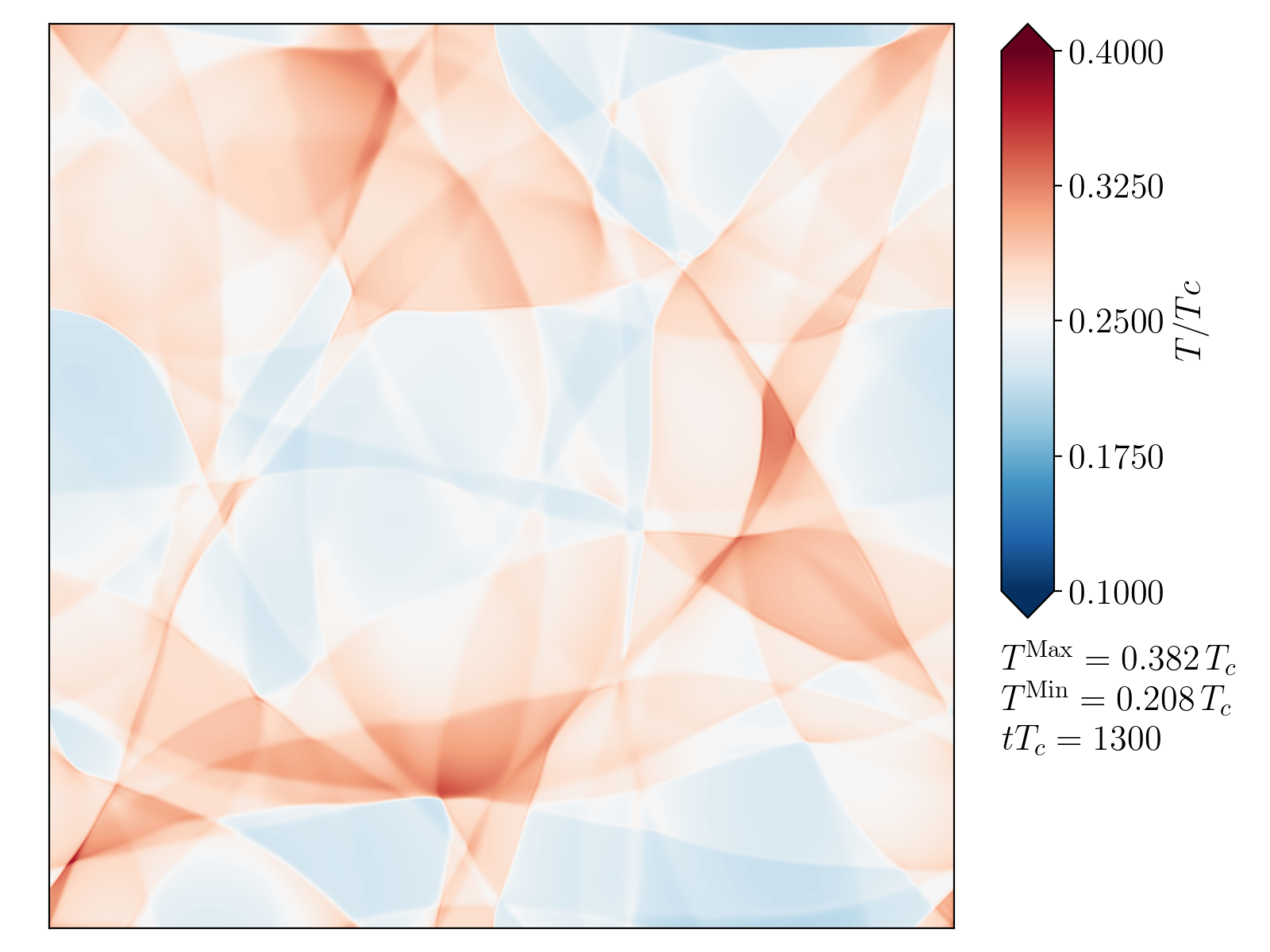}}
  \hfill
  \\
  \hfill
  \subfigure
  {\includegraphics[width=0.32\textwidth,clip]{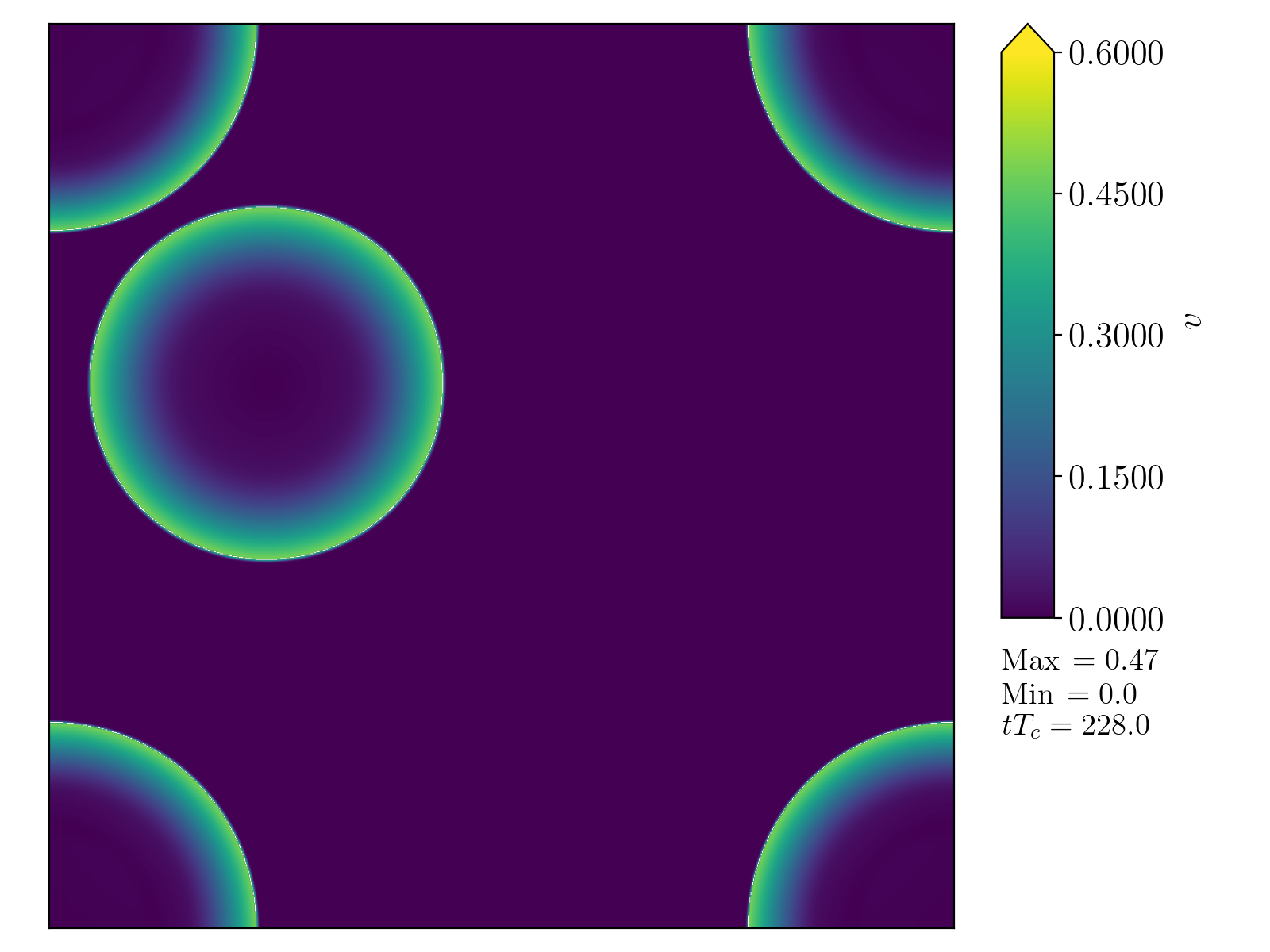}}
  \hfill
  \subfigure
  {\includegraphics[width=0.32\textwidth,clip]{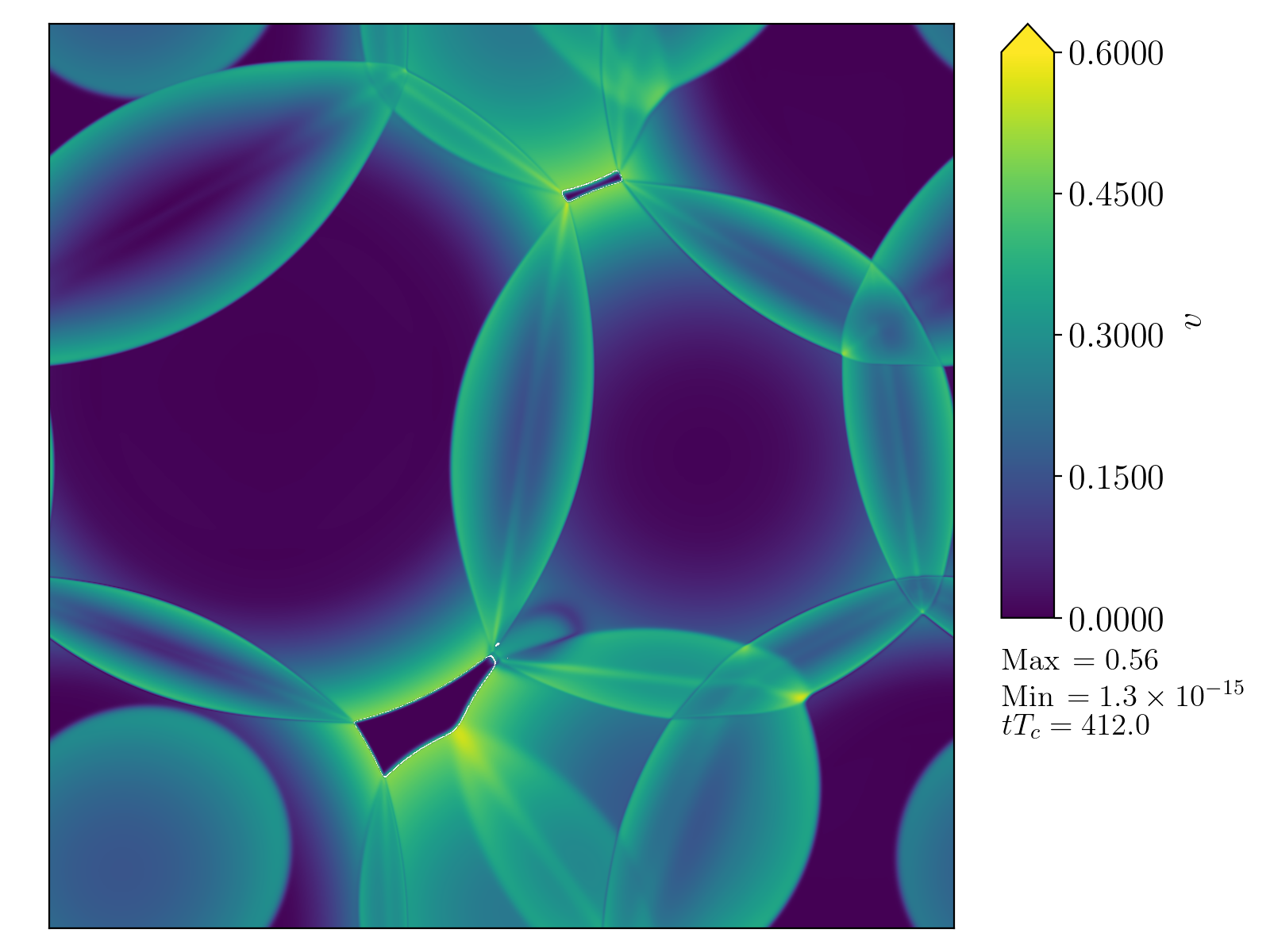}}
  \hfill
  \subfigure
  {\includegraphics[width=0.32\textwidth,clip]{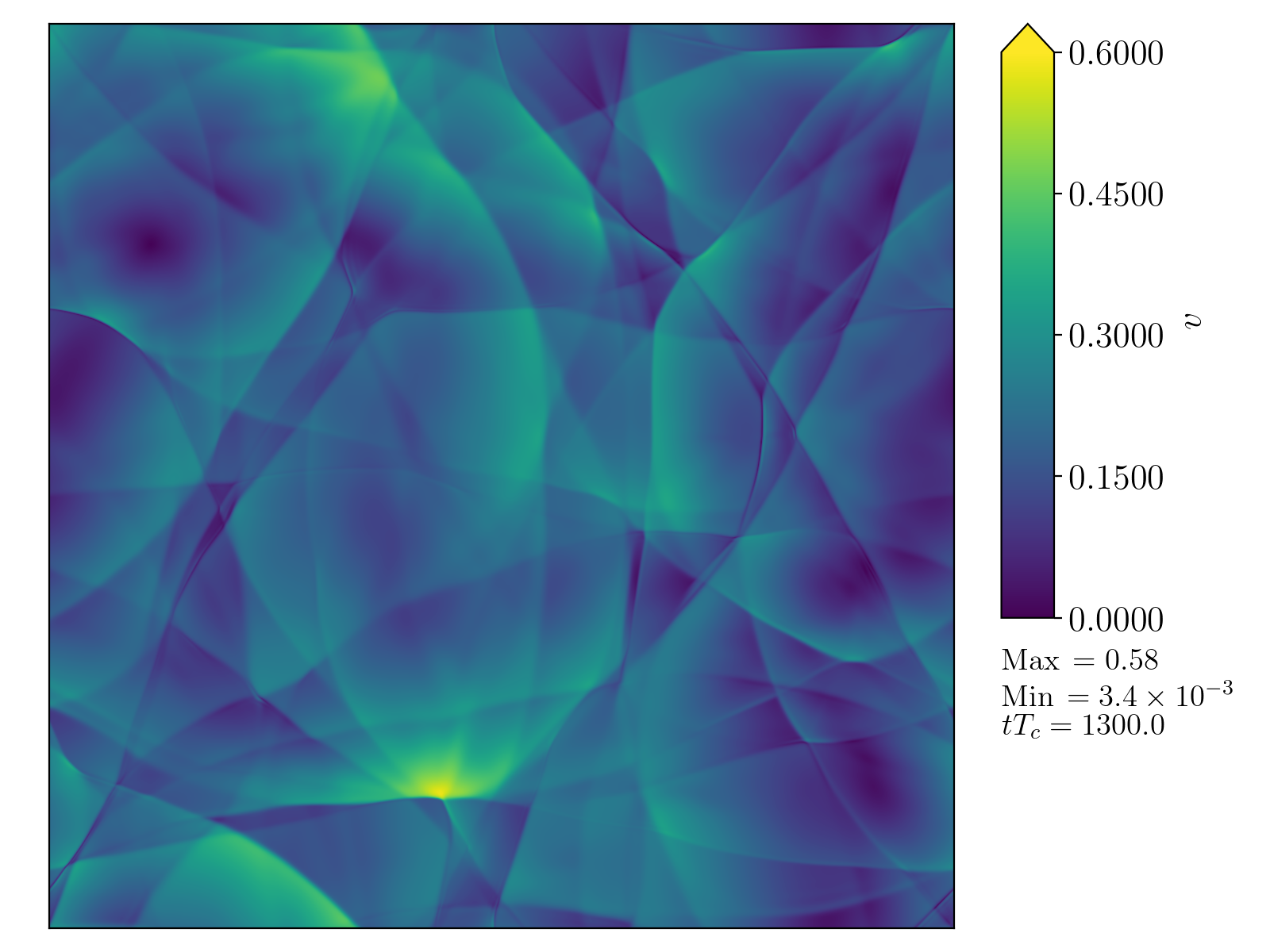}}
  \hfill
  \\
  \hfill
  \subfigure
  {\includegraphics[width=0.32\textwidth,clip]{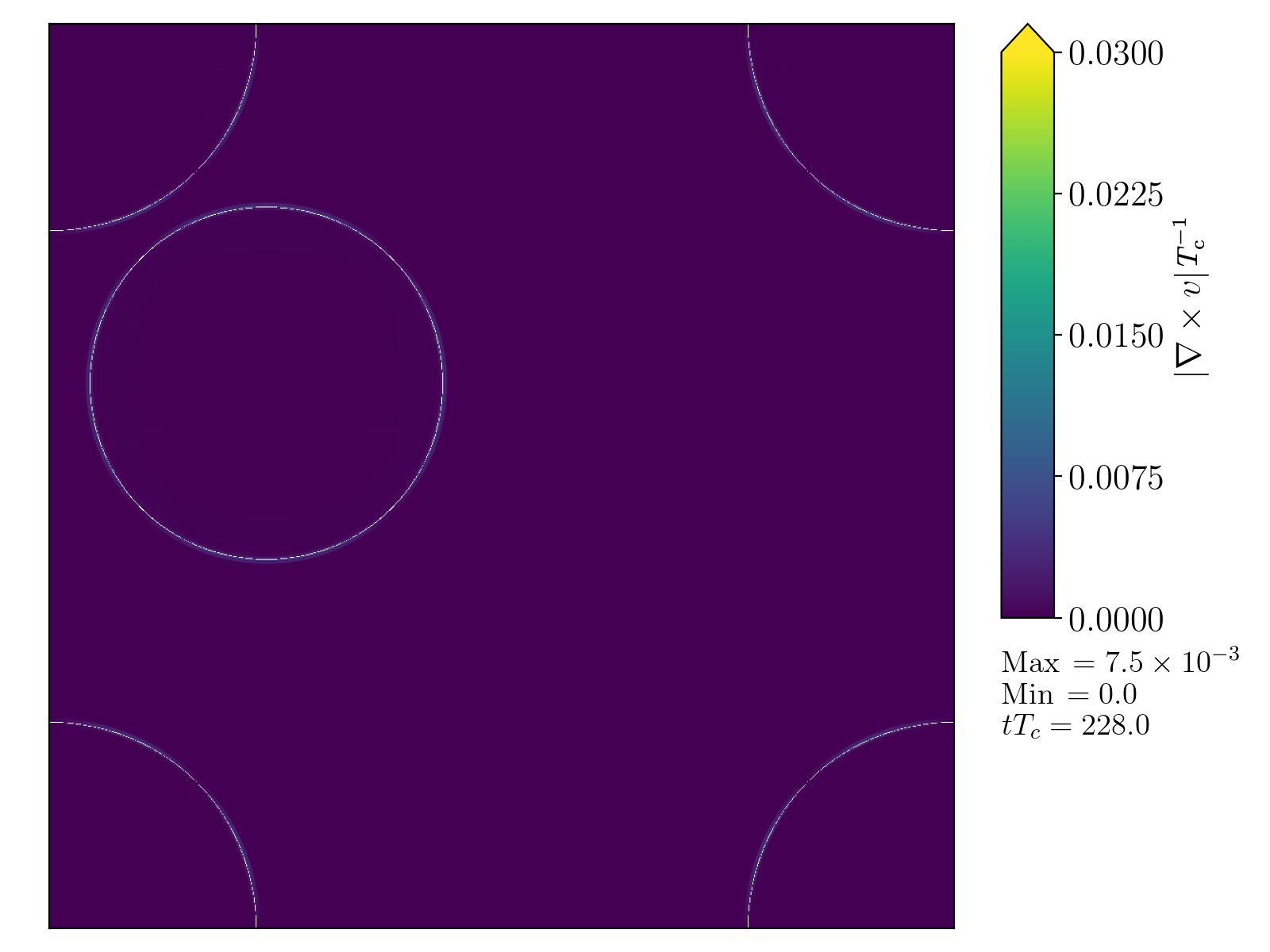}}
  \hfill
  \subfigure
  {\includegraphics[width=0.32\textwidth,clip]{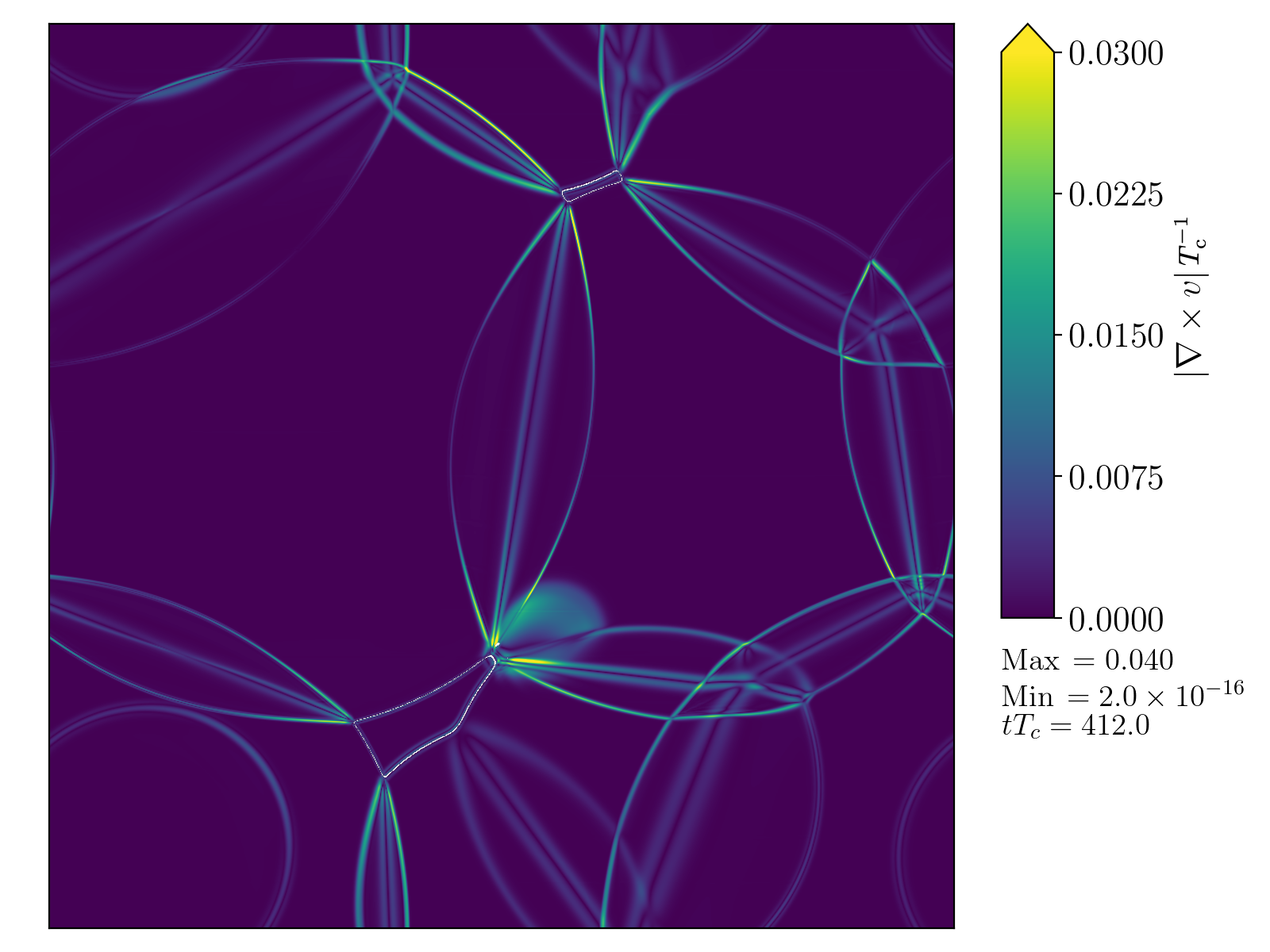}}
  \hfill
  \subfigure
  {\includegraphics[width=0.32\textwidth,clip]{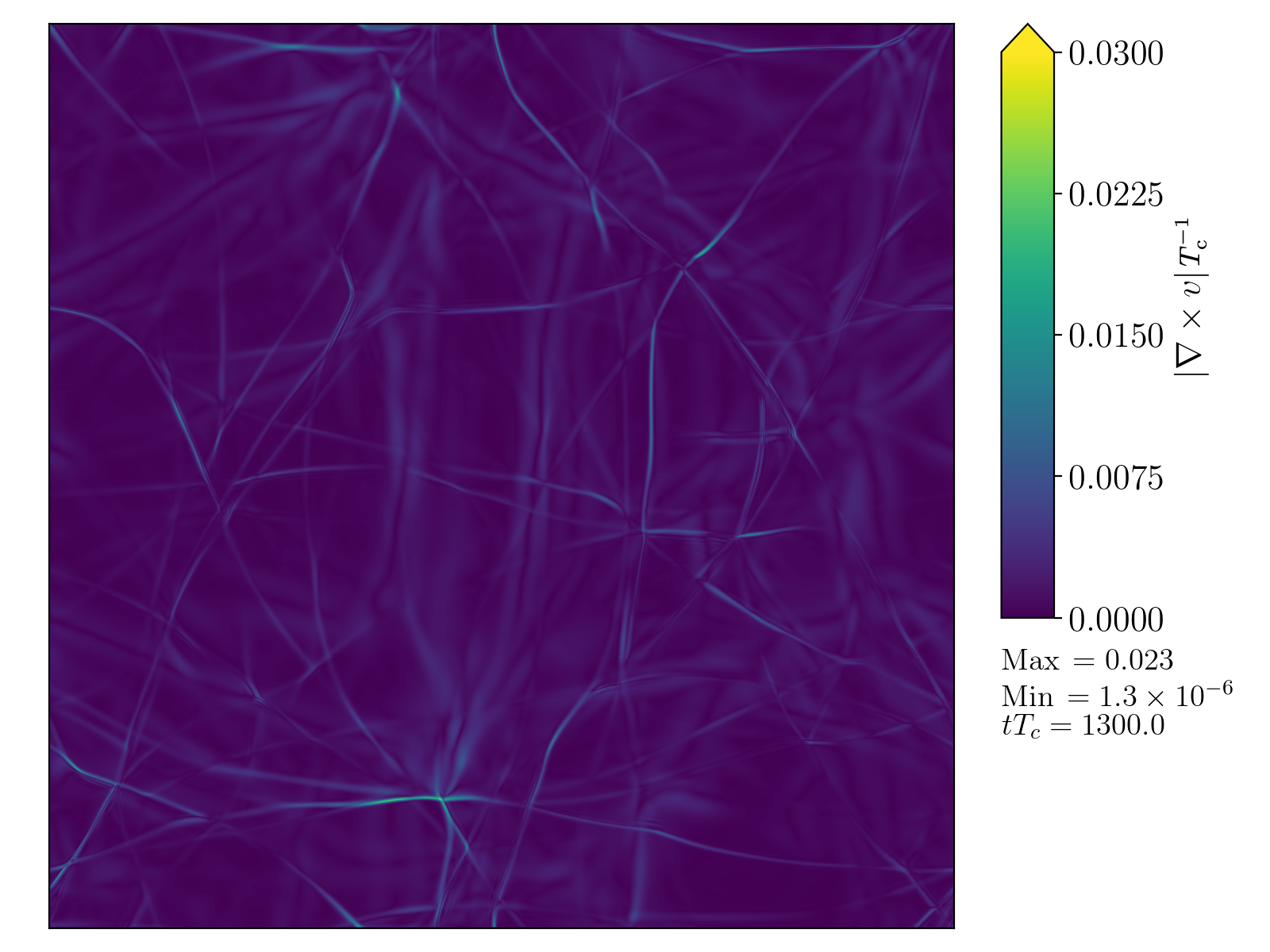}}
  \hfill
  \caption{Slices through $(0,y,z)$ for a simulation with $\vwass=0.92$,
    $\StrParB=0.5$, corresponding to a detonation.  In the top row we
    plot the temperature $T/\Tc$. The midpoint of this colormap
    corresponds to $\TN$. The middle row shows the fluid velocity
    $v$. The bottom row shows the vorticity
    $\left| \nabla \times v\right|$. The bubble walls are shaded in
    black for the top row, and white for the middle and bottom row.}
  \label{fig:slices-det-supl}
\end{figure}

In this supplemental material we include various stills taken from
movies of our simulations of strong phase transitions in the early
universe, which can be seen in Fig.~\ref{fig:slices-def-supl} and
Fig.~\ref{fig:slices-det-supl}. The movies these stills have been
taken from can be found in a Vimeo album \cite{external}.

\onecolumngrid

\twocolumngrid

\onecolumngrid

\twocolumngrid

\FloatBarrier

\begin{widetext}
\subsection{{Simulation parameters and measurements}}

\FloatBarrier

\onecolumngrid

\begin{table}
  \centering
  \resizebox{\textwidth}{!}{
    \begin{ruledtabular}
  \begin{tabular}{D{.}{.}{1.2} D{.}{.}{1.4} D{.}{.}{1.2}  D{.}{.}{1.5} D{.}{.}{1.3} D{.}{.}{1.4} D{.}{.}{1.4}
    D{.}{.}{1.6} D{.}{.}{1.6}  D{.}{.}{1.6} D{.}{.}{1.6}}
    \multicolumn{1}{c}{$\vwass$} & \multicolumn{1}{c}{$\StrParB$} 
    & \multicolumn{1}{c}{$\TN/\Tc$} & \multicolumn{1}{c}{$t_\mathrm{fin}\Tc$}
    & \multicolumn{1}{c}{$\eta/\Tc$}
    & \multicolumn{1}{c}{$\Ubfmax$} & \multicolumn{1}{c}{$\UbfExp$}
    & \multicolumn{1}{c}{$\VbPerp^2$} & \multicolumn{1}{c}{$\VbTot^2$}  
    & \multicolumn{1}{c}{$\overline{\left(\dfrac{\OmGW}{\HN t}\right)}\dfrac{1}{ \HN \Rbc }$}
    & \multicolumn{1}{c}{$\dfrac{\OmGWExp}{\HN t}\dfrac{1}{ \HN \Rbc }$} \\
    \hline
0.24 & 0.050 & 0.45 &  4.8 \times 10^{3}  & 1.2 & 0.036 & 0.039 &  6.4 \times 10^{-5}  &  1.3 \times 10^{-3}  &  4.6 \times 10^{-8}  &  1.2 \times 10^{-7} \\
0.24 & 0.073 & 0.41 &  4.8 \times 10^{3}  & 1.3 & 0.048 & 0.055 &  1.5 \times 10^{-4}  &  2.3 \times 10^{-3}  &  1.4 \times 10^{-7}  &  5.0 \times 10^{-7} \\
0.24 & 0.11 & 0.37 &  4.8 \times 10^{3}  & 1.5 & 0.063 & 0.082 &  3.7 \times 10^{-4}  &  4.1 \times 10^{-3}  &  4.1 \times 10^{-7}  &  2.4 \times 10^{-6} \\
0.24 & 0.16 & 0.33 &  4.8 \times 10^{3}  & 1.8 & 0.074 & 0.12 &  9.1 \times 10^{-4}  &  5.7 \times 10^{-3}  &  6.0 \times 10^{-7}  &  9.5 \times 10^{-6} \\
0.24 & 0.23 & 0.30 &  4.8 \times 10^{3}  & 2.4 & 0.075 & 0.16 &  2.0 \times 10^{-3}  &  5.9 \times 10^{-3}  &  4.6 \times 10^{-7}  &  3.3 \times 10^{-5} \\
0.24 & 0.34 & 0.28 &  4.8 \times 10^{3}  & 5.1 & 0.066 & 0.22 &  2.9 \times 10^{-3}  &  4.4 \times 10^{-3}  &  2.3 \times 10^{-7}  &  1.2 \times 10^{-4} \\
 \hline
0.44 & 0.0050 & 0.79 &  2.4 \times 10^{3}  & 0.34 & 0.0083 & 0.0069 &  1.7 \times 10^{-7}  &  6.9 \times 10^{-5}  &  1.0 \times 10^{-10}  &  1.2 \times 10^{-10} \\
0.44 & 0.016 & 0.59 &  2.4 \times 10^{3}  & 0.52 & 0.025 & 0.021 &  1.4 \times 10^{-6}  &  6.0 \times 10^{-4}  &  9.1 \times 10^{-9}  &  1.1 \times 10^{-8} \\
0.44 & 0.050 & 0.45 &  2.4 \times 10^{3}  & 0.66 & 0.066 & 0.061 &  1.6 \times 10^{-5}  &  4.3 \times 10^{-3}  &  5.6 \times 10^{-7}  &  7.2 \times 10^{-7} \\
0.44 & 0.073 & 0.41 &  2.4 \times 10^{3}  & 0.71 & 0.090 & 0.085 &  4.0 \times 10^{-5}  &  7.8 \times 10^{-3}  &  2.0 \times 10^{-6}  &  2.8 \times 10^{-6} \\
0.44 & 0.11 & 0.37 &  2.4 \times 10^{3}  & 0.81 & 0.12 & 0.12 &  1.1 \times 10^{-4}  &  1.3 \times 10^{-2}  &  7.0 \times 10^{-6}  &  1.0 \times 10^{-5} \\
0.44 & 0.16 & 0.33 &  2.4 \times 10^{3}  & 0.94 & 0.15 & 0.16 &  3.0 \times 10^{-4}  &  2.1 \times 10^{-2}  &  2.0 \times 10^{-5}  &  3.7 \times 10^{-5} \\
0.44 & 0.23 & 0.30 &  2.4 \times 10^{3}  & 1.2 & 0.18 & 0.22 &  1.0 \times 10^{-3}  &  3.2 \times 10^{-2}  &  4.3 \times 10^{-5}  &  1.3 \times 10^{-4} \\
0.44 & 0.34 & 0.28 &  2.4 \times 10^{3}  & 1.7 & 0.18 & 0.30 &  3.6 \times 10^{-3}  &  3.6 \times 10^{-2}  &  5.2 \times 10^{-5}  &  4.2 \times 10^{-4} \\
0.44 & 0.50 & 0.25 &  2.4 \times 10^{3}  & 3.5 & 0.19 & 0.39 &  8.0 \times 10^{-3}  &  4.0 \times 10^{-2}  &  5.8 \times 10^{-5}  &  1.2 \times 10^{-3} \\
 \hline
0.56 & 0.050 & 0.45 &  2.8 \times 10^{3}  & 0.53 & 0.080 & 0.075 &  1.3 \times 10^{-5}  &  5.6 \times 10^{-3}  &  1.1 \times 10^{-6}  &  1.7 \times 10^{-6} \\
0.56 & 0.073 & 0.41 &  2.8 \times 10^{3}  & 0.59 & 0.10 & 0.10 &  3.5 \times 10^{-5}  &  9.6 \times 10^{-3}  &  3.6 \times 10^{-6}  &  5.3 \times 10^{-6} \\
0.56 & 0.11 & 0.37 &  2.8 \times 10^{3}  & 0.67 & 0.14 & 0.14 &  9.2 \times 10^{-5}  &  1.6 \times 10^{-2}  &  1.2 \times 10^{-5}  &  2.0 \times 10^{-5} \\
0.56 & 0.16 & 0.33 &  2.8 \times 10^{3}  & 0.76 & 0.18 & 0.19 &  2.4 \times 10^{-4}  &  2.7 \times 10^{-2}  &  3.6 \times 10^{-5}  &  6.4 \times 10^{-5} \\
0.56 & 0.23 & 0.30 &  2.8 \times 10^{3}  & 0.90 & 0.22 & 0.25 &  5.8 \times 10^{-4}  &  4.3 \times 10^{-2}  &  9.3 \times 10^{-5}  &  2.0 \times 10^{-4} \\
0.56 & 0.34 & 0.28 &  2.8 \times 10^{3}  & 1.2 & 0.27 & 0.33 &  1.5 \times 10^{-3}  &  6.4 \times 10^{-2}  &  2.1 \times 10^{-4}  &  6.2 \times 10^{-4} \\
0.56 & 0.50 & 0.25 &  2.8 \times 10^{3}  & 1.7 & 0.28 & 0.43 &  5.2 \times 10^{-3}  &  7.8 \times 10^{-2}  &  3.1 \times 10^{-4}  &  1.8 \times 10^{-3} \\
0.56 & 0.67 & 0.23 &  2.8 \times 10^{3}  & 2.9 & 0.30 & 0.51 &  1.1 \times 10^{-2}  &  9.0 \times 10^{-2}  &  3.0 \times 10^{-4}  &  3.7 \times 10^{-3} \\
 \hline
0.82 & 0.0050 & 0.79 &  2.8 \times 10^{3}  & 0.11 & 0.0064 & 0.0066 &  2.3 \times 10^{-7}  &  4.0 \times 10^{-5}  &  4.8 \times 10^{-11}  & 1.0  \times 10^{-10} \\
0.82 & 0.016 & 0.59 &  2.8 \times 10^{3}  & 0.16 & 0.019 & 0.020 &  2.2 \times 10^{-6}  &  3.6 \times 10^{-4}  &  4.3 \times 10^{-9}  &  9.1 \times 10^{-9} \\
0.82 & 0.050 & 0.45 &  2.8 \times 10^{3}  & 0.18 & 0.055 & 0.061 &  1.5 \times 10^{-5}  &  2.8 \times 10^{-3}  &  2.9 \times 10^{-7}  &  7.6 \times 10^{-7} \\
0.82 & 0.073 & 0.41 &  2.8 \times 10^{3}  & 0.19 & 0.076 & 0.088 &  2.6 \times 10^{-5}  &  5.2 \times 10^{-3}  &  1.1 \times 10^{-6}  &  3.2 \times 10^{-6} \\
0.82 & 0.11 & 0.37 &  2.8 \times 10^{3}  & 0.20 & 0.11 & 0.13 &  4.7 \times 10^{-5}  &  9.4 \times 10^{-3}  &  4.1 \times 10^{-6}  &  1.4 \times 10^{-5} \\
0.82 & 0.16 & 0.33 &  2.8 \times 10^{3}  & 0.22 & 0.15 & 0.18 &  8.4 \times 10^{-5}  &  1.6 \times 10^{-2}  &  1.2 \times 10^{-5}  &  5.5 \times 10^{-5} \\
 \hline
0.92 & 0.0050 & 0.79 &  2.4 \times 10^{3}  & 0.053 & 0.0051 & 0.0051 &  3.2 \times 10^{-7}  &  2.6 \times 10^{-5}  &  2.0 \times 10^{-11}  &  3.6 \times 10^{-11} \\
0.92 & 0.016 & 0.59 &  2.4 \times 10^{3}  & 0.086 & 0.015 & 0.016 &  3.4 \times 10^{-6}  &  2.4 \times 10^{-4}  &  1.9 \times 10^{-9}  &  3.6 \times 10^{-9} \\
0.92 & 0.050 & 0.45 &  2.4 \times 10^{3}  & 0.099 & 0.045 & 0.049 &  2.2 \times 10^{-5}  &  1.9 \times 10^{-3}  &  1.5 \times 10^{-7}  &  3.0 \times 10^{-7} \\
0.92 & 0.073 & 0.41 &  2.4 \times 10^{3}  & 0.10 & 0.064 & 0.070 &  3.6 \times 10^{-5}  &  3.7 \times 10^{-3}  &  6.0 \times 10^{-7}  &  1.3 \times 10^{-6} \\
0.92 & 0.11 & 0.37 &  2.4 \times 10^{3}  & 0.10 & 0.087 & 0.10 &  5.8 \times 10^{-5}  &  6.9 \times 10^{-3}  &  2.4 \times 10^{-6}  &  6.0 \times 10^{-6} \\
0.92 & 0.16 & 0.33 &  2.4 \times 10^{3}  & 0.11 & 0.12 & 0.14 &  8.8 \times 10^{-5}  &  1.2 \times 10^{-2}  &  8.4 \times 10^{-6}  &  2.3 \times 10^{-5} \\
0.92 & 0.23 & 0.30 &  2.4 \times 10^{3}  & 0.11 & 0.16 & 0.20 &  1.4 \times 10^{-4}  &  2.0 \times 10^{-2}  &  2.6 \times 10^{-5}  &  8.0 \times 10^{-5} \\
0.92 & 0.34 & 0.28 &  2.4 \times 10^{3}  & 0.12 & 0.21 & 0.27 &  2.5 \times 10^{-4}  &  3.2 \times 10^{-2}  &  8.1 \times 10^{-5}  &  2.9 \times 10^{-4} \\
0.92 & 0.50 & 0.25 &  2.4 \times 10^{3}  & 0.13 & 0.28 & 0.36 &  4.6 \times 10^{-4}  &  4.9 \times 10^{-2}  &  2.2 \times 10^{-4}  &  9.3 \times 10^{-4} \\
0.92 & 0.60 & 0.24 &  2.4 \times 10^{3}  & 0.14 & 0.32 & 0.42 &  6.3 \times 10^{-4}  &  5.8 \times 10^{-2}  &  3.5 \times 10^{-4}  &  1.6 \times 10^{-3} \\
0.92 & 0.67 & 0.23 &  2.4 \times 10^{3}  & 0.15 & 0.34 & 0.45 &  7.8 \times 10^{-4}  &  6.5 \times 10^{-2}  &  4.5 \times 10^{-4}  &  2.2 \times 10^{-3} \\
\hline
  \end{tabular}
\end{ruledtabular}
}
\caption{Key simulation parameters and measured quantities used to generate the graphs in this
  paper. Note that $\overline{\left(\dfrac{\OmGW}{\HN t}\right)}$
  signifies that we average the quantity inside the brackets over the final
  $\Delta t= 2 \Rbc$ of the simulation.}
\label{table:alldata}
\end{table}
\twocolumngrid
\end{widetext}

\end{document}